%% file: wp2.tex
\documentclass[final,times,5p]{elsarticle}

\usepackage{placeins}
\usepackage[english]{babel}
\usepackage[utf8]{inputenc}
\usepackage{amsmath}
\usepackage{amssymb}
\usepackage{amsthm}
\usepackage{eurosym}
\usepackage{bm}  %
\usepackage{color}

\usepackage[colorlinks]{hyperref}  %

\usepackage[capitalise]{cleveref}
\usepackage{caption}  %
\Crefname{figure}{Fig.}{Figs.}
\Crefname{equation}{Eq.}{Eqs.}
\usepackage{longtable}
\usepackage[T1]{fontenc}  %
\usepackage{booktabs}
\usepackage{multicol}  %
\usepackage{multirow}  %
\usepackage{nomencl}  %
\usepackage{makecell}  %
\usepackage{lipsum}
\usepackage{subfig}
\usepackage[nohyperlinks, nolist]{acronym}  %
\usepackage[export]{adjustbox} %
\usepackage[percent]{overpic}

\makenomenclature
\renewcommand*\nompreamble{\begin{multicols}{2}}  %
\renewcommand*\nompostamble{\end{multicols}}    %

\usepackage{etoolbox}

\renewcommand\nomgroup[1]{
	\item[\bfseries \itshape
	\ifstrequal{#1}{A}{Acronyms / abbreviations / superscripts}{
		\ifstrequal{#1}{Y}{Symbols}{
			\ifstrequal{#1}{I}{Indices / sets}{
	}}}]}

\newcommand{\scen}[1]{{\texttt{#1}}}
\newcommand{\scenbf}[1]{{\fontfamily{qag} \textbf{#1}}}

\usepackage{framed}  %
\usepackage[final]{pdfpages}  %
\usepackage{microtype}  %
\allowdisplaybreaks

\usepackage{tikz}
\usetikzlibrary{shapes, arrows.meta, positioning}

\usepackage{pifont}
\newcommand{\cmark}{\ding{51}}
\newcommand{\xmark}{\color{gray!20} \ding{55}}

\newcommand{\RomanNumeralCaps}[1]{\MakeUppercase{\romannumeral #1}}

\usepackage{enumitem}  %
\setlist{nosep}

\journal{Applied Energy}

\begin{document}

\captionsetup[figure]{labelfont={bf},labelformat={default},labelsep=period,name={Fig.}}
\captionsetup[table]{labelfont={bf},labelsep=newline}

\pdfinfo{
	/Author(Markus Fleschutz)
	/Title(wp2)
	/Subject(wp2)
}

\input{chapter/abstract}  %
\input{chapter/nomencl}

\input{chapter/1_intro}

\input{chapter/2_case}
\input{chapter/3_model}

\input{chapter/4_eval}
\input{chapter/5_scens}
\input{chapter/6_res}

\input{chapter/7_concl}

\input{chapter/credits}

\bibliographystyle{elsarticle-num}
\bibliography{ref,from_citavi}

\appendix
\input{chapter/appendix}

\end{document}

%% file: chapter/abstract.tex
\begin{frontmatter}

\title{From prosumer to flexumer: Case study on the value of flexibility in decarbonizing the multi-energy system of a manufacturing company}

\author[HKA,MTU]{Markus Fleschutz}
\author[HKA]{Markus Bohlayer}
\author[HKA]{Marco Braun}
\author[MTU]{Michael D. Murphy}
\ead{michaeld.murphy@mtu.ie}
\address[HKA]{Institute of Refrigeration, Air-Conditioning, and Environmental Engineering, University of Applied Sciences Karlsruhe, Moltkestraße 30, 76133 Karlsruhe, Germany}
\address[MTU]{Department of Process, Energy and Transport Engineering, Munster Technological University, Bishopstown, Cork, Ireland}

\date{\today}
\definecolor{green}{RGB}{2, 89, 0}

\begin{abstract}
\noindent
Digitalization and sector coupling enable companies to turn into flexumers.
By using the flexibility of their multi-energy system (MES), they reduce costs and carbon emissions while stabilizing the electricity system.
However, to identify the necessary investments in energy conversion and storage technologies to leverage demand response (DR) potentials, companies need to assess the value of flexibility.
Therefore, this study quantifies the flexibility value of a production company's MES by optimizing the synthesis, design, and operation of a decarbonizing MES considering self-consumption optimization, peak shaving, and integrated DR based on hourly prices and carbon emission factors (CEFs).
The detailed case study of a beverage company in northern Germany considers vehicle-to-X of powered industrial trucks, power-to-heat on multiple temperatures, wind turbines, photovoltaic systems, and energy storage systems (thermal energy, electricity, and hydrogen).
We propose and apply novel data-driven metrics to evaluate the intensity of price-based and CEF-based DR.
The results reveal that flexibility usage reduces decarbonization costs (by 19--80\% depending on electricity and carbon removal prices), total annual costs, operating carbon emissions, energy-weighted average prices and CEFs, and fossil energy dependency.
The results also suggest that a net-zero operational carbon emission MES requires flexibility, which, in an economic case, is provided by a combination of different flexible technologies and storage systems that complement each other.
While the value of flexibility depends on various market and consumer-specific factors such as electricity or carbon removal prices, this study highlights the importance of demand flexibility for the decarbonization of MESs.

{\centering
\noindent\includegraphics[width=510px]{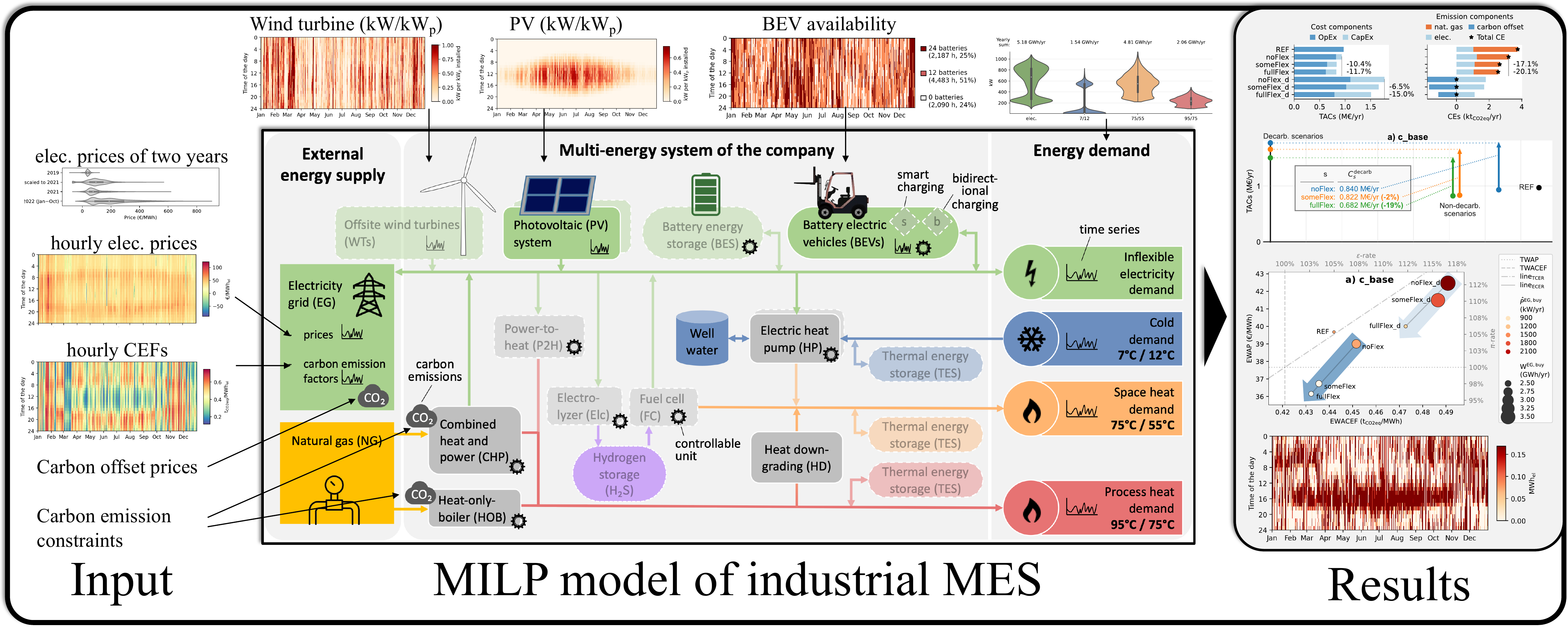}}
\end{abstract}

\end{frontmatter}

\section*{Highlights}
\begin{itemize}\setlength \itemsep{0.1em}
	\item Detailed case study of optimal net-zero multi-energy system design using MILP.
	\item Quantified the flexibility value of an distributed energy system for decarbonization.
	\item Novel metrics to evaluate price-based and emission-based demand response.
	\item Considered power-to-hydrogen, vehicle-to-X, and multi-temperature power-to-heat.
	\item Compare the value of using existing versus new flexibility.
	\item Flexibility of distributed energy resources reduced decarbonization costs by 19--80\%.
\end{itemize}

%% file: chapter/nomencl.tex
\nomenclature[A]{BES}{Battery energy storage}
\nomenclature[A]{BEV}{Battery electric vehicle}
\nomenclature[A]{C\&I}{Commercial and industrial}
\nomenclature[A]{CapEx}{Capital expenditures}
\nomenclature[A]{capx/capn}{Nominal capacity of existing/new assets}
\nomenclature[A]{CE}{Carbon emission}
\nomenclature[A]{CEF}{Carbon emission factor}
\nomenclature[A]{CHP}{Combined heat and power}
\nomenclature[A]{cond}{Condensation}
\nomenclature[A]{COP}{Coefficient of performance}
\nomenclature[A]{DAC}{Direct air capture}
\nomenclature[A]{DR}{Demand response}
\nomenclature[A]{DRAF}{Demand Response Analysis Framework}
\nomenclature[A]{ECER}{Energy-based cost-emission ratio}
\nomenclature[A]{EG}{Electricity grid}
\nomenclature[A]{Elc}{Electrolyzer}
\nomenclature[A]{elec}{Electricity}
\nomenclature[A]{eva}{Evaporation}
\nomenclature[A]{EWACEF}{Energy-weighted average carbon emission factor}
\nomenclature[A]{EWAP}{Energy-weighted average price}
\nomenclature[A]{FC}{Fuel cell}
\nomenclature[A]{H\textsubscript{2}}{Hydrogen}
\nomenclature[A]{H\textsubscript{2}S}{Hydrogen storage}
\nomenclature[A]{HD}{Heat downgrading}
\nomenclature[A]{HOB}{Heat-only boiler}
\nomenclature[A]{HP}{Electrical heat pump}
\nomenclature[A]{inv}{Investment}
\nomenclature[A]{KPI}{Key performance indicator}
\nomenclature[A]{MEF}{Marginal emission factor}
\nomenclature[A]{MES}{Multi-energy system}
\nomenclature[A]{MILP}{Mixed-integer linear programming}
\nomenclature[A]{NF}{Network fees}
\nomenclature[A]{NG}{Natural gas}
\nomenclature[A]{OC}{Own-consumption}
\nomenclature[A]{OpEx}{Operating expenses}
\nomenclature[A]{P2H}{Power-to-heat}
\nomenclature[A]{PV}{Photovoltaic}
\nomenclature[A]{RMI}{Repair, maintenance, and inspection}
\nomenclature[A]{TAC}{Total annualized cost}
\nomenclature[A]{TCER}{Time-based cost-emission ratio}
\nomenclature[A]{TES}{Thermal energy storage}
\nomenclature[A]{TWACEF}{Time-weighted average carbon emission factor}
\nomenclature[A]{TWAP}{Time-weighted average price}
\nomenclature[A]{V2X}{Vehicle-to-everything}
\nomenclature[A]{vRES}{Variable renewable energy source}
\nomenclature[A]{WT}{Wind turbine}
\nomenclature[A]{XEF}{Grid-mix emission factor}
\nomenclature[I]{$b \in \mathcal{B}$}{Types of battery electric vehicle}
\nomenclature[I]{$c \in \mathcal{C}$}{Condensing temperature levels}
\nomenclature[I]{$h \in \mathcal{H}$}{Heating temperature levels}
\nomenclature[I]{$i \in \mathcal{I}$}{Technology components}
\nomenclature[I]{$n \in \mathcal{N}$}{Cooling temperature levels}
\nomenclature[I]{$t \in \mathcal{T}$}{Time steps}
\nomenclature[Y]{$\dot{Q}$}{Thermal energy flow}
\nomenclature[Y]{$\eta$}{Efficiency coefficient}
\nomenclature[Y]{$\gamma$}{Coefficient related to capacity}
\nomenclature[Y]{$\pi_s$, $\varepsilon_s$, $\omega_s$}{Normalized flexibility metrics}
\nomenclature[Y]{$\varepsilon$}{Specific carbon emissions}
\nomenclature[Y]{$C, c$}{Annual/specific costs}
\nomenclature[Y]{$c$}{Specific costs}
\nomenclature[Y]{$E$}{Electrical energy}
\nomenclature[Y]{$n$}{Economic life in years}
\nomenclature[Y]{$P$}{Electrical power}
\nomenclature[Y]{$R$}{Annual revenue}
\nomenclature[Y]{$r$}{Discount rate}
\nomenclature[Y]{$Y$, $y$}{Binary variable}
\nomenclature[Y]{$z$}{Switch for flexibility or investment option}

\begin{table*}[!t]
	\begin{framed}
		\printnomenclature[1.4cm]
	\end{framed}
\end{table*}

%% file: chapter/1_intro.tex
\section{Introduction}

\subsection{Motivation}

In the course of the European Green Deal~\cite{EUGreenDeal}, the European Union is aiming for climate neutrality by 2050.
As an interim target, a 55\% carbon emission (CE) reduction by 2030 compared to 1990 levels was set.
According to the Emissions Gap Report 2021 of the United Nations Environment Programme~\cite{UnepGapReport2021}, these targets are not sufficient to meet the 1.5\,\textdegree C target of the Paris Agreement.
However, they are still ambitious and just one example of the global pursuit of deep decarbonization.

To meet these challenging targets and to stay within sustainable biomass limits, a high degree of electrification and hydrogen (H\textsubscript{2}) integration is needed in the energy system~\cite{Mortensen2020} and the industry sector~\cite{Rissman.2020}.
Sector coupling such as the electrification of heat and mobility enables sectors that traditionally relied on fossil fuels, to benefit from the decarbonization possibilities of the electricity sector, e.g. photovoltaic (PV) and wind power~\cite{Knobloch.2020}.
In turn, the sector coupling allows transferring operational flexibility from other energy sectors (heat, mobility, H\textsubscript{2}) to the electricity sector~\cite{Arteconi.2016}.
This operational flexibility is necessary to compensate for the increasing shares of fluctuating renewable energy and the decommissioning of controllable thermal power plants.
Thus, carbon-free electricity systems require four times the operational flexibility of conventional systems~\cite{IEA2021}.
Demand response (DR) programs activate demand-side flexibility through different incentive systems.
Electrification gives rise to multi-energy systems (MESs) that can provide existing and newly created flexibility that can be used through DR~\cite{Mancarella.2014}.
Besides private households and heavy industry, also the commercial and industrial (C\&I) sector exhibits so far untapped flexibility potential~\cite{Papadaskalopoulos.2018}.

\begin{figure}[ht]
	\includegraphics[trim=2.8cm 1.3cm 0 0,clip,width=1.00\linewidth]{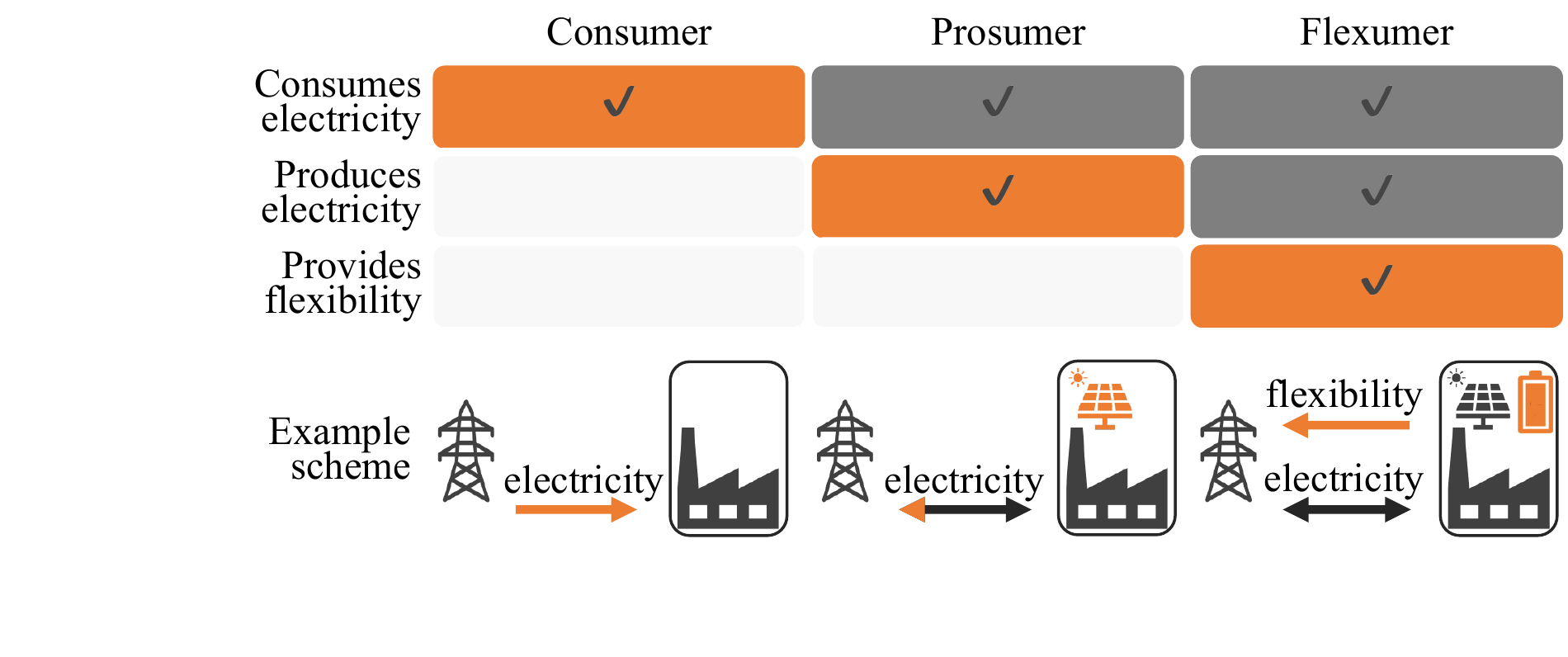}
	\caption{Differentiation of the terms consumer, prosumer, and flexumer.
		\label{fig:flexumer}}
\end{figure}

By untapping demand-side flexibility potentials, an electricity prosumer turns into a ``flexumer''~\cite{Jee2022}.
While a prosumer draws power from the grid and feeds in surpluses, a flexumer additionally controls the energy profiles through smart use of energy conversion and storage technologies.
\Cref{fig:flexumer} illustrates a comparison of the terms consumer, prosumer, and flexumer.

\begin{figure}[ht]
	\centering
	\includegraphics[width=1\linewidth]{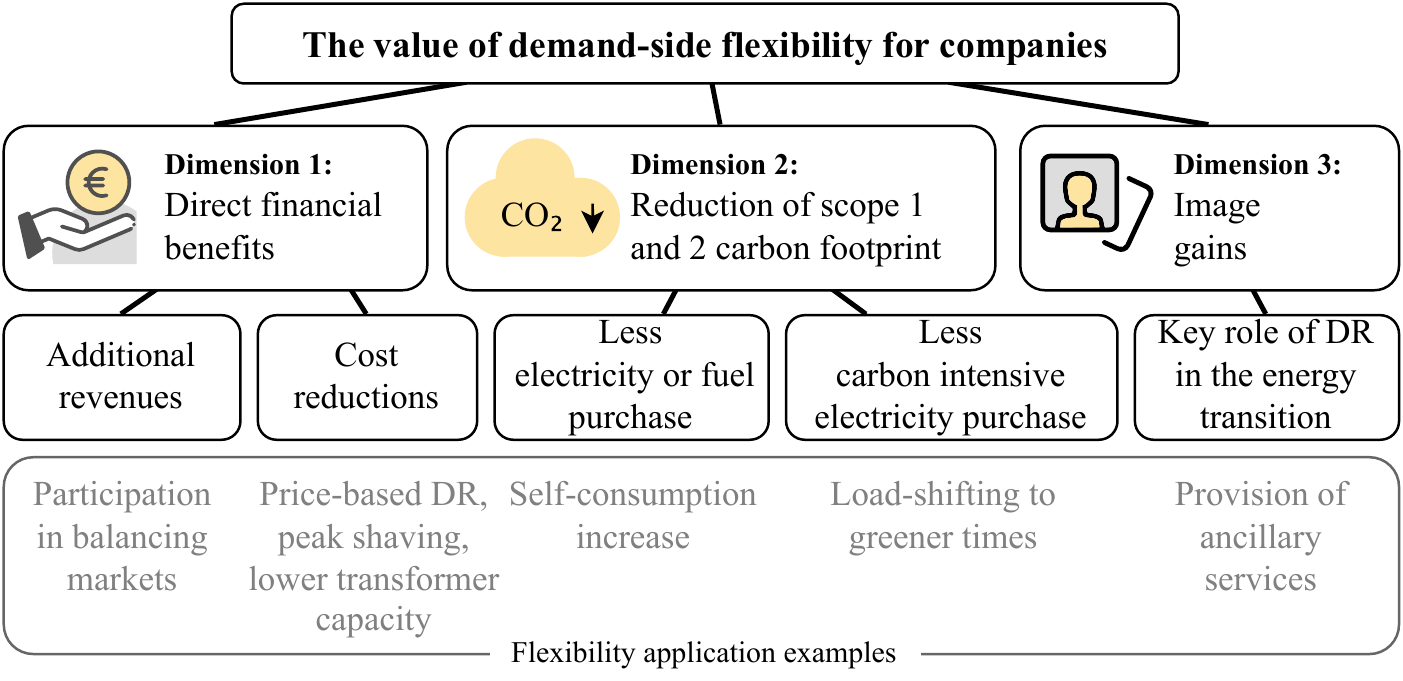}
	\caption{The value of demand-side flexibility for companies.
		\label{fig:value_of_flex}}
\end{figure}

\begin{figure}[ht]
	\includegraphics[width=1.00\linewidth]{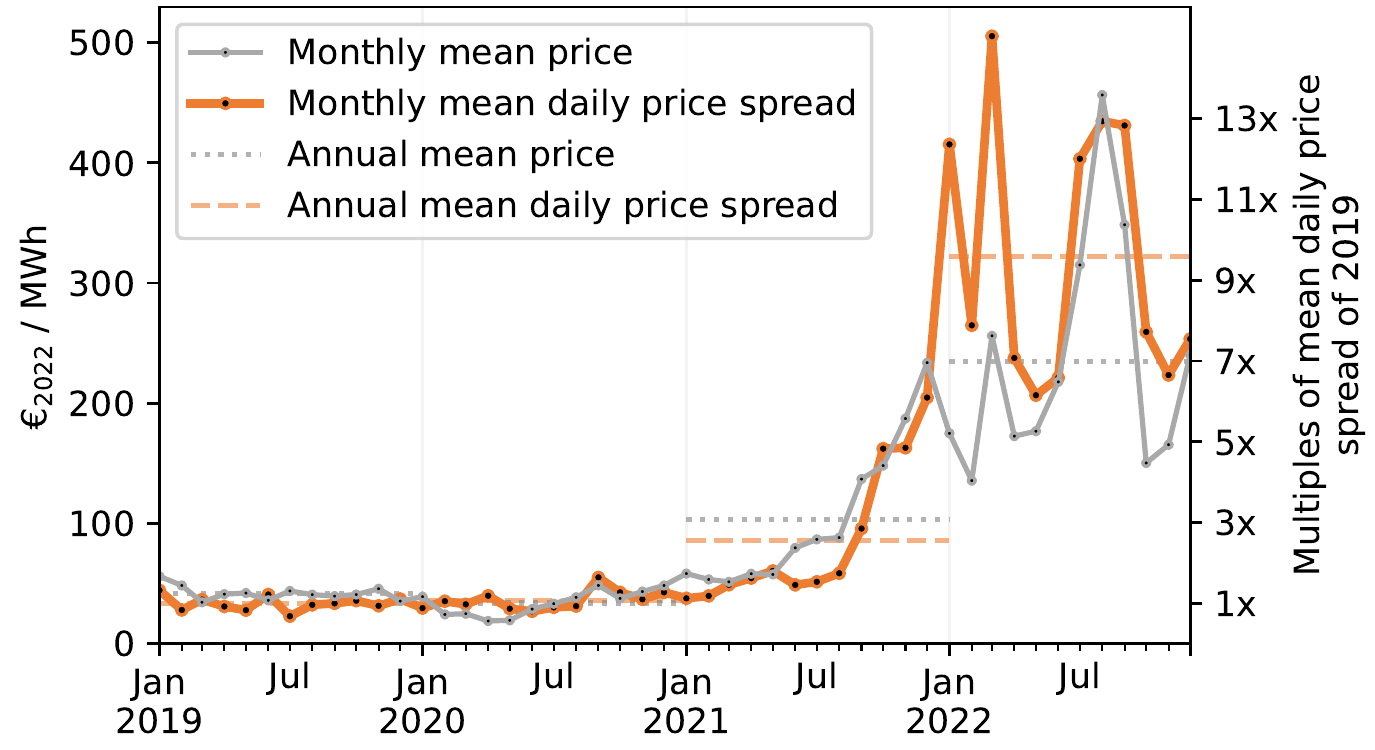}
	\caption{Day-ahead spot market prices and daily price spreads of Germany and Luxembourg for 2019--2022.
	Prices sourced from \cite{ENTSOE_TP}.
	\label{fig:spreads}}
\end{figure}

\subsection{Background}
Despite the worldwide expansion of variable renewable energy sources (vRESs), the wholesale electricity prices of the last decades offered few incentives for flexibility~\cite{Mays2021}.
However, since the second half of the year 2021, European electricity prices skyrocketed as a result of increased carbon and fuel prices.
The natural gas price were especially driven by the COVID-19 pandemic and the Russian invasion of Ukraine, but also by technical issues, longer winter, and lower wind and hydro generation~\cite{Aspenia2022}.
Additionally, in countries with substantial vRES shares, not only the level of wholesale electricity prices has increased, but also the price fluctuation due to the merit order effect~\cite{Sensfuss2008meritorder}.
For example, in Germany, mean daily price spreads (i.e., difference between lowest and highest price of each day) increased almost ten-fold between 2019 and 2022, see \Cref{fig:spreads}.
Interestingly, in 2022, on average, the daily price spread were 1.37x the price itself which means that the load shift of a unit of energy from the most expensive hour of the day to the cheapest paid for the average generation of 1.37 times that energy.

Despite the existence of fixed electricity price components such as network charges and taxes that hinder flexibility, demand-side energy flexibility can provide value for the electricity system but also for companies~\cite{Fleschutz2022igu}. %
There are several flexibility applications, e.g., ancillary service provision, energy-only market optimization, peer-to-peer energy trading, congestion management, self-consumption optimization, and peak-shaving~\cite{Buhl2021,Hamwi2021}.
The value of demand-side flexibility for a company can be divided into three dimensions, see \Cref{fig:value_of_flex}.
Dimension 1 is direct financial benefits through additional revenue streams or cost reductions.
Dimension 2 represents the reduction of the company's Scope \RomanNumeralCaps{1} and \RomanNumeralCaps{2} carbon footprint by reducing the purchased energy's volume or carbon-intensity or both.
Dimension 3 includes corporate image gains, since DR helps stabilize the electricity system, prevents blackouts, and integrates vRESs in the electricity system~\cite{Huang2019}.
With increasing renewable energy shares, DR and energy flexibility in general will gain reputation as key components of the energy transition~\cite{Li2022}.
By participating in DR, production companies provide an important overarching contribution that they can publicly communicate~\cite{Lashmar2022}.
Note that the environmental and the financial benefits may interact, e.g., financial benefits (Dimension 1) could finance clean technologies and reduce carbon footprint.
Analogously, carbon footprint reductions (Dimension 2) may reduce the need for expensive carbon removal and, therefore, imply financial benefits.  %

\subsection{Literature review}

There is a growing body of literature that evaluate the potentials from demand-side flexibility.
Muche et al.~\cite{Muche.2016} used mixed-integer linear programming (MILP) to analyze the revenue of participation of combined heat and power (CHP) plants in control reserve markets.
In~\cite{Wang.2018}, an intelligent energy management framework based on MILP with DR capability was proposed for industrial facilities.
Besides thermostatically controlled loads and battery electric vehicles (BEVs) also production processes and renewable generation were analyzed.
However, in both studies, CEs were not considered.
Kelley et al.~\cite{Kelley.2018} used MILP to model DR of air separation units.
They provided a novel scheduling framework that accounts for plant dynamics, however, only focussing on one DR application.
Scholz et al.~\cite{Scholz2022} analyzed the energy flexibility of forklift trucks within a MES but without considering investment decisions or vehicle-to-everything (V2X).

Summerbell et al.~\cite{Summerbell.2017} studied the cost and CE reduction potentials of a cement plant through price-based DR using real-time pricing.
The CE reduction potential was calculated from dynamic CEFs, however, investments were not considered.
A scheduling simulation resulted in electricity costs and electricity-derived CE reductions of 4\% each.
Alabi et al.~\cite{Alabi2021} proposed a multi-objective design and operation optimization model for carbon neutral MESs.
In a case study, they analyzed the effects of considering energy storage aging and DR.
However, the methodology was not applied to the C\&I sector but to a residential district in Hong Kong.
Ahmarinejad~\cite{Ahmarinejad2021}, presented a multi-objective planning and operation MILPs model considering DR.
However, in contrast to our study, DR was not evaluated in a real-world case study.
Petkov and Gabrielli~\cite{Petkov2020} presented a multi-objective design optimization formulation for MESs considering H\textsubscript{2} as a seasonal energy storage.
They concluded that for valid seasonal storage evaluations, optimization must be performed over at least 2,000 hourly resolved time steps.
Also the results indicated that a MES can only reach net zero operational CEs using power-to-H\textsubscript{2}.
Mansouri et al.~\cite{Mansouri2022} proposed an investment and operation optimization framework for MES considering DR.
The results indicate a 15.1\% reduction of operating expenses (OpEx) through price-based load shifting.
However, only Scope \RomanNumeralCaps{1} CEs were considered and with fixed carbon emission factors (CEFs).
In~\cite{Cremi.2020}, a design and operation optimization MILP problem was formulated and applied to the data of an exhibition center in south Italy.
Energy flexibility was considered through peak shaving with a BES.
CEs were assessed, but only with annual CEFs, and carbon neutrality was not analyzed.
Baumgärtner et al.~\cite{Baumgartner.2019} proposed an optimal planning and operation model considering hourly CEFs, price-based DR and carbon neutrality.
Costs and CEs were calculated comparing the use of annual versus hourly CEFs.
The significant differences in resulting costs and CEs highlight the importance of hourly CEFs.
The model was applied within a real-world case study of a chemical industry MES.
However, the value of flexibility was not explicitly quantified.
Furthermore, Jordehi~\cite{Jordehi.2019} concluded a comprehensive review on DR that more research effort should be placed on the C\&I sector, on more realistic optimization problems, and the analysis of environmental effects.

Recently several studies proposed flexibility metrics to quantify and characterize price-based DR.
In~\cite{Jee2022}, four metrics (price responsivity score, consistency score, flexible amount, and response time score) were proposed to identify and classify flexumers based on their electricity price and load.
However, CEFs were not considered and the metrics were not designed with real-time pricing in mind but with time-of-use tariffs.
Zhengyi et al.~\cite{Luo2022} reviewed DR quantification indicators for residential energy flexibility and classified them into direct and indirect indicators.
While direct indicators are directly related to the features of a building, indirect indicators also consider economic and environmental factors.
The mentioned indirect indicators are operation cost savings, operation cost reduction ratio, cost flexibility factor and carbon emission reductions.
However weighted averages based on electricity price and load are not mentioned.
The carbon emission reduction indicator is based on time-independent CEFs.
In~\cite{Zhang2019a}, several flexibility metrics are proposed, however, focusing only on incentive-based DR, so they cannot be applied to price or CEF-based DR.
Very recently, this study~\cite{Li} reviewed 156 articles and identified 48 data-driven energy flexibility key performance indicators (KPIs) for operational buildings.
The KPIs could be grouped based on whether a baseline energy demand (penalty-ignorant operation) is required.
For baseline-required KPIs, building performance data in both flexible and reference scenarios are necessary, while baseline-free KPIs can be calculated without a reference scenario.
Also, the development of new baseline-free KPIs and the consideration of non-engineering factors such as costs and CEs was identified as major research gaps.
The energy-weighted average electricity price -- mostly referred to as load-weighted or demand-weighted average price -- is a wide-spread metric, e.g., used in~\cite{Joung2013,Hirth2015b,Zappa2021}. %
However, to the best of the authors' knowledge, it has been used in the context of large-scale electricity systems and not to evaluate the DR intensity of local MESs.
Demand-weighted CEFs, in contrast, have only been used since hourly CEFs became widely available~\cite{Yang2022}.
However, in this study, we use energy-weighted average electricity prices to evaluate the effectiveness of price-based and CEF-based DR of an industrial company's MES and to compare it between scenarios.

In summary, the inclusion and evaluation of energy flexibility are gaining interest also for the C\&I sector.
However, the presented studies often do not consider investments, dynamic CEFs, or carbon neutrality as we do.
Others do not explicitly evaluate DR with metrics designed for this purpose.
Thus, to the best of the authors' knowledge, no study evaluates the flexibility of a production company's MES in a detailed case study considering investments and evaluating price-based and CEF-based DR under carbon neutrality conditions.
Also, there is a lack of studies on the development of metrics to evaluate price-based and CEF-based DR.

\subsection{Study contribution}

In this paper, we quantify the value of existing and new flexibility of a production company's MES in the context of decarbonization.
To this end, we apply a synthesis, design, and operation optimization MILP model to the real-world data of a beverage plant in Germany for different scenarios and contexts.
We propose and apply several flexibility metrics to evaluate DR based on prices and hourly CEFs.
The detailed case study considers H\textsubscript{2} storage, hourly electricity CEFs and prices, power-to-mobility, power-to-heat on multiple temperature levels and more and focuses on the flexibility applications self-consumption optimization, price-based and CEF-based voluntary DR, and peak shaving.

The remainder of this paper is organized as follows.
We first explain the case study in \Cref{sec:case}.
Subsequently, the optimization model is described in \Cref{sec:model} and the evaluation metrics defined in \Cref{sec:metrics}.
The scenarios and contexts are defined in \Cref{sec:scens} before we present and discuss the results of the analysis in \Cref{sec:res}.
Finally, we conclude in \Cref{sec:concl}.

%% file: chapter/2_case.tex
\section{Case study} \label{sec:case}
\begin{figure*}[!htb]
	\centering
	\includegraphics[trim=0 6.2cm 4.9cm 0, clip, width=0.95\linewidth]{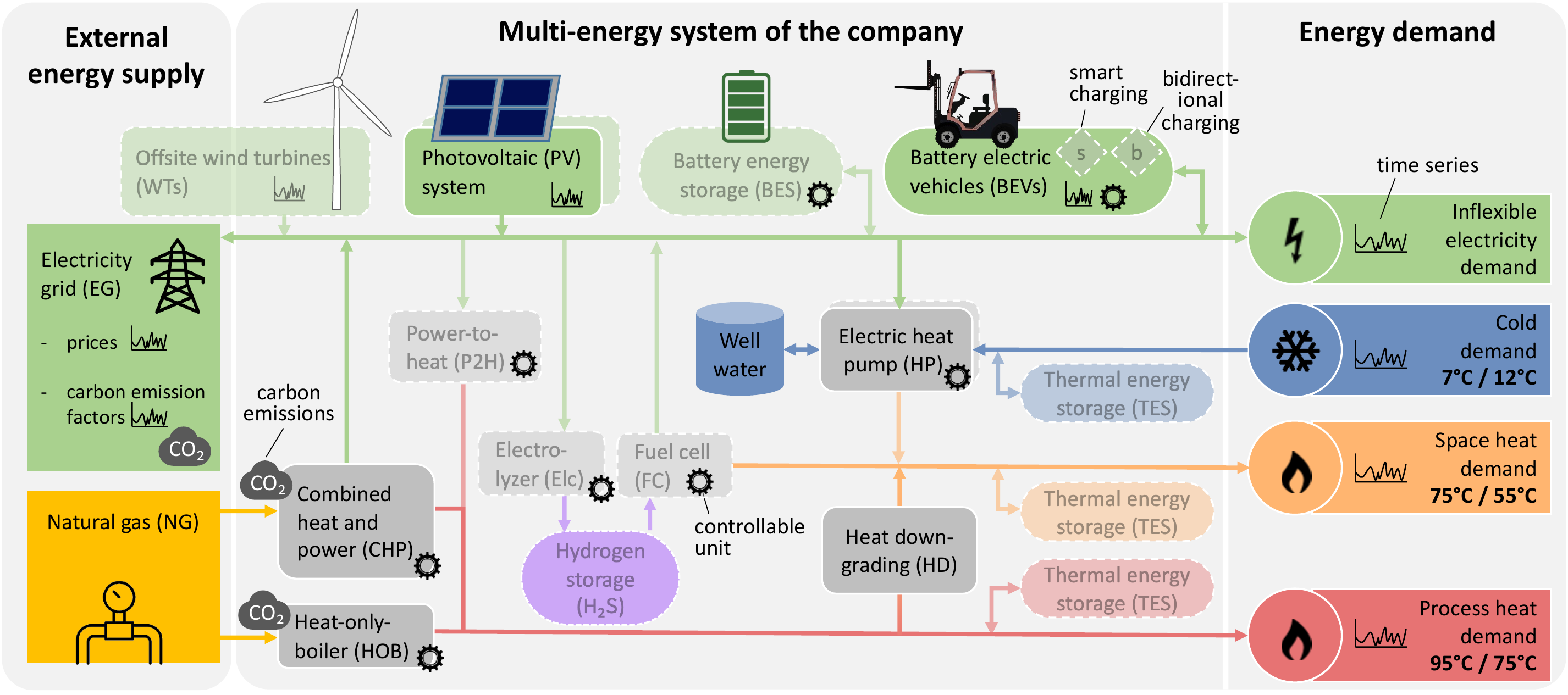}
	\caption{Overview scheme of the case study.
		It shows the external energy supply (left), the company's multi-energy system (MES) (center) including conversion and storage technologies, the energy demand of the company (right), and the energy flows between the components. 
		Greyed-out energy flows or components such as the thermal energy storage (TES) or the power-to-heat (P2H) do currently not exist but are possible to implement.
		In the status quo of the MES, the electricity demand is provided by a photovoltaic (PV) system, a combined heat and power (CHP) system based on natural gas, and the electricity grid (EG).
		There are no energy storage systems except the battery electric vehicles (BEVs) but they are charged unidirectionally with full power until full.
		The cold demand is served by the cooling machine (electric heat pump (HP) without using the hot side to serve heat demands).
		The heat demands on both temperature levels are met by the CHP and a natural gas fueled heat-only boiler (HOB).
		\label{fig:casestudy}}
\end{figure*}

We consider the MES of a beverage company in northern Germany.
\Cref{{fig:casestudy}} shows a schematic depiction of the case study.
This company has already invested in renewable infrastructure and is working towards carbon neutrality.

\begin{figure}[!htb]
	\begin{overpic}[width=1.0\linewidth]{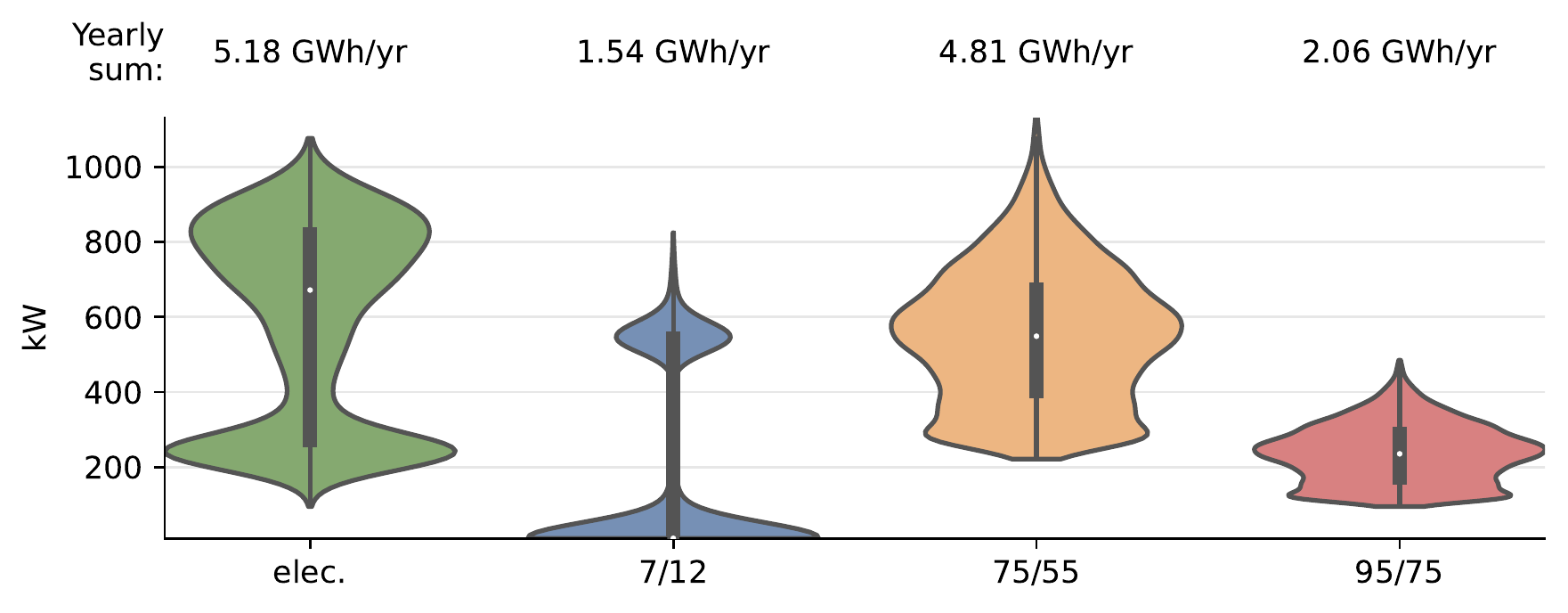} \put(0,36){\textbf{a)}} \put(0,27){\textbf{b)}} \end{overpic}
	\begin{overpic}[trim=0 0.6cm 0 0, clip, width=1.0\linewidth]{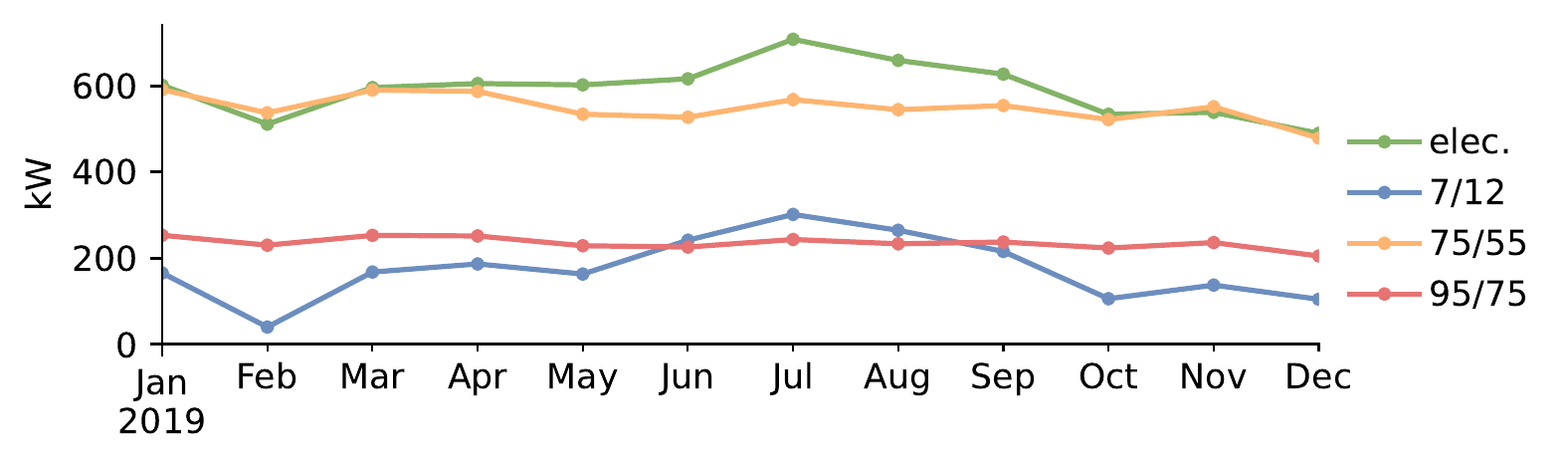} \put(0,22){\textbf{c)}} \end{overpic}
	\begin{overpic}[width=1.0\linewidth]{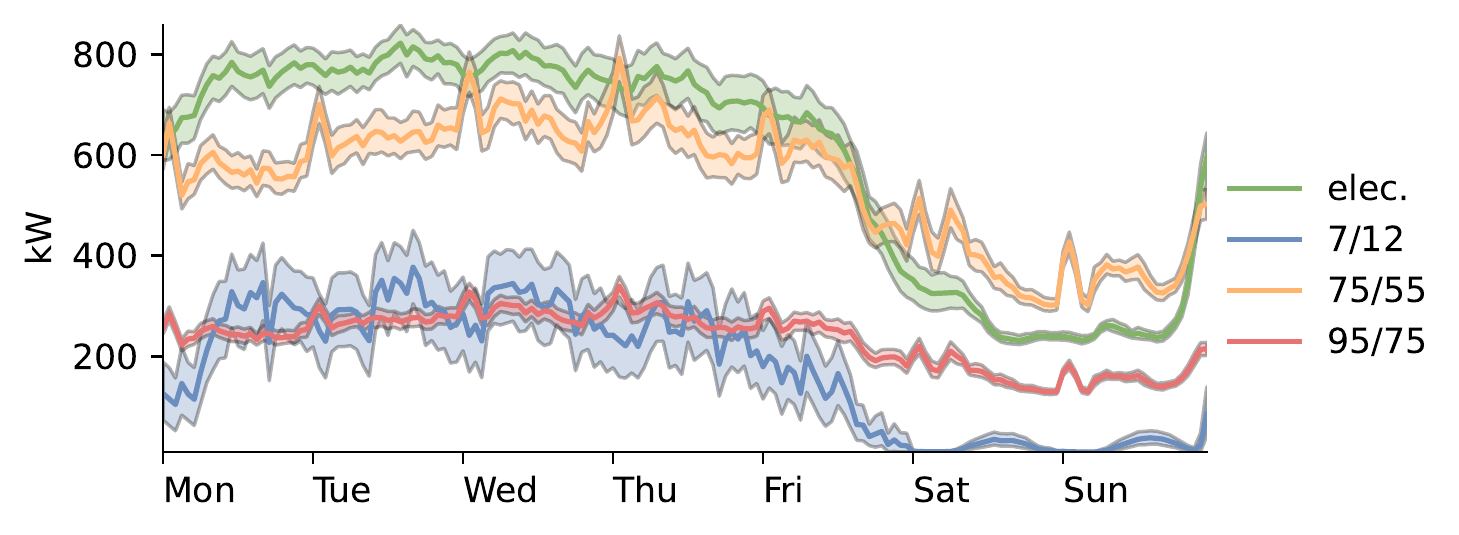} \put(0,27){\textbf{d)}} \end{overpic}
	\caption{Description of the electricity (elec.) and thermal energy (75/55, 95/75, 7/12) demand time series.
		\textbf{a)} Annual energy sums,
		\textbf{b)} violin plots showing annual distribution,
		\textbf{c)} monthly average demand,
		\textbf{d)} average week and 95\% confidence interval.
		\label{fig:dem_violin}}
\end{figure}

\subsection{Energy demands}
The company has an electricity demand and three different thermal energy demands with the following flow/return temperature levels in \textdegree C: Cooling demand (7/12), space heat demand (75/55), and process heat demand (95/75).
We calculated the hourly inflexible electricity demand that is independent of flexible operation and stays the same for all scenarios:
\begin{equation}
	P_t^\mathrm{eDem} = \tilde{P}_t^\mathrm{EG,buy} + \tilde{P}_t^\mathrm{PV,OC} - \tilde{P}_t^\mathrm{BEV,drive} - \tilde{P}_t^\mathrm{CM} \quad \forall t \in \mathcal{T}
\end{equation}
where $\tilde{P}_t^\mathrm{EG,buy}$, $\tilde{P}_t^\mathrm{PV,OC}$, and $\tilde{P}_t^\mathrm{BEV,drive}$ are the historic data for the purchased electricity from the electricity grid (EG), the PV energy consumed from the onsite PV system, and the electricity consumption of the BEVs, respectively. $\tilde{P}_t^\mathrm{CM}$ is an approximation of the cooling machine.
\Cref{fig:dem_violin} presents key metrics of the three thermal demands and the inflexible electricity demand.
It can be seen that the cold demand has a seasonal pattern peaking in summer, while the heat demands are more stable throughout the year.
Electricity demand is slightly higher during the summer months due to the increased beverage production.

\subsection{Existing technologies}
The company runs an onsite 307\,kW$_\mathrm{p}$ PV system.
For the analyses, we used the historic measured electricity generation of the year 2019, see \Cref{sec:convTec}.
Based on available roof space, an additional PV capacity of up to 2,770\,kW$_\mathrm{p}$ can be built with the same profile and efficiency.
A cooling machine with 860\,kW cooling power exists on site that can recool with 10\textdegree C well water.
The company owns a CHP with 410\,kW$_\mathrm{th}$ nominal heating power and a 1,620\,kW$_\mathrm{th}$ heat-only boiler (HOB).
Both run on natural gas and supply heat on the 97/75 temperature level.
There are 24 Li-ion industrial truck batteries with a net capacity of 81.1\,kWh each. %

%% file: chapter/3_model.tex
\section{Optimization model} \label{sec:model}

\subsection{Overview}
\begin{figure}[!htb]
	\includegraphics[trim=0.0cm 13.5cm 5.8cm 0cm,clip,width=1.00\linewidth]{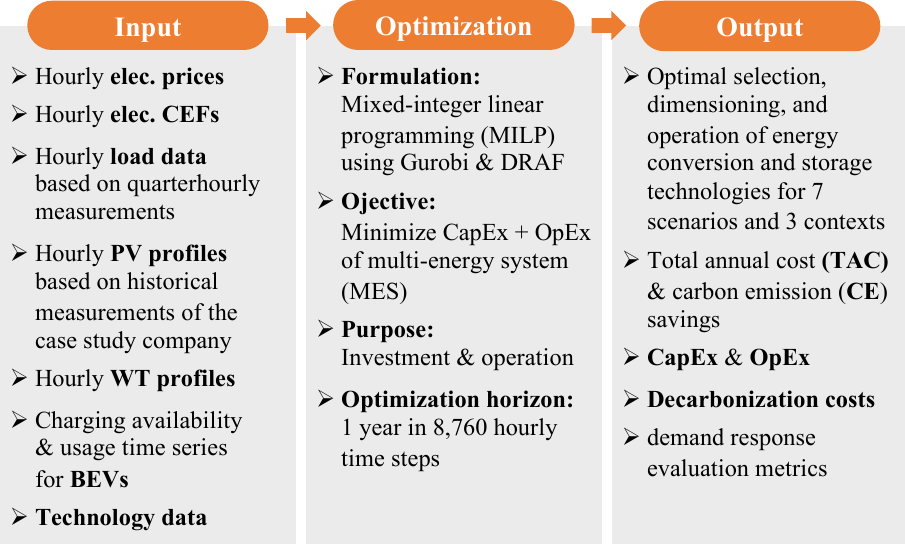}
	\caption{Main technical model features
		\label{fig:io}} %
\end{figure}

To quantify the value of flexibility for this company, we formulate a deterministic MILP using technology components of version v0.3.0~\cite{Fleschutz_draf_zenodo} of the open-source Python Demand Response Analysis Framework (DRAF), a modular tool for economic and environmental evaluation of DR~\cite{Fleschutz2022}.
An overview of the model is shown in \Cref{fig:io}.
DRAF and most used technology components are detailed in~\cite{Fleschutz2022}.
However, model parts that are crucial (cost and carbon balances, HP, BEV) or not described in~\cite{Fleschutz2022} such as wind turbine (WT), pressurized H\textsubscript{2} storage (H\textsubscript{2}S), proton exchange membrane fuel cell (FC), proton exchange membrane electrolyzer (Elc), and direct air capture (DAC) are described in this section.

We model one year with 8,760 hourly time steps $t \in \mathcal{T}$, so $\Delta t=1\,\mathrm{h}$.
We assume perfect foresight.
Optimization variables are indicated in bold and are non-negative continuous variables if not specified.

The objective function of the MILP problem is the minimization of the total annualized cost ($\bm{\mathrm{TAC}}$) in k\euro/yr and the penalty term $\bm{X}^\mathrm{penalty}$ to model conventional charging.
The $\bm{\mathrm{TAC}}$ include capital expenditures ($\bm{\mathrm{CapEx}}$) and operating expenses ($\bm{\mathrm{OpEx}}$), see \Cref{eq:TAC}.
\begin{align}
	\text{minimize} \quad & \bm{\mathrm{TAC}} + \bm{X}^\mathrm{penalty} \label{eq:obj}\\
	& \bm{\mathrm{TAC}} = \bm{\mathrm{CapEx}} + \bm{\mathrm{OpEx}} \label{eq:TAC}\\
	& \bm{\mathrm{TAC}}, \bm{\mathrm{OpEx}} \in \mathbb{R} \nonumber
\end{align}

The $\bm{\mathrm{CapEx}}$ represent the annualized investment costs for all technical components in the technology set $\mathcal{I}$:
\begin{align}
	& \bm{\mathrm{CapEx}} = \sum_{i \in \mathcal{I}} a(r, n_i) \, c^\mathrm{inv}_i \, \bm{P}^\mathrm{capn}_i
\end{align}
where $c^\mathrm{inv}_i$ are the specific investment costs and $\bm{P}^\mathrm{capn}_i$ the new nominal capacity of technology $i \in \mathcal{I}$.
$a(r, n_i)$ is the annuity factor of component $i$ as function of the discount rate $r$ and the economic life $n_i$ in years, see \Cref{eq:annuity}.
\begin{equation}
	a(r, n_i) = \frac{r(1+r)^{n_i}}{(1+r)^{n_i}-1} \label{eq:annuity}
\end{equation}

The $\bm{\mathrm{OpEx}}$ represent the operating costs of the first year
\begin{align}
	\bm{\mathrm{OpEx}} = \, & \bm{C}^\mathrm{RMI} + \bm{C}^\mathrm{EG,buy} - \bm{R}^\mathrm{EG,sell} + \bm{C}^\mathrm{EG,NF} \nonumber \\
	& + \bm{C}^\mathrm{WT,NF} + \bm{C}^\mathrm{NG} + \bm{C}^\mathrm{NG,tax} + \bm{C}^\mathrm{DAC} \\
	& \bm{C}^\mathrm{EG,sell}, \bm{C}^\mathrm{EG,sell} \in \mathbb{R} \nonumber
\end{align}
where $\bm{C}^\mathrm{RMI}$ are the repair, maintenance, and inspection (RMI) costs,
$\bm{C}^\mathrm{EG,buy}$ are the costs of purchased electricity,
$\bm{R}^\mathrm{EG,sell}$ are the revenue of sold electricity,
$\bm{C}^\mathrm{EG,NF}$ and $\bm{C}^\mathrm{WT,NF}$ are the network fees of purchased electricity and self-consumed electricity from the WT, respectively,
$\bm{C}^\mathrm{NG}$ are the costs of purchased natural gas,
$\bm{C}^\mathrm{NG,tax}$ are the carbon tax costs due to the purchased natural gas,
and $\bm{C}^\mathrm{DAC}$ are the costs for carbon removal.

The total operating CEs of the first year are calculated through:
\begin{equation}
  \bm{\mathrm{CE}} = \underbrace{\Delta t\,\varepsilon^\mathrm{NG} \sum_{t \in \mathcal{T}} \bm{P}^\mathrm{NG,buy}_t}_\text{Scope \RomanNumeralCaps{1} (direct)} + \underbrace{\Delta t \sum_{t \in \mathcal{T}} \varepsilon^\mathrm{elec}_t \bm{P}^\mathrm{EG,buy}_t}_\text{Scope \RomanNumeralCaps{2} (indirect)} - \underbrace{\bm{\mathrm{CE}}^\mathrm{DAC}}_\text{carbon removal}
\end{equation}
where $\varepsilon^\mathrm{NG}$ is the CEF for natural gas,
$\varepsilon^\mathrm{elec}_t$ denote the hourly electricity CEFs
, and $\bm{\mathrm{CE}}^\mathrm{DAC}$ are the removed CEs.

\subsection{Electricity grid (EG)}
\begin{figure}[!htb]
	\includegraphics[trim= 0 0.9cm 0 0.2cm,clip,width=1.00\linewidth]{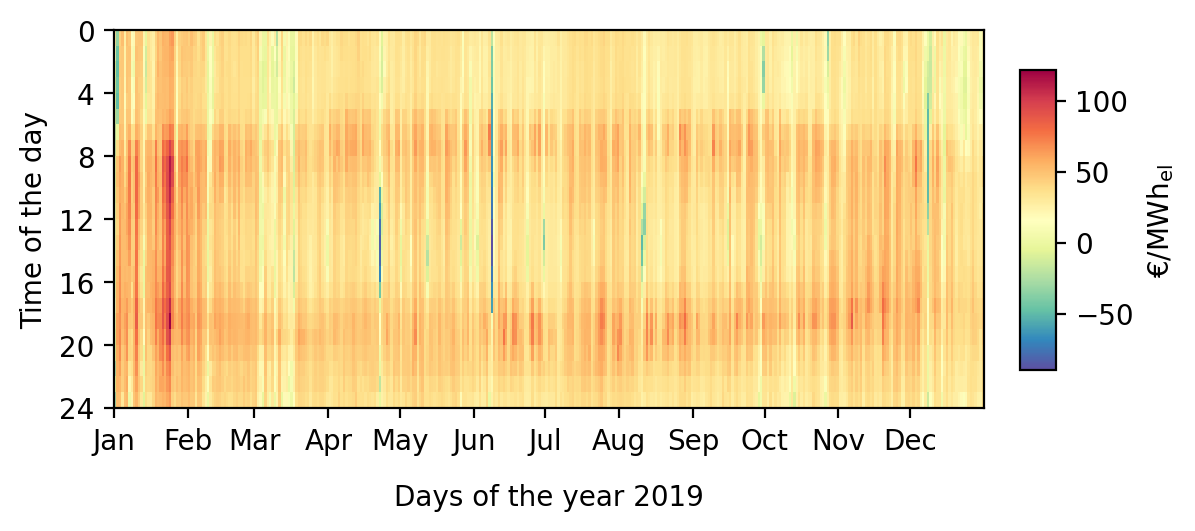}
	\includegraphics[trim= 0 0.2cm 0 0.2cm,clip,width=1.00\linewidth]{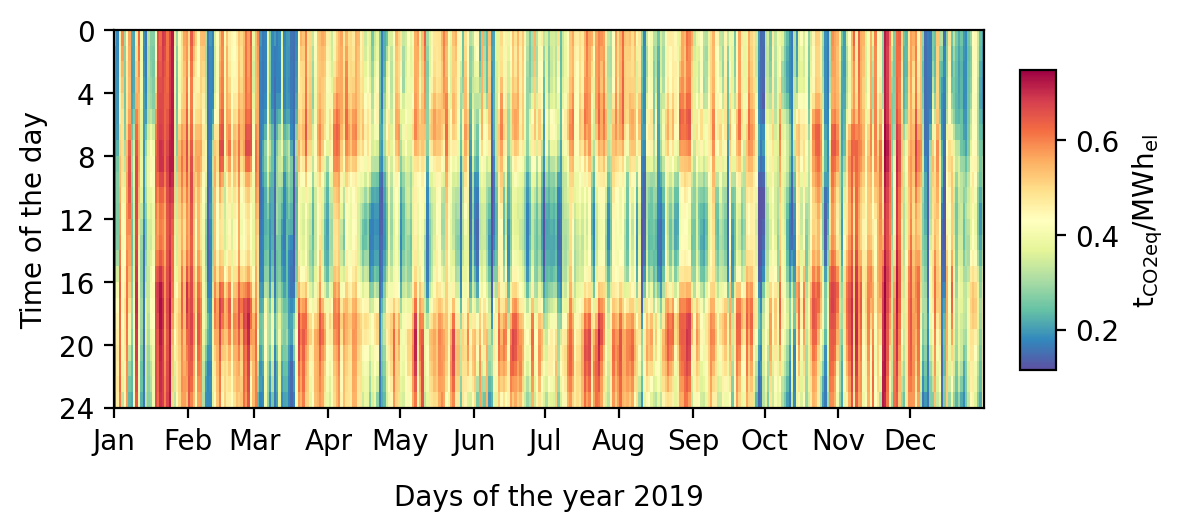}
	\caption{Day-ahead spot market prices ($c^\mathrm{elec,2019}_t$, top) and XEFs ($\varepsilon_t^\mathrm{elec}$, bottom) for the German electricity market of the year 2019.
		\label{fig:cefs_and_prices}}
\end{figure}

\paragraph{Electricity tariff}
We used historic hourly day-ahead electricity prices $c^\mathrm{elec}_t$ of the year 2019 ($c^\mathrm{elec,2019}_t$), as shown in \Cref{fig:cefs_and_prices} top, which were sourced from the ENTSO-E transparency platform~\cite{ENTSOE_TP} via the Python package elmada~\cite{Fleschutz2021_elmada}.
The purchased electricity $P^\mathrm{EG,buy}_t$ is limited to $P^\mathrm{EG,buy,max}$ of 20\,MW and evaluated with a real-time pricing tariff consisting of $c^\mathrm{elec}_t$ and a fixed price add-on $c^\mathrm{EG,addon}$ of \euro62.28\,MWh$^{-1}$ for taxes and levies, see \Cref{eq:egbuy,eq:phategbuy}.
Additionally, \euro70 fixed network fees $\hat{c}^\mathrm{EG,buy}$ are charged per kW of the highest demand of the year $\bm{\hat{P}}^\mathrm{EG,buy}$, see \Cref{eq:egnf}.
The sold electricity $\bm{P}^\mathrm{EG,sell}_t$ is limited to $P^\mathrm{EG,sell,max}$ of 20\,MW and compensated with $c^\mathrm{elec}_t$, see \Cref{eq:egsell,eq:maxegsell}.
\begin{align}
	&\bm{C}^\mathrm{EG,buy} = \Delta t \sum_{t \in \mathcal{T}} \bm{P}^\mathrm{EG,buy}_t \big(c^\mathrm{elec}_t + c^\mathrm{EG,addon} \big) \label{eq:egbuy}\\
	& \bm{P}^\mathrm{EG,buy}_t \leq \bm{\hat{P}}^\mathrm{EG,buy} \leq P^\mathrm{EG,buy,max} \quad \forall t \in \mathcal{T} \label{eq:phategbuy}\\
	&\bm{C}^\mathrm{EG,NF} = \bm{\hat{P}}^\mathrm{EG,buy} \hat{c}^\mathrm{EG,buy} \label{eq:egnf}\\
	&\bm{R}^\mathrm{EG,sell} = \Delta t \sum_{t \in \mathcal{T}} \bm{P}^\mathrm{EG,sell}_t c^\mathrm{elec}_t \label{eq:egsell}\\
	& \bm{P}^\mathrm{EG,sell}_t \leq P^\mathrm{EG,sell,max}  \quad \forall t \in \mathcal{T} \label{eq:maxegsell}
\end{align}

\paragraph{Electricity carbon emission factors (CEFs)}
National dynamic electricity grid-mix CEFs (XEFs), shown in \Cref{fig:cefs_and_prices} bottom, were calculated using the \texttt{XEF\_EP} method of elmada~\cite{Fleschutz2021_elmada}.
This method uses fuel type-specific generation data from the ENTSO-E transparency platform~\cite{ENTSOE_TP} and fuel type-specific CE intensities from~\cite{Quaschning2019}.
Electricity purchased from the EG was evaluated with XEFs, while the negative CEs through grid feed-in from own vRESs were not compensated.

Our analysis includes short-term load shifting, for which marginal emission factors (MEFs) should be used as demonstrated in~\cite{Fleschutz.2021} if the impact on CEs is explicitly examined.
However, since the environmental DR potentials in this analysis are modeled implicitly and inseparable from the investment decisions we take an attributional approach, thus using XEFs for $\varepsilon_t^\mathrm{EG}$.

\subsection{Natural gas grid}
Natural gas can be purchased for \euro43\,MWh$^{-1}$ with 0.240\,t$_\mathrm{CO2eq}$ /MWh~\cite{GEG9_2020} carbon emission intensity.
Additionally, a carbon tax $c^\mathrm{NG,tax}$ of \euro55\,t$_\mathrm{CO2eq}^{-1}$ has to be paid.

\subsection{Technologies}
\Cref{fig:cs_scheme} depicts the conversion and storage technologies used in our case study and an overview of their main parameters is given in \Cref{tab:tech_params}.

\begin{table*}[!htb]
	\let\tsc\textsuperscript
	\let\mc\multicolumn
	\centering
	\caption{Technology parameters \label{tab:tech_params}}
	\resizebox{0.90\linewidth}{!}{\begin{tabular}{cc|cccccccc}
		\toprule
		\mc{2}{c|}{Technology} & Specific inv. costs& Base & RMI costs & Economic life & Input & Efficiency & Output & C rate\\
		\mc{2}{c|}{~} & $c^\mathrm{inv}$&  & $c^\mathrm{RMI}$ & $n$ &  & $\eta$ &  & $\lambda$ \\
		\mc{2}{c|}{~} & (\euro/base)&  & (\%\tsc{a}/yr) & (yr) & & (\%) & & (kW/kWh)\\
		\midrule
		\multirow{10}{*}{\rotatebox{0}{Conversion}} & CHP & 589.46~\cite{ASUE_2011}&kW$_\mathrm{el}$ & 18 & 25 & natural gas & 40 & elec.& -\\
		& --\textquotedbl-- &-&-&-&-&natural gas& 45 & heat& -\\
		& Elc\tsc{b} & 1,295\,\cite{Petkov2020} &kW$_\mathrm{el}$ & 3.8 & 14 & elec. & 71 & $\mathrm{H}_2$ & -\\
		& FC\tsc{b} & 1,684\,\cite{Petkov2020} &kW$_\mathrm{el}$ & 3.8 & 14 &  $\mathrm{H}_2$ & 50 & elec.& -\\
		& --\textquotedbl-- &--&&-&-& $\mathrm{H}_2$ & 34 & heat& -\\
		& HOB & 57.13\,\cite{VIESSMANN} &kW$_\mathrm{th}$ & 18 & 15 &natural gas& 90 & heat & -\\
		& HP & 387\,\cite{Petkov2020} &kW$_\mathrm{th,cond}$ & 2.5 & 18 & elec., heat & 50\tsc{b} & heat & -\\
		& P2H & 100\,\cite{SMT_2018} &kW$_\mathrm{th}$ & 0 & 30 & elec. & 90 & heat & -\\
		& PV & 460\,\cite{Vartiainen_2019} &kW$_\mathrm{p}$ & 2 & 25 & \mc{2}{c}{- given profile -} & elec.& -\\
		& WT & 1,682\,\cite{WINDEUROPE2022} &kW$_\mathrm{p}$ & 1 & 20 & \mc{2}{c}{- given profile -} & elec.& -\\
		\midrule
		\multirow{4}{*}{\rotatebox{0}{Storage}} & BES & 600\,\cite{Figgener_2021} &kWh$_\mathrm{el}$ & 2 & 20 & elec. & 95$_\mathrm{cycle}$\tsc{c} & elec.& 0.7 \\
		& BEV & - & - & - & - & elec. & 95$_\mathrm{cycle}$ & elec. & 0.7\\
		& H\textsubscript{2}S & 10\,\cite{Petkov2020} &kWh$_{\mathrm{H}_2}$ & 0 & 23 &  $\mathrm{H}_2$ & 90$_\mathrm{cycle}$ &  $\mathrm{H}_2$ & - \tsc{d} \\
		& TES & 28.71\,\cite{FFE_2016} &kWh$_\mathrm{th}$ & 0.1 & 30 & heat & 99.5$_\mathrm{time}$ & heat & 0.5 \\
		\bottomrule
		\mc{9}{l}{\tsc{a} Based on investment costs} \\
		\mc{9}{l}{\tsc{b} Exergy efficiency $\eta^\mathrm{HP,Carnot}$ (ratio of reaching ideal Carnot COP)} \\
		\mc{9}{l}{\tsc{c} Additionally, $\eta^\mathrm{time}$=99.998\% applies every hour due to self-discharge~\cite{Redondo_2016}} \\
		\mc{9}{l}{\tsc{d} The C rate of H\textsubscript{2}S is determined by the Elc and FC capacities.}
	\end{tabular}}
\end{table*}

\subsubsection{Conversion technologies} \label{sec:convTec}
The conversion technologies CHP, Elc, FC, HOB, and P2H are modeled with a linear input-output relationship
\begin{align}
	& P^\mathrm{in}_t = \eta \, P^\mathrm{out}_t \quad \forall t \in \mathcal{T} \\
	& P^\mathrm{out}_t \leq P^\mathrm{cap} \quad \forall t \in \mathcal{T}
\end{align}
where some technologies have multiple outputs, see also \Cref{tab:tech_params}.

\paragraph{Electrical heat pump (HP)}
The HP has two heat sources $n \in \mathcal{N}$ (cold demand with 7\,\textdegree C and the well water with 10\,\textdegree C) and two heat sinks $c \in \mathcal{C}$ (the well water and the space heat demand with 75\,\textdegree C).
This results in three possible combinations of source-sink temperature levels defining the operating mode: Cold demand to well, cold demand to heat demand, and well to heat demand.
The coefficient of performance (COP) is calculated for each time step $t$ and operating mode:
\begin{align}
	& \mathrm{COP}^\mathrm{HP}_{t,c,n} = \eta^\mathrm{HP,Carnot} \underbrace{\frac{\vartheta^\mathrm{HP,cond}_{t,c} + 273.15}{\vartheta^\mathrm{HP,cond}_{t,c} - \vartheta^\mathrm{HP,eva}_{t,n}}}_{\mathrm{COP}_\mathrm{Carnot}} \quad \forall t \in \mathcal{T}, c \in \mathcal{C}, n \in \mathcal{N}.
\end{align}
The evaporation and condensation temperatures ($\vartheta^\mathrm{HP,eva}_{t,n}$, $\vartheta^\mathrm{HP,cond}_{t,c}$) were calculated assuming a 5\,\textdegree C temperature difference to consider heat exchange for both evaporator and condenser.
Following \cite{Bohlayer.2018}, a decision variable $\bm{Y}^\mathrm{HP}_{t,c,n}$ is used to select the operating mode.
Multiple HPs are considered if $n^\mathrm{HP,modes} > 1$:
\begin{align}
	& \bm{\dot{Q}}^\mathrm{HP}_{t,c,n} \leq \bm{Y}^\mathrm{HP}_{t,c,n} \dot{Q}^\mathrm{HP,max} \quad \forall t \in \mathcal{T}, c \in \mathcal{C}, n \in \mathcal{N}\\
	& \sum_{c \in \mathcal{C}}\sum_{n \in \mathcal{N}}\bm{Y}^\mathrm{HP}_{t,c,n} \leq n^\mathrm{HP,modes} \quad \forall t \in \mathcal{T}
\end{align}

\begin{figure}[!htb]
	\includegraphics[trim= 0 0.9cm 0 0.2cm,clip,width=1.00\linewidth]{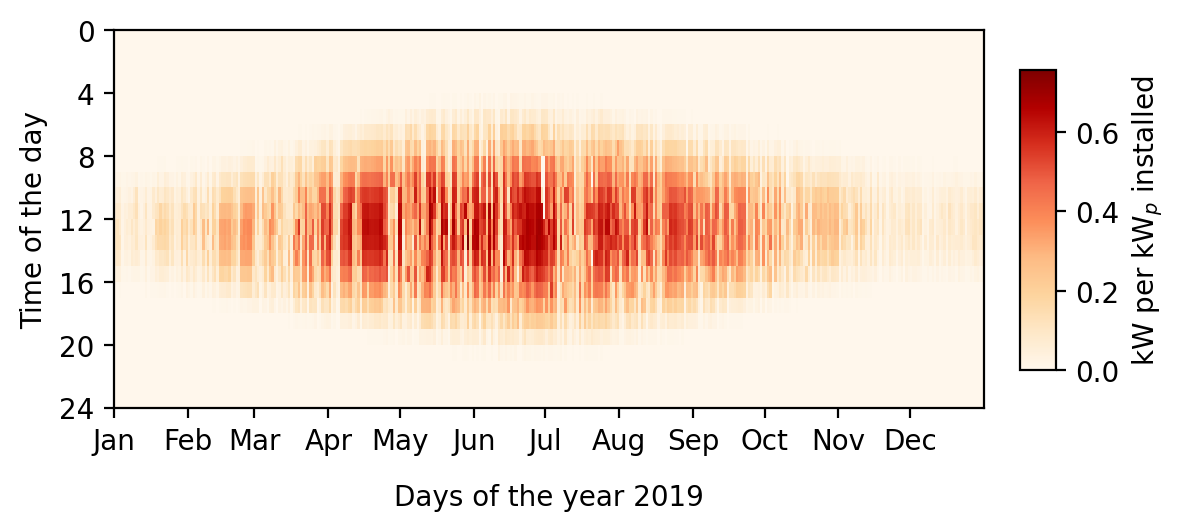}
	\includegraphics[trim= 0 0.2cm 0 0.2cm,clip,width=1.00\linewidth]{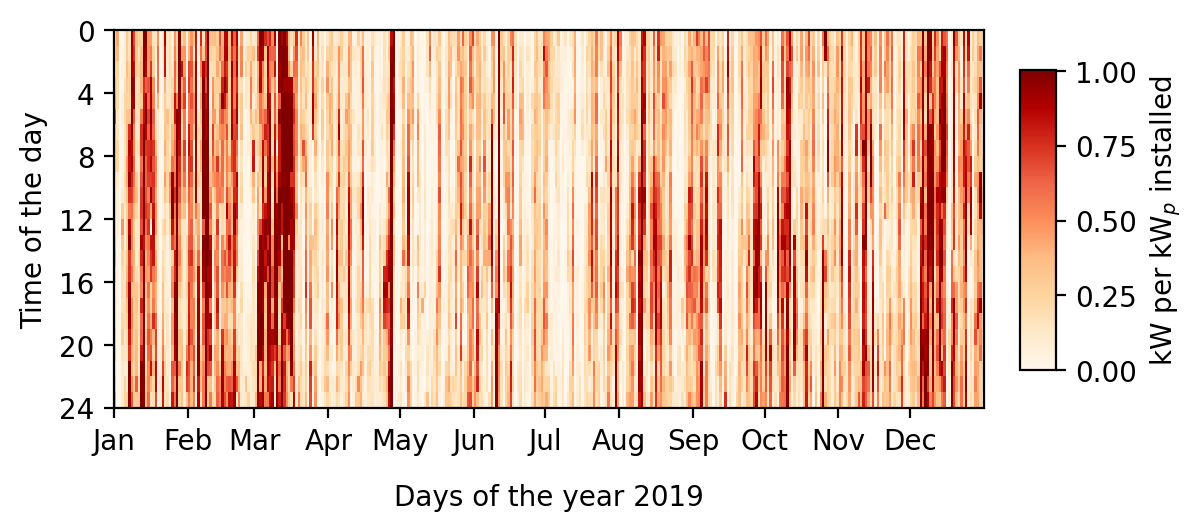}
	\caption{Historic electricity generation profile per kW\textsubscript{p} installed capacity of the case study PV system (top) and the Kelmarsh 6 WT (bottom). \label{fig:pv_wind_heatmap}}
\end{figure}

\paragraph{Photovoltaic (PV)}
The PV electricity generation time series per kW\textsubscript{p} installed capacity is an exogenous variable.
We used historic measurements of the case study company for the year 2019 in an aggregated form, see \Cref{fig:pv_wind_heatmap} top.
The capacity factor for 2019 is 0.10.

\paragraph{Wind turbine (WT)}
The wind electricity generation time series per kWh\textsubscript{p} installed capacity is an exogenous variable, see \Cref{fig:pv_wind_heatmap} bottom.
The hourly profile for 2019 is aggregated from the open data of the Kelmarsh 6~\cite{KelmarshData2022}, a 2.04\,MW Senvion MM92 WT located in the Kelmarsh wind farm in the United Kingdom which is used as a proxy for northern Germany.
The capacity factor for 2019 is 0.33.
For used WT electricity, an electricity price addon $c^\mathrm{EG,addon}$ is paid.
Unused WT electricity is sold at the day-ahead market price assuming that the guarantees of origins are sold for a negligible price, so there are no negative CEs.
Investment in partial WTs (shareholder) are possible.

\subsubsection{Storage technologies}

\paragraph{Battery energy storage (BES)}
The constraints for the BES are:
\begin{align}
	& \bm{E}_{t} = \eta^\mathrm{time}
	\begin{cases}
	  k^\mathrm{ini} \bm{E}^\mathrm{capn} & \text{if } t = t_0 \\
	  \bm{E}_{t-1} & \text{otherwise}
	\end{cases} \nonumber\\
	& \quad \quad \quad + \Delta t \, \Bigg( \eta^\mathrm{ch} \bm{P}^\mathrm{in}_{t} - \frac{\bm{P}^\mathrm{out}_t}{\eta^\mathrm{dis}} \Bigg) \quad \forall t \in \mathcal{T} \\
	& \bm{P}^\mathrm{in}_{t} \leq \lambda^\mathrm{in} \bm{E}^\mathrm{capn} \quad \forall t \in \mathcal{T} \\
	& \bm{P}^\mathrm{out}_{t} \leq \lambda^\mathrm{out} \bm{E}^\mathrm{capn} \quad \forall t \in \mathcal{T} \\
	& \bm{E}_{t} \leq \bm{E}^\mathrm{capn} \quad \forall t \in \mathcal{T} \\
	& \bm{E}_{t_{|\mathcal{T}|}} = k^\mathrm{ini} \bm{E}^\mathrm{capn}
\end{align}

\paragraph{Hydrogen storage (H\textsubscript{2}S) and thermal energy storage (TES)}
H\textsubscript{2}S and TES are formulated analogously to BES.

\paragraph{Battery electric vehicle (BEV)}
In our case study we model multiple powered industrial trucks as special form of BEVs.
BEVs are modeled as multiple energy storage systems $b \in \mathcal{B}$ similar to BESs but with restricted availability and two types of discharging: $P^\mathrm{drive}_t$ and $\bm{P}^\mathrm{V2X}_t$.
While $P^\mathrm{drive}_t$ is given exogenously through historic measurements, $\bm{P}^\mathrm{V2X}_t$ can be activated ($z^\mathrm{V2X}$=1) to allow V2X, i.e., smart discharging for non-driving purposes.
Smart charging can be deactivated ($z^\mathrm{smart}$=0) by penalizing the sum of charges weighted by the according time steps, see \Cref{eq:BEV_penalty} and the objective function \Cref{eq:obj}.
Battery degradation was limited by constraining the minimum and maximum state of charge ($k^\mathrm{empty}_{b}$=0.15, $k^\mathrm{full}_{b}$=0.85) and the C rate ($\lambda^\mathrm{in}_{b}$ = $\lambda^\mathrm{out}_{b}$ = 0.7).
Please see~\cite{Leippi.2022} for a recent review regarding battery degradation in DR scenarios.
The BEV constraints are:
\begin{align}
	& \bm{E}_{t,b}  = \eta^\mathrm{time}
	\begin{cases}
	  k^\mathrm{ini}_{b} E^\mathrm{capx}_{b} & \text{if } t = t_0\\
	  \bm{E}_{t-1,b} & \text{otherwise} \\
	\end{cases} \nonumber \\
	& \quad \quad \quad + \Delta t \,\Bigg( \eta^\mathrm{ch} \bm{P}^\mathrm{in}_{t,b} - \frac{P^\mathrm{drive}_{t,b} + \bm{P}^\mathrm{V2X}_{t,b}}{\eta^\mathrm{dis}} \Bigg) \quad \forall t \in \mathcal{T}, b \in \mathcal{B} \\
	& k^\mathrm{empty}_{b} E^\mathrm{capx}_{b} \leq \bm{E}_{t,b} \leq k^\mathrm{full}_{b} E^\mathrm{capx}_{b} \quad \forall t \in \mathcal{T}, b \in \mathcal{B} \\
	& \bm{P}^\mathrm{in}_{t,b} \leq y^\mathrm{avail}_{t,b} \lambda^\mathrm{in}_{b} E^\mathrm{capx}_{b} \quad \forall t \in \mathcal{T}, b \in \mathcal{B} \\
	& \bm{P}^\mathrm{V2X}_{t,b} \leq z^\mathrm{V2X} y^\mathrm{avail}_{t,b} \lambda^\mathrm{V2X}_{b} E^\mathrm{capx}_{b} \quad \forall t \in \mathcal{T}, b \in \mathcal{B} \\
	& \bm{E}_{t_{|\mathcal{T}|},b} = k^\mathrm{ini}_{b} E^\mathrm{capx}_{b}  \quad \forall b \in \mathcal{B} \\
	& \bm{X}^\mathrm{penalty} = (1 - z^\mathrm{smart}) \sum_{t \in \mathcal{T}} \sum_{b \in \mathcal{B}} t \, \bm{P}^\mathrm{in}_{t,b} \label{eq:BEV_penalty}
\end{align}

\begin{figure}[!htb]
	\includegraphics[width=1.0\linewidth]{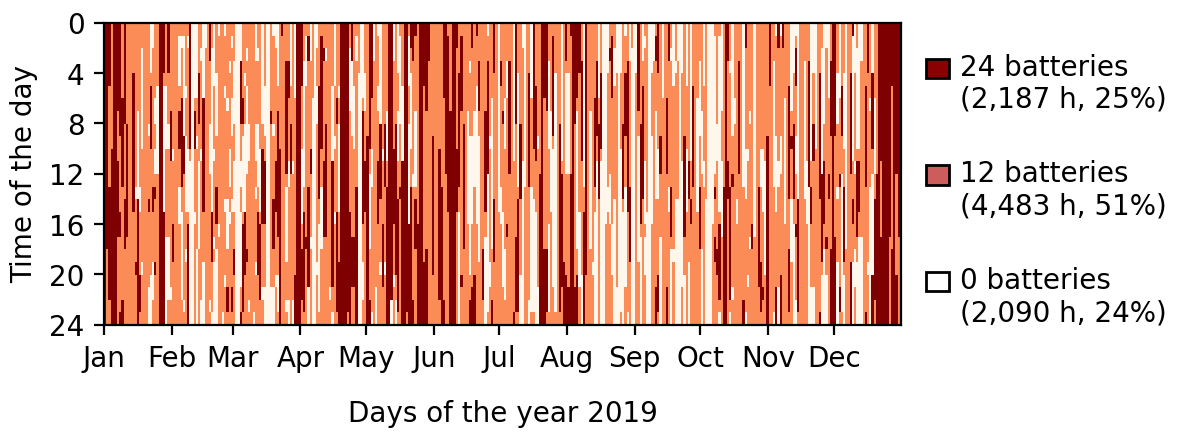}
	\caption{Number of available batteries throughout the year 2019.
	The numbers and percentages in brackets indicate how often an availability appears.
	While all batteries are available in 2,187 hours (25\% of the year), no battery is available in 2,090 hours (24\%).
		\label{fig:bev_avail_heatmap}}
\end{figure}

To reduce model complexity, we aggregated the capacity of 12 batteries that had similar availability patterns resulting in two batteries with 973.2\,kWh capacity each and used two representative availability time series based on the real empirical data from the battery energy management system.
\Cref{fig:bev_avail_heatmap} indicates the number of available batteries for charging across the year.

\subsection{Limitations}
The flexibility potentials of the applications spot market optimization, self-consumption optimization, and peak shaving in this paper are overestimated due to the following simplifications:
(1) Perfect foresight is assumed for all input time series.
(2) Battery degradation is limited by constraining the minimum and maximum state of charge but not explicitly modeled.
(3) Batteries of powered industrial trucks are aggregated.

%% file: chapter/4_eval.tex
\section{Evaluation metrics} \label{sec:metrics}

\subsection{Evaluation of DR based on prices and CEFs} \label{sec:eval_DR}

\begin{figure}[!htb]
	\centering
	\includegraphics[width=1\linewidth]{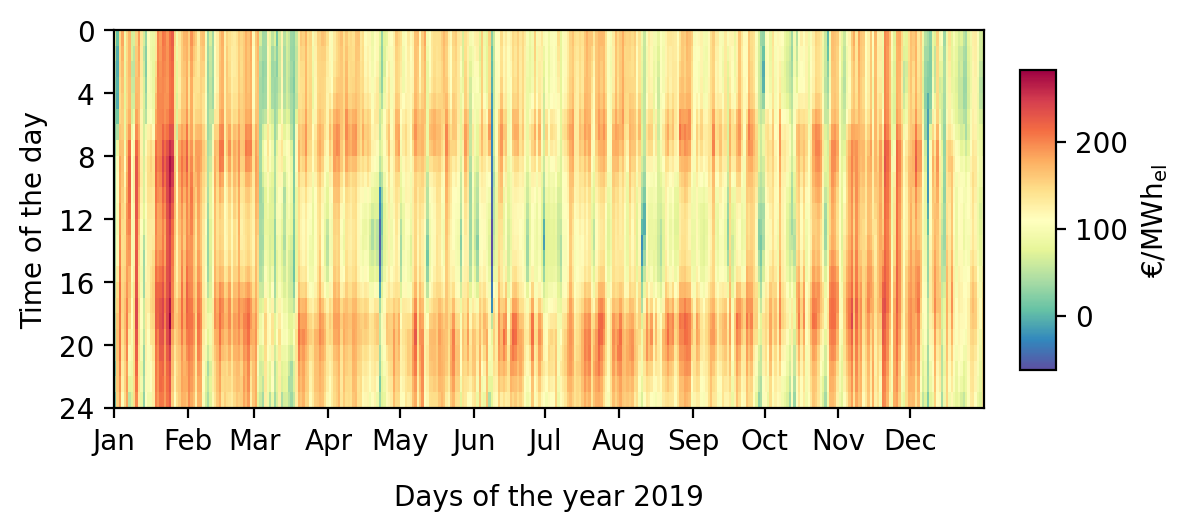}
	\caption{Financial DR signal that consists of electricity prices of \Cref{fig:cefs_and_prices} plus the CEFs of \Cref{fig:cefs_and_prices} weighted by a carbon (removal) price of 222\euro/t\textsubscript{CO2eq}.
		\label{fig:EAP_222_heatmap}}
\end{figure}

Through penalizing or constraining operating CEs in a MES model considering price-based DR, CEFs can work as an additional DR incentive.
E.g., if Scope \RomanNumeralCaps{1} CEs are forced to be zero and Scope \RomanNumeralCaps{2} CEs are penalized with a carbon price of \euro222\,t$_\mathrm{CO2eq}^{-1}$, the effective DR incentive is the sum of electricity prices and carbon price-weighted CEFs, presented in \Cref{fig:EAP_222_heatmap}.

To evaluate DR based on prices and CEFs, we define multiple related metrics.
The energy-weighted average price (EWAP$_s$) and energy-weighted average CEF (EWACEF$_s$) are defined in \Cref{eq:ewap,eq:ewapcef}.
For a scenario $s$, they measure the price and CEF of the average unit of purchased electricity respectively.
While the EWAP is the total purchased electricity costs (excluding taxes and levies) divided by the total purchased electrical energy, the EWACEF is the quotient of the total CEs due to purchased electricity and the total purchased electrical energy.
EWAP and EWACEF can be interpreted as the inability of the MES to react to the electricity prices and CEFs, respectively.
\begin{align}
	& \underbrace{\mathrm{EWAP}_s}_{\text{in \euro/MWh}} = \frac{\sum_{t \in \mathcal{T}} c^\mathrm{elec}_t \bm{P}^\mathrm{EG,buy}_{t,s} \Delta t }{\sum_{t \in \mathcal{T}} \bm{P}^\mathrm{EG,buy}_{t,s} \Delta t} \label{eq:ewap} \\
	& \underbrace{\mathrm{EWACEF}_s}_{\text{in t$_\mathrm{CO2eq}$/MWh}} = \frac{\sum_{t \in \mathcal{T}} \varepsilon^\mathrm{elec}_t \bm{P}^\mathrm{EG,buy}_{t,s} \Delta t}{\sum_{t \in \mathcal{T}} \bm{P}^\mathrm{EG,buy}_{t,s} \Delta t} \label{eq:ewapcef}
\end{align}

For comparison, we define the time-weighted average price (TWAP) and the time-weighted average CEF (TWACEF):
\begin{align}
	& \underbrace{\mathrm{TWAP}}_{\text{in \euro/MWh}} = \frac{\sum_{t \in \mathcal{T}} c^\mathrm{elec}_t \Delta t }{\sum_{t \in \mathcal{T}} \Delta t} \label{eq:twap}\\
	& \underbrace{\mathrm{TWACEF}}_{\text{in t$_\mathrm{CO2eq}$/MWh}} = \frac{\sum_{t \in \mathcal{T}} \varepsilon^\mathrm{elec}_t \Delta t}{\sum_{t \in \mathcal{T}} \Delta t} \label{eq:twacef}
\end{align}

In simple terms, TWAP is the average price, EWAP is the average paid price, TWACEF is the average CEF, and EWACEF is the average accounted CEF.
While TWAP and TWACEF only look at the electricity market results $c^\mathrm{elec}_t$ and $\varepsilon^\mathrm{elec}_t$, EWAP and EWACEF also look at the profile of the demand-side electricity purchase $\bm{P}^\mathrm{EG,buy}_t$.

Next, we define the $\pi_s$-rate and the $\varepsilon_s$-rate of scenario $s$ by normalizing EWAP$_s$ with TWAP and EWACEF$_s$ with TWACEF:
\begin{align}
	& \pi_s = \frac{\mathrm{EWAP}_s}{\mathrm{TWAP}} \label{eq:pi}\\
	& \varepsilon_s = \frac{\mathrm{EWACEF}_s}{\mathrm{TWACEF}} \label{eq:epsilon}
\end{align}
The $\pi$-rate has three advantages over EWAP.
Firstly, it considers the price profile $c^\mathrm{elec}_t$ independently of the units, i.e., a European case study using \euro\,can be compared to an American using \$.
Secondly, it is relative to the average price allowing comparison between time frames of different price levels, e.g., the years 2019 and 2021.
Thirdly, it tells if the EWAP$_s$ is over the TWAP ($\pi_s$>100\%) or under ($\pi_s$<100\%).
The same advantages apply analogously to the $\varepsilon$-rate.

\begin{figure}[ht]
	\includegraphics[width=1.00\linewidth]{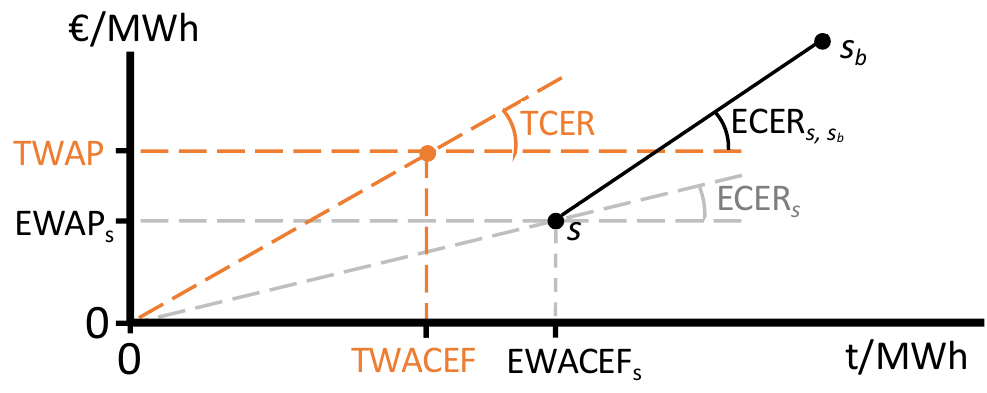}
	\caption{Energy-specific price-emission diagram to illustrate flexibility metrics.
	\label{fig:metrics_scheme}}
\end{figure}

The time-based cost–emission ratio (TCER), is the scenario-independent ratio of the average electricity prices to the average CEF:
\begin{equation}
	\underbrace{\text{TCER}}_{\text{in \euro/t$_\mathrm{CO2eq}$}} = \frac{\text{TWAP}}{\text{TWACEF}} \label{eq:tcer}
\end{equation}
A TCER of \euro100\,t$_\mathrm{CO2eq}^{-1}$, e.g., means that if a load is reduced equally over all time steps, \euro100 is saved for each reduced ton of CEs.
On an energy-specific price-emission diagram with flexibility metrics \Cref{fig:metrics_scheme}, TCER is the gradient of the line between the origin and the intersection between the TWAP and the TWACEF.

In contrast, the energy-based cost-emission ratio (ECER) of scenario $s$ is the ratio of EWAP$_s$ to EWACEF$_s$:
\begin{equation}
	\underbrace{\text{ECER}_s}_{\text{in \euro/t$_\mathrm{CO2eq}$}} = \frac{\text{EWAP}_s}{\text{EWACEF}_s} \label{eq:ecer}
\end{equation}
On an EWAP-EWACEF plot, the ECER$_\mathrm{s}$ is the gradient of the line between the origin and the scenario $s$, see \Cref{fig:metrics_scheme}.

Next, we define the ECER for two scenarios, $s$ and $s_b$, where $s_b$ is the baseline scenario without flexibility:
\begin{equation}
	\underbrace{\text{ECER}_{s,s_b}}_{\text{in \euro/t$_\mathrm{CO2eq}$}} = \frac{\text{EWAP}_{s_b}-\text{EWAP}_{s}}{\text{EWACEF}_{s_b}-\text{EWACEF}_{s}} \label{eq:ecer_2}
\end{equation}
ECER$_{s,s_b}$ can be also interpreted as the ratio of electrical energy cost savings to CE savings of scenario $s$ compared to the baseline scenario $s_b$.
E.g., an ECER$_\mathrm{a,b}$ of \euro150\,t$_\mathrm{CO2eq}^{-1}$ means that on average \euro150 is saved for each reduced ton of CEs.
On the EWAP-EWACEF plot, ECER$_{s,s_b}$ represents the gradient of the line between $s$ and $s_b$, see \Cref{fig:metrics_scheme}.
When assessing DR, i.e., the modification of $\bm{P}^\mathrm{EG,buy}_t$, ECER$_{s,s_b}$ can be interpreted as how strongly cost reduction is weighted relative to CE reduction.

Finally, based on the previous metrics, we define the normalized gradients $\omega_s$ and $\omega_{s,s_b}$:
\begin{align}
	& \omega_s = \frac{\pi_s}{\varepsilon_s} = \frac{\text{ECER}_s}{\text{TCER}} \label{eq:omega}\\
	& \omega_{s,s_b} = \frac{\pi_{s}-\pi_{s_b}}{\varepsilon_{s}-\varepsilon_{s_b}} \label{eq:omega_2}
\end{align}
They have the same advantages over ECER$_s$ then the $\pi$-rate has over EWAP.

The metrics defined in \Cref{sec:eval_DR} can be applied to any price and load data set.
Also, they can be applied to a subset of data, e.g., months or weekdays.
It is worth mentioning that while the EG feed-in also impacts the stability of the EG and the electricity markets, the evaluation of EG feed-in is beyond the scope of this paper.

\subsection{Evaluation of decarbonization costs}
To evaluate the additional cost of achieving net-zero CEs, we calculate the decarbonization costs $C^\mathrm{decarb}_s$:
\begin{equation}
	C^\mathrm{decarb}_{s} = \mathrm{TAC}_{s_d} - \mathrm{TAC}_{s}
\end{equation}
where the scenarios $s_d$ and $s$ represent the cost optimal MES with and without enforcing net-zero CEs, respectively.

%% file: chapter/5_scens.tex
\section{Scenario and context definition} \label{sec:scens}

\subsection{Scenario definition}
\begin{figure}[!htb]
	\let\tsc\textsuperscript
	\let\mc\multicolumn
	\raisebox{-0.5\height}{\includegraphics[trim=0 0.11cm 0 0, clip, width=0.628\linewidth]{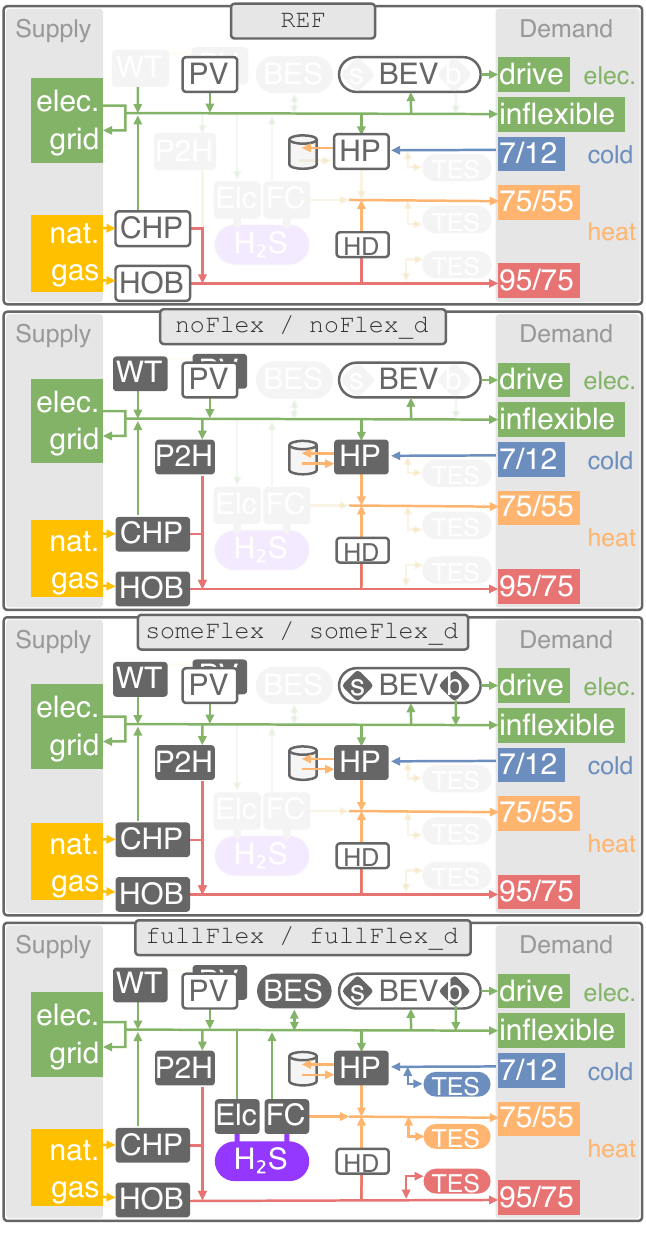}}
	\raisebox{-0.5\height}{\includegraphics[trim=0.02cm 0.15cm 0.05cm 0, clip, width=0.361\linewidth]{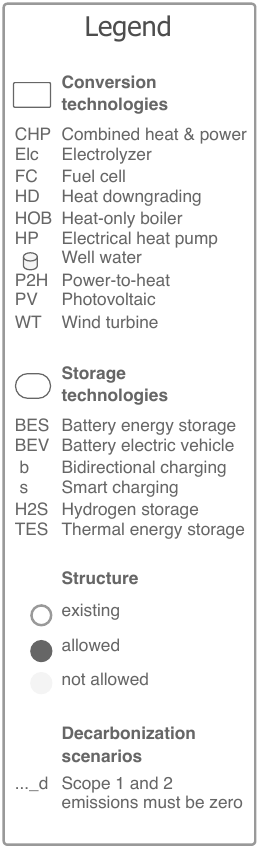}}
	\vskip 2mm
	\resizebox{1\linewidth}{!}{\begin{tabular}{l|cccc}
		\toprule
		\mc{1}{r|}{Scenarios $\rightarrow$} &  & \scen{noFlex}, & \scen{someFlex}, & \scen{fullFlex},\\ 
		$\downarrow$ Possible options & \scen{REF} & \scen{noFlex\_d} & \scen{someFlex\_d} & \scen{fullFlex\_d}\\ 
		\midrule
		Conversion tech. investments\tsc{a} & \xmark & \cmark & \cmark & \cmark \\
		Use of existing flexibility\tsc{b} & \xmark & \xmark & \cmark & \cmark \\
		Storage tech. investments\tsc{c} & \xmark & \xmark & \xmark & \cmark \\
		\bottomrule
		\mc{5}{l}{\tsc{a} WT, PV, HP, and P2H} \\
		\mc{5}{l}{\tsc{b} smart/bidirectional BEV charging, fuel switching, and smart HP mode selection} \\
		\mc{5}{l}{\tsc{c} BES, TES (on 	all temperature levels), and H\textsubscript{2}S (including Elc and FC)}
	\end{tabular}}
	\caption{Definition of scenarios \label{fig:cs_scheme}}
\end{figure}

We model seven scenarios:
One unoptimized reference scenario (\scen{REF}),
three optimized non-decarbonization scenarios (\scen{noFlex}, \scen{someFlex}, \scen{fullFlex}),
and three optimized decarbonization scenarios (\scen{noFlex\_d}, \scen{someFlex\_d}, \scen{fullFlex\_d}).
The scenario definition is summarized in \Cref{fig:cs_scheme} and described in the following:
\begin{itemize}
	\item \scen{REF}:
	The reference scenario represents the unoptimized status quo, also described in \Cref{fig:casestudy}.
	Electricity demand is covered by the EG, the PV, and the CHP.
	BEVs are charged unidirectionally with full power until full.
	Heat demands are met by the CHP and the HOB.
	The cold demand is served by the cooling machine which was modeled as a HP with 1,147\,kW on the hot side and in a restricted mode so that heat in the condenser is only transferred to the well water.
	We assume that the cooling machine only exists in the REF scenario and must be replaced in other scenarios, to make the results comparable between them.
	There are no energy storage systems.
	Investments are deactivated.

	\item \scen{noFlex}:
	Investments in the conversion technologies WT, PV, HP, P2H are allowed.
	According capacities are optimally designed.
	The smart HP mode selection is deactivated by disallowing heat transfer from cold to heat demand.
	Fuel switching is deactivated by fixing the CHP operation to the CHP operation in \scen{REF}.
	Smart/bidirectional BEV charging and investments in storage systems are deactivated.
	
	\item \scen{someFlex}:
	Additionally to \scen{noFlex}, some flexibility sour\-ces can be used:
	BEVs can be charged smartly/\-bidirection\-ally.
	For each time step heat supply can be chosen to be gas-based (HOB, CHP) and/or electricity-based (HP, P2H), since the restriction to the CHP operation is removed.
	However, storage investments are still deactivated.
	
	\item \scen{fullFlex}:
	Additionally, investments in the following storage technologies are allowed: BES, TES (on all three temperature levels), and H\textsubscript{2}S including Elc and FC with waste heat utilization on the 75/55 temperature level.

	\item \scen{noFlex\_d}, \scen{someFlex\_d}, \scen{fullFlex\_d}:
	Decarbonization scenarios based on \scen{noFlex}, \scen{someFlex}, and \scen{fullFlex}, respectively.
	Scope 1 and 2 carbon neutrality is enforced.
	Natural gas is strictly forbidden to prevent investments in and operation of fossil-based infrastructure.
	Electricity-based CEs must be removed with the specific carbon removal costs $c^\mathrm{DAC}$.
\end{itemize}

\begin{figure}[h]
	\includegraphics[width=1.00\linewidth]{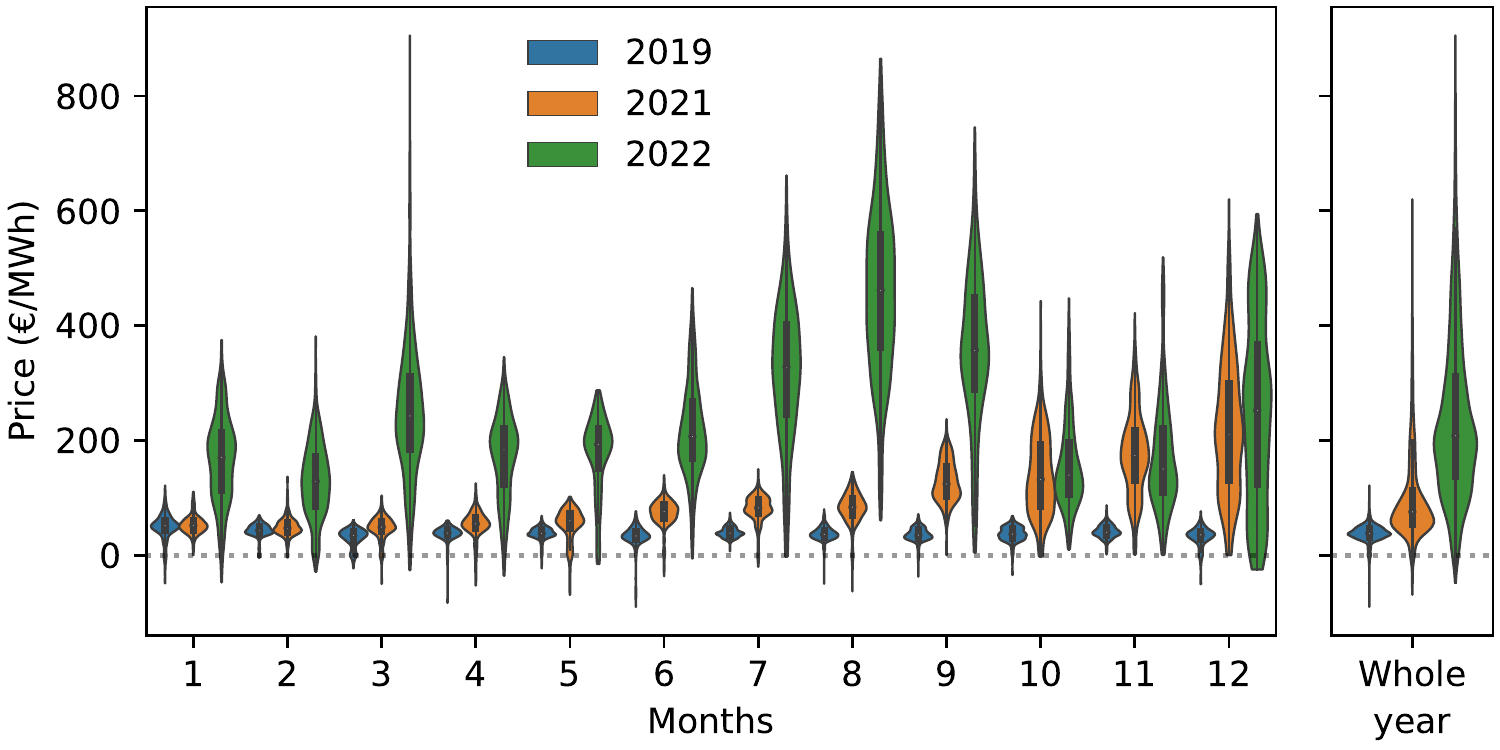}
	\includegraphics[width=0.832\linewidth]{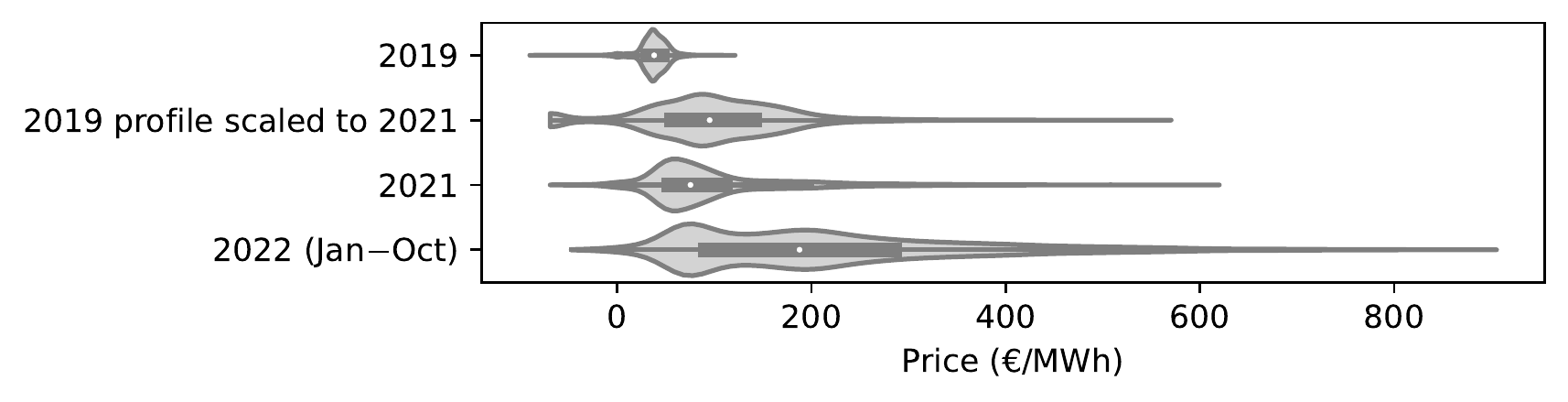}
	\hspace{-0.36cm}
	\includegraphics[trim=0.2cm -1.54cm 0 0,clip,width=0.185\linewidth]{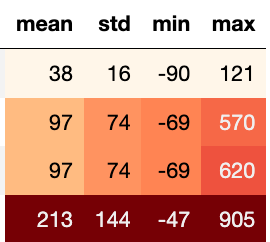}
	\caption{Distribution of day-ahead electricity market prices of 2019, 2021, and 2022 (top, until October) and the used 2019 profile that was scaled to 2021 levels (bottom).
		\label{fig:violins2021}}
\end{figure}

\subsection{Context definition}

\begin{table}[!htb]
	\centering
	\caption{Definition of contexts \label{tab:contexts}}
	\resizebox{1\linewidth}{!}{\begin{tabular}{l|cc}
		\toprule
		Context & Carbon removal price $c^\mathrm{DAC}$& Electricity price $c^\mathrm{elec}_t$ \\
		\midrule
		\scen{c\_base} & \euro222\,t\,$_\mathrm{CO2eq}^{-1}$ & 2019 data \\
		\scen{c\_strict} & \euro10.000\,t\,$_\mathrm{CO2eq}^{-1}$ & 2019 data \\
		\scen{c\_scaled} & \euro222\,t\,$_\mathrm{CO2eq}^{-1}$ & 2019 data scaled to 2021 \\
		\bottomrule
	\end{tabular}}
\end{table}

We model all scenarios for the three contexts defined in \Cref{tab:contexts}.
\scen{c\_base} is the base context with electricity price data of 2019 and specific carbon removal costs $c^\mathrm{DAC}$ of \euro222\,t\,$_\mathrm{CO2eq}^{-1}$ -- the specific costs of direct air capturing~\cite{Fasihi2019}.
In \scen{c\_strict}, electricity price data of 2019 is used but $c^\mathrm{DAC}$ is set to \euro10k\,t\,$_\mathrm{CO2eq}^{-1}$.
If this high price is paid, the onsite decarbonization is not viable.  
This context applies when a company does not plan to rely on carbon removal or on the carbon reductions of the electricity grid.
In \scen{c\_scaled}, $c^\mathrm{DAC}$ is \euro222\,t\,$_\mathrm{CO2eq}^{-1}$ but the electricity prices are scaled to 2021 levels.
By picking the year 2019 as base year, the quantified value of flexibility is underestimated due to the relatively low price spreads on the day-ahead market, see \Cref{fig:spreads}, however, energy demands don't feature anomalies from the COVID-19 crisis.
Therefore, in \scen{c\_scaled}, we scaled the day-ahead market prices of the year 2019 to roughly match the mean, the standard deviation, and the extreme points of the prices of 2021, see~\Cref{fig:violins2021}.
All other time series such as energy demands, vRES profiles, and electricity grid CEFs are kept unchanged.
By avoiding the combination of time series from different years, correlations between day-ahead market prices, thermal loads, and renewable energy generation are maintained.
Also we do not use the 2022 price distribution since we assume that they are overestimating the potential due to short-term consequences of the Russian invasion of Ukraine.

%% file: chapter/6_res.tex
\section{Results} \label{sec:res}

In this section, the results of the three contexts are described sequentially and discussed at the end.
The MILP models were solved to global optimality (MIP gap=0) using Gurobi 9.5.1.
The calculation time for one optimization run with seven scenarios was between 10 and 80 minutes depending on the context.
The discount rate $r$ was assumed to be 10\% for all calculations.

\subsection{Results of the base context \scen{c\_base}}
\begin{figure}[!t]
	\begin{overpic}[width=1.0\linewidth]{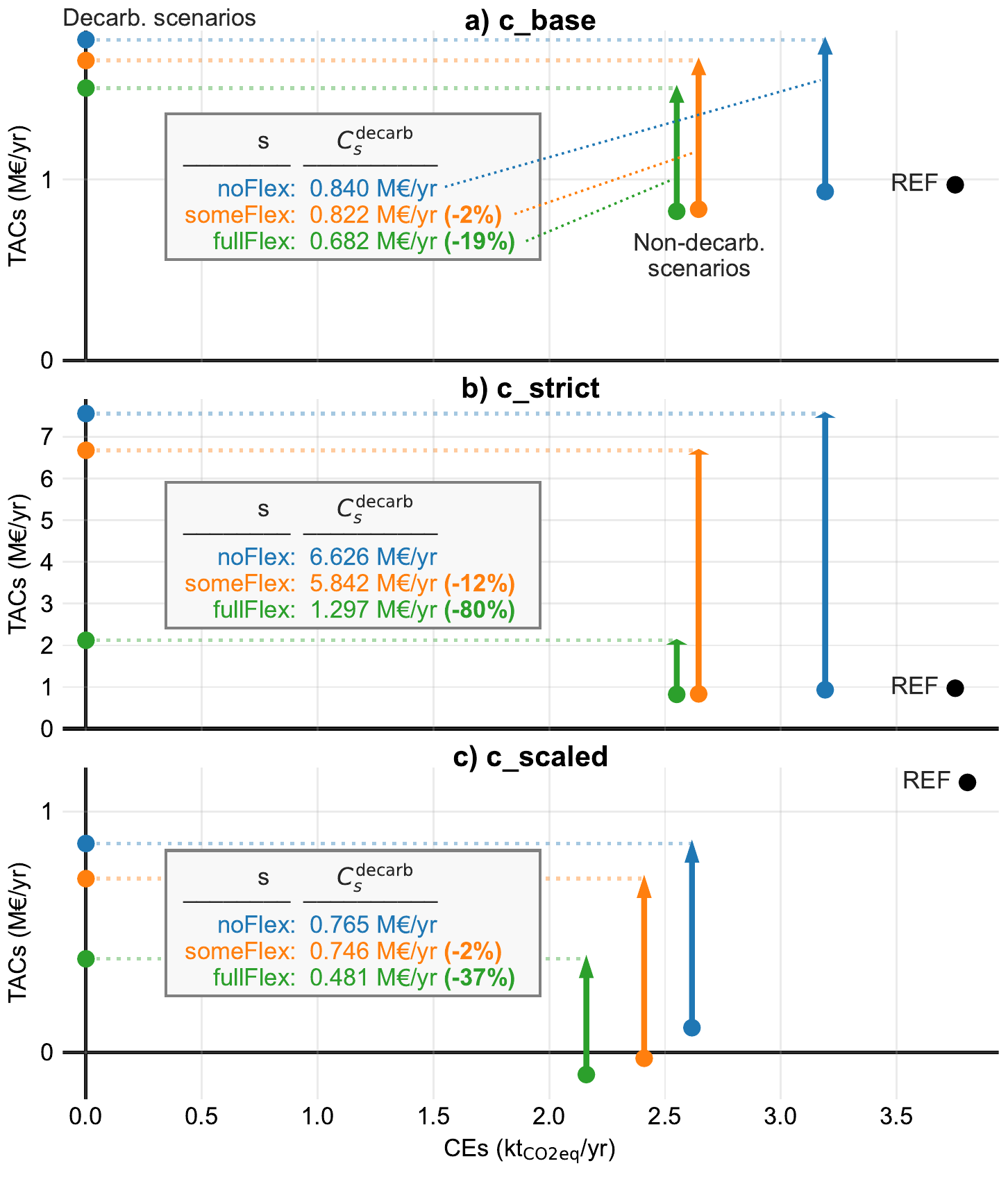}
\end{overpic}
	\caption{Pareto plots and decarbonization costs $C^\mathrm{decarb}_s$ for the three contexts.
		Note, the results of the non-decarbonization scenarios for \scen{c\_base} and \scen{c\_strict} are identical.
		Percentages are based on \scen{noFlex}.
		\label{fig:pareto}}
\end{figure}

\begin{figure}[!htb]
	\centering
	\includegraphics[width=1\linewidth]{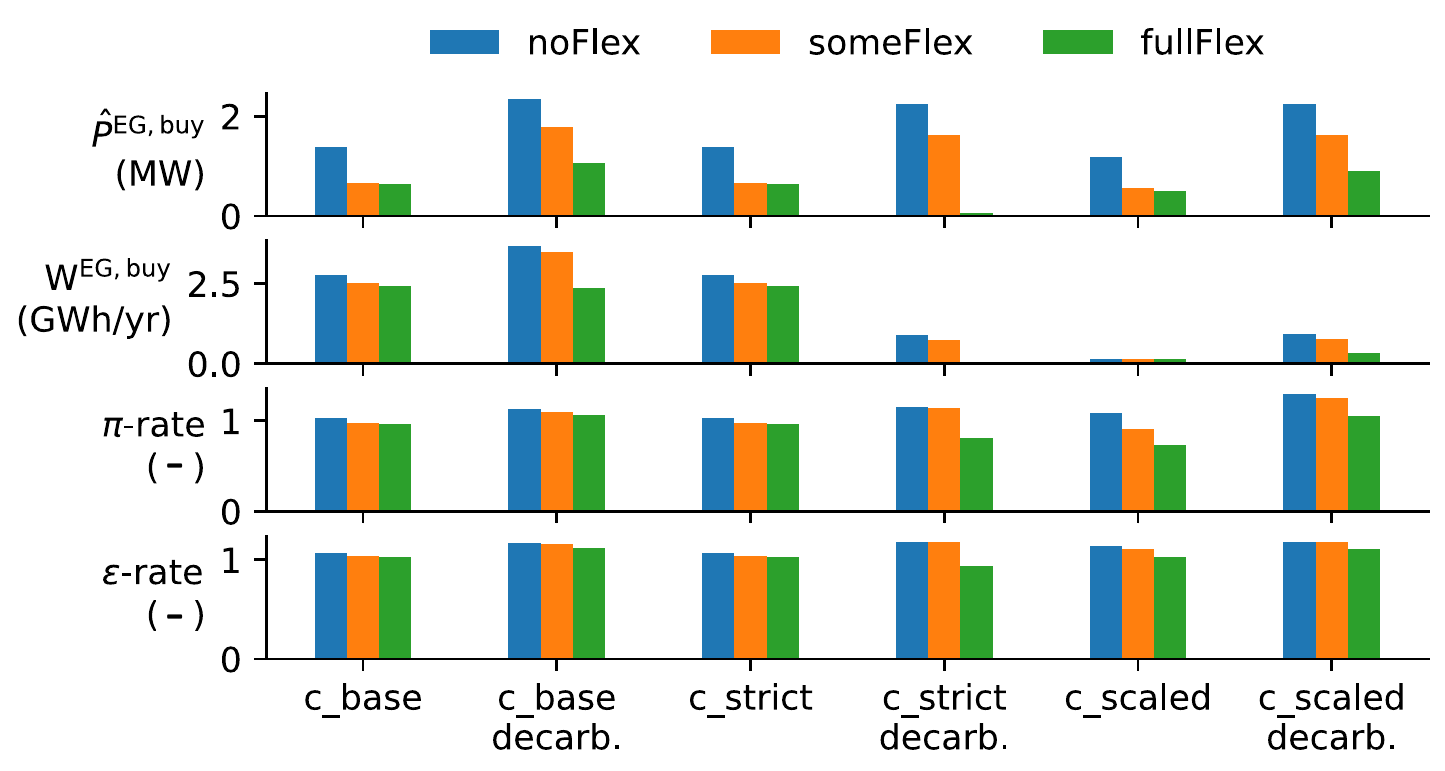}
	\caption{
		Comparison of $\hat{P}^\mathrm{EG,buy}$, W$^\mathrm{EG,buy}$, $\pi$-rate, and $\varepsilon$-rate of all optimized scenarios of all contexts.
		Here, decarb. denote the decarbonization scenarios \scen{noFlex\_d}, \scen{someFlex\_d}, and \scen{fullFlex\_d}.
		\label{fig:metric_bars}}
\end{figure}

\begin{figure}[!t]
	\centering
	\includegraphics[trim=-1.7cm 0 -7cm 0,clip,width=1\linewidth]{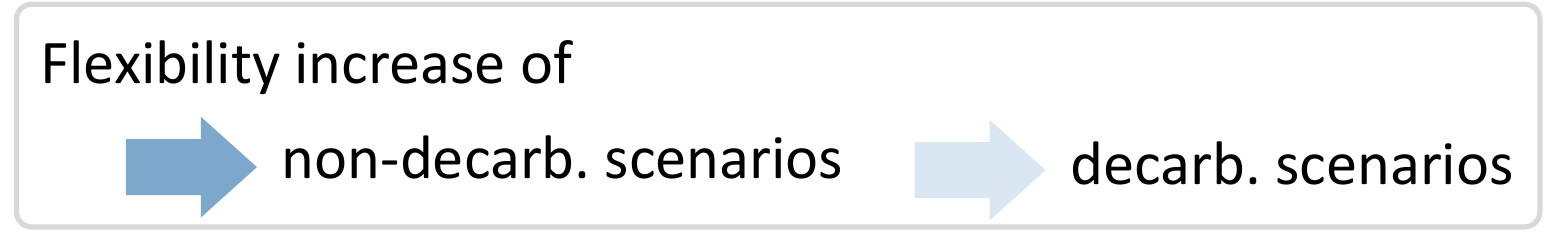}
	\includegraphics[width=1.0\linewidth]{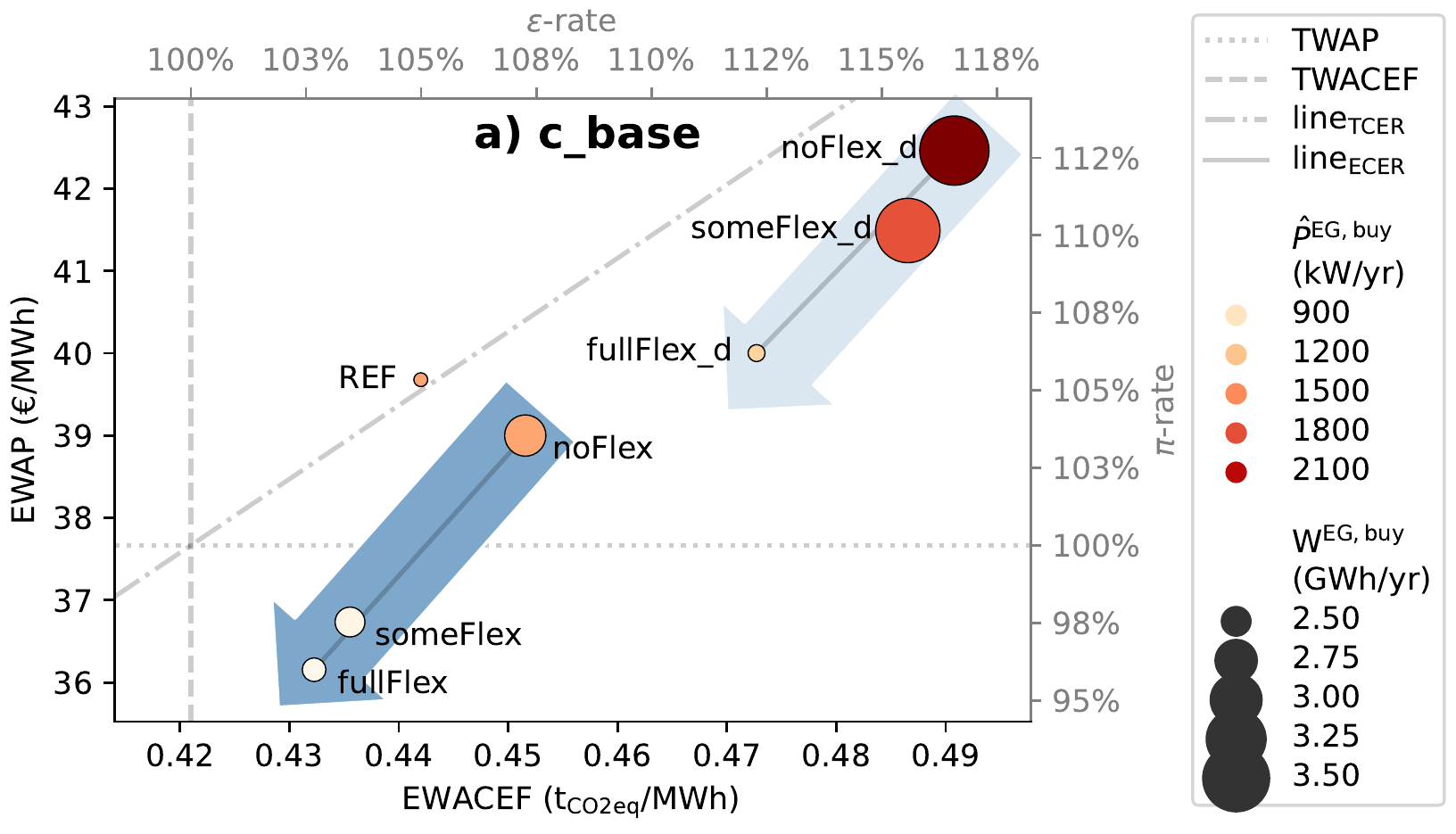}
	\includegraphics[width=1.0\linewidth]{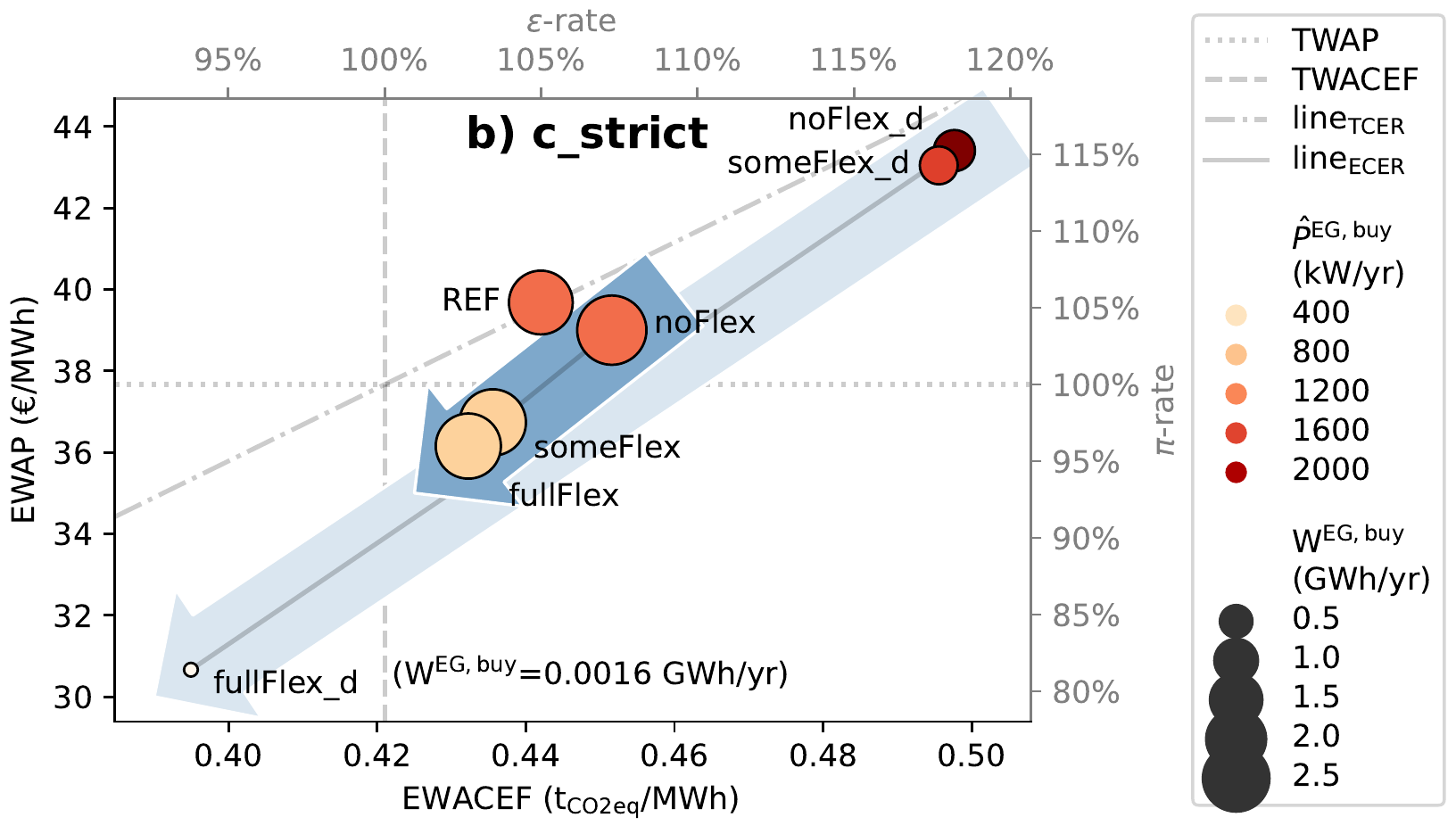}
	\includegraphics[width=1.0\linewidth]{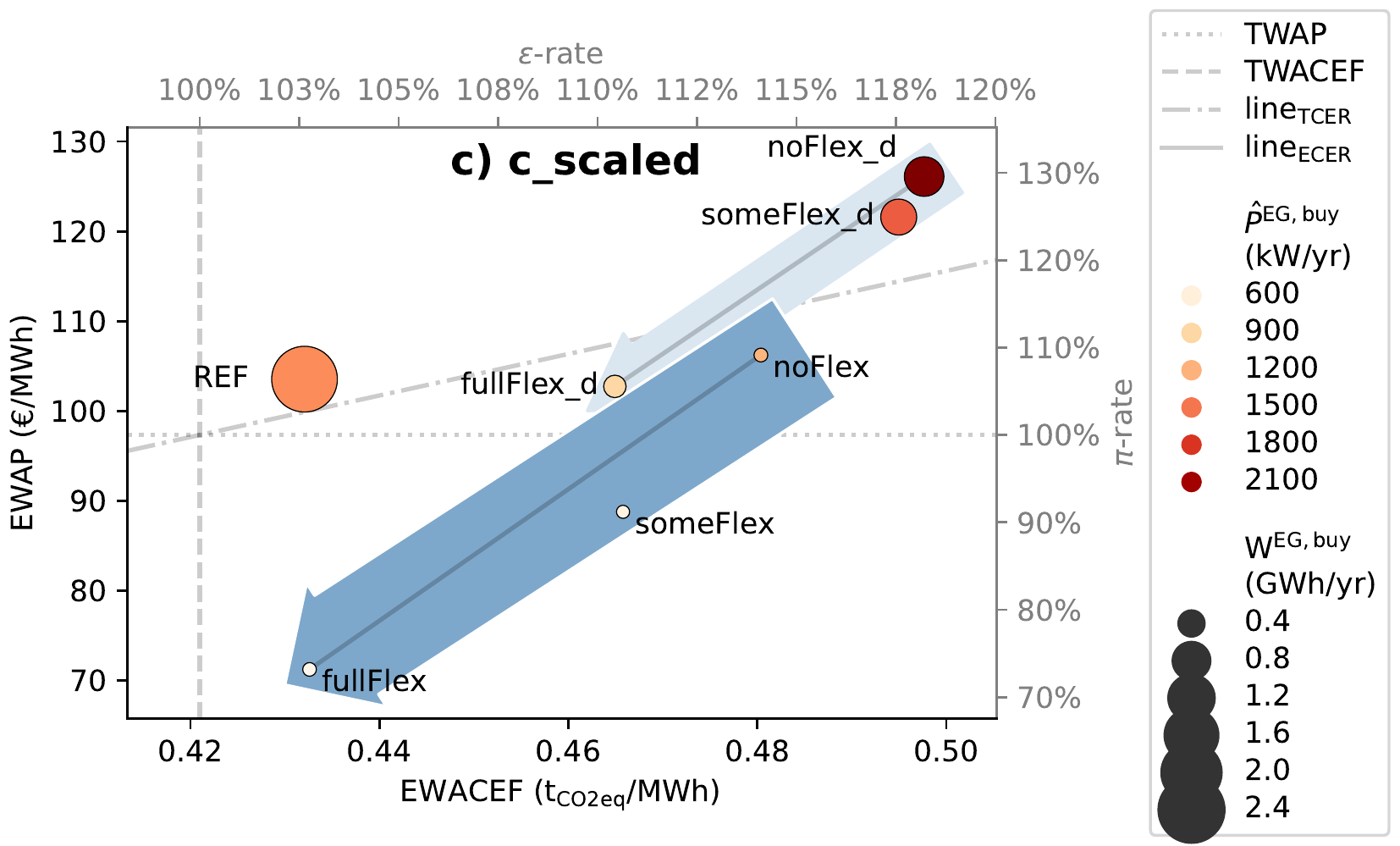}
	\caption{Bubble plot of EWAP and EWACEF of purchased electricity for the three contexts.
		Dark blue and light blue arrows indicate the flexibility increase of non-decarbonization and decarbonization scenarios, respectively.
		Solid grey lines inside the arrows connect \scen{noFlex} and \scen{fullFlex} and have the gradient ECER$_\mathrm{fullFlex(\_d),noFlex(\_d)}$.
		Dotted and dashed lines indicate the context-independent TWAP and TWACEF, respectively.
		Bubble color correspond to electricity peak purchase power $\hat{P}^\mathrm{EG,buy}$.
		Bubble size represents the total electricity purchase volume $W^\mathrm{EG,buy}$.
		Note, the results of the non-decarbonization scenarios for \scen{c\_base} and \scen{c\_strict} are identical.
		\label{fig:WAP}}
\end{figure}

\begin{figure}[h]
	\includegraphics[width=1.00\linewidth]{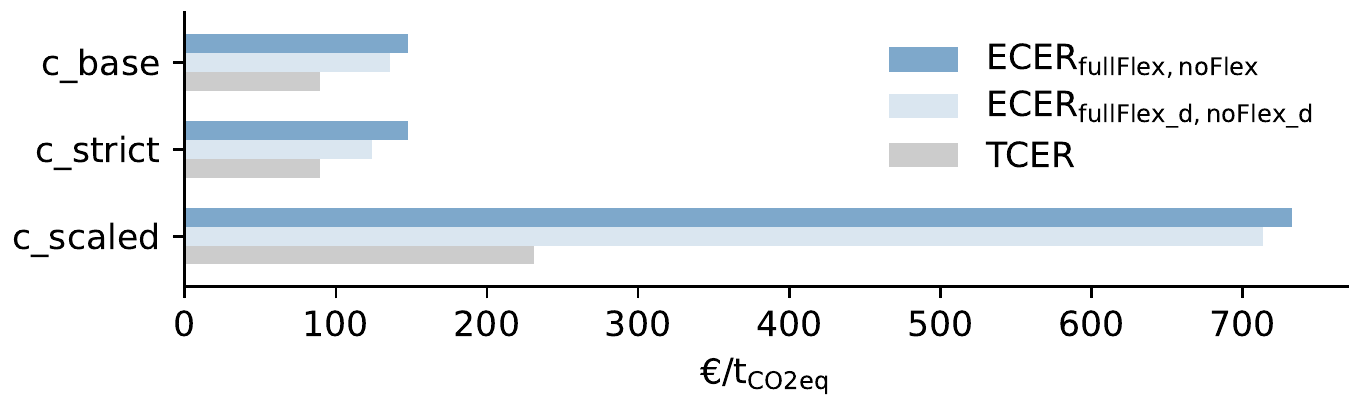}
	\includegraphics[width=1.00\linewidth]{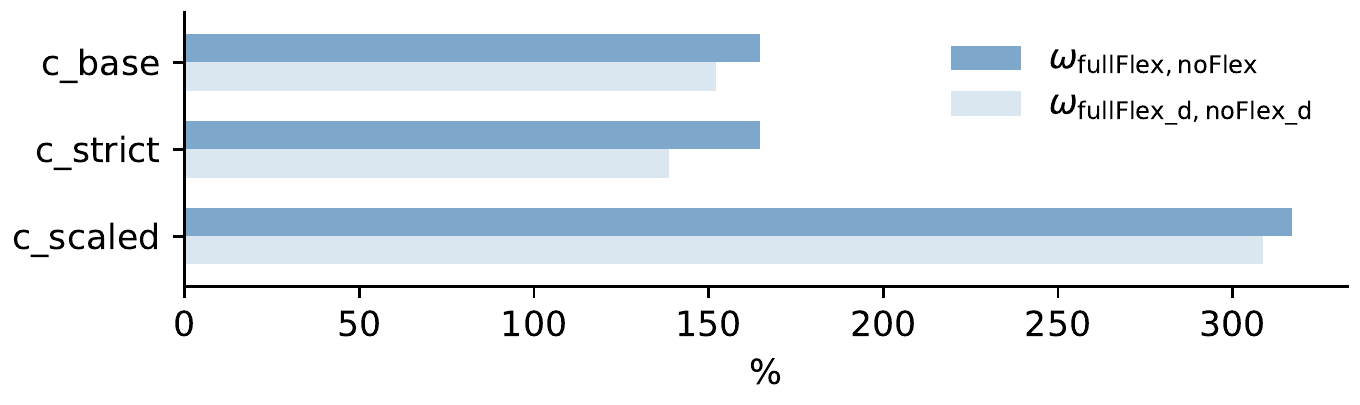}
	\caption{
		Comparison of the gradients of the lines in \Cref{fig:WAP}.
		Top: Absolute gradients (in \euro/t\textsubscript{CO2eq}) with TCER as comparison (i.e., gradient of line\textsubscript{TCER} \Cref{fig:WAP}).
		Bottom: Gradients relative to TCER (in \%).
		\label{fig:wap_slope}}
\end{figure}

\begin{figure}[!htb]
	\centering
	\scenbf{c\_base}\par\vskip 2mm
	\begin{overpic}[width=0.99\linewidth]{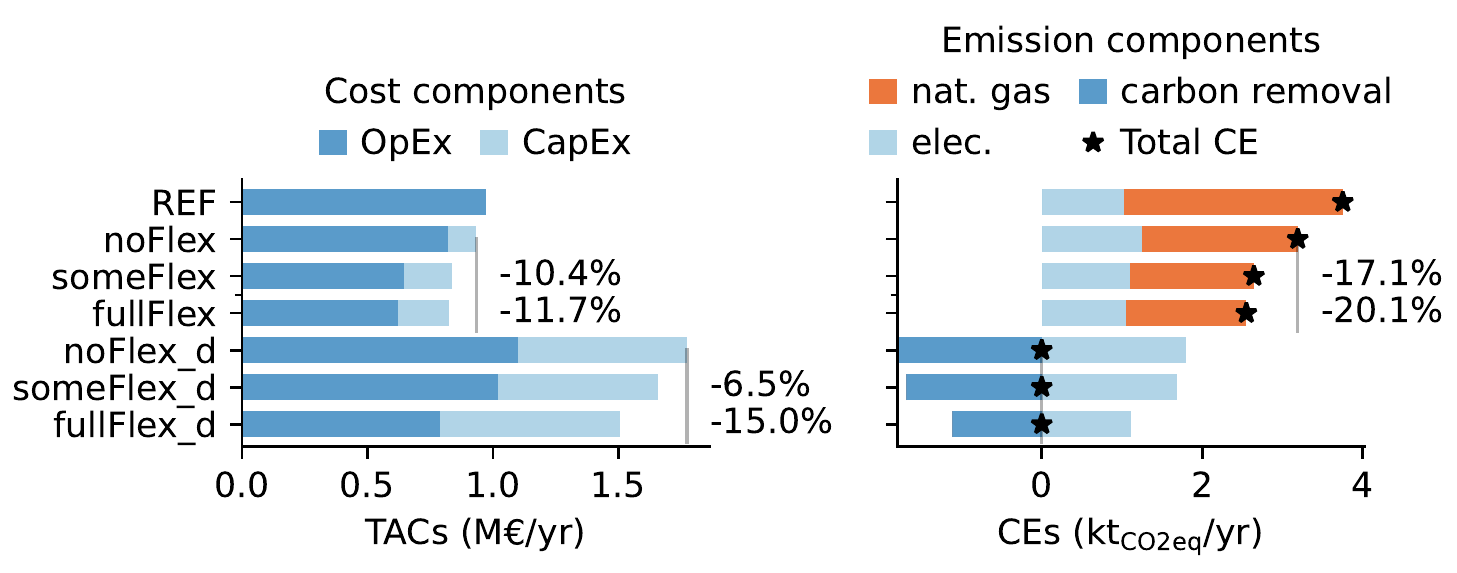} \put(1,35){\textbf{a)}} \put(55,35){\textbf{b)}} \end{overpic}
	\vskip 1mm {\color{lightgray}\hrule} \vskip 1mm
	\begin{overpic}[width=0.99\linewidth]{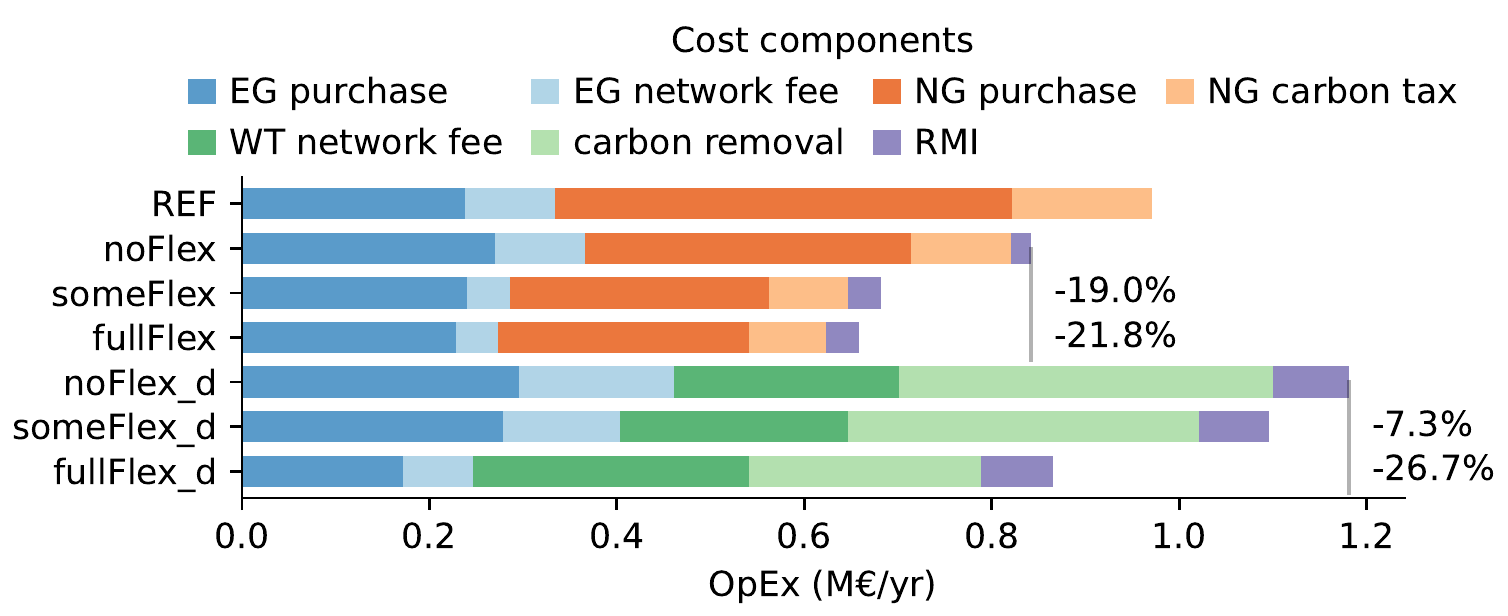} \put(1,36){\textbf{c)}} \end{overpic}
	\vskip 1mm {\color{lightgray}\hrule} \vskip 1mm
	\begin{overpic}[width=0.99\linewidth]{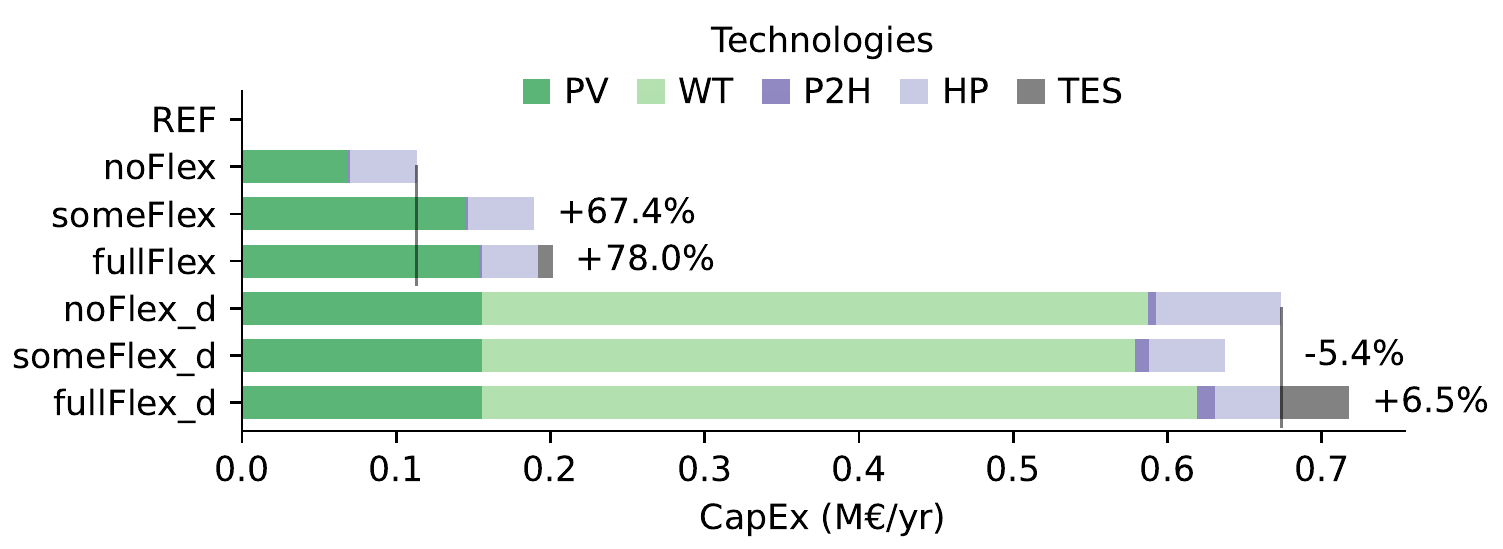} \put(1,32){\textbf{d)}} \end{overpic}
	\vskip 1mm {\color{lightgray}\hrule} \vskip 1mm
	\begin{overpic}[width=0.99\linewidth]{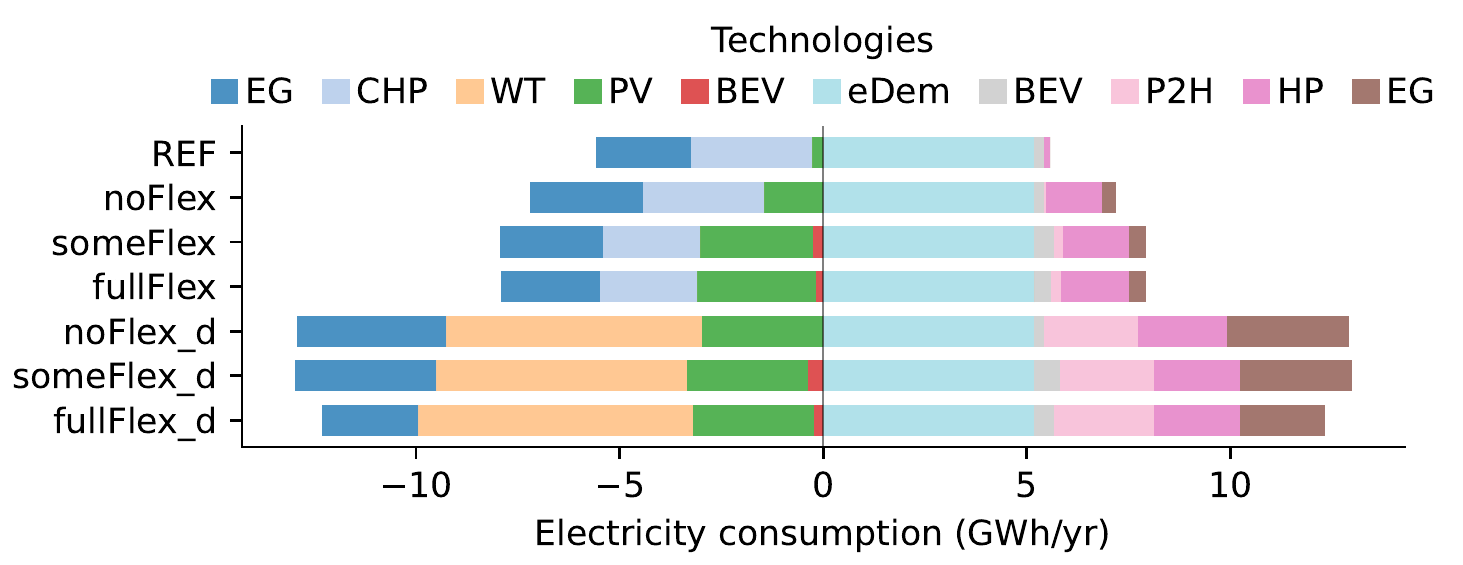} \put(1,34){\textbf{e)}} \end{overpic}
	\vskip 1mm {\color{lightgray}\hrule} \vskip 1mm
	\begin{overpic}[width=0.99\linewidth]{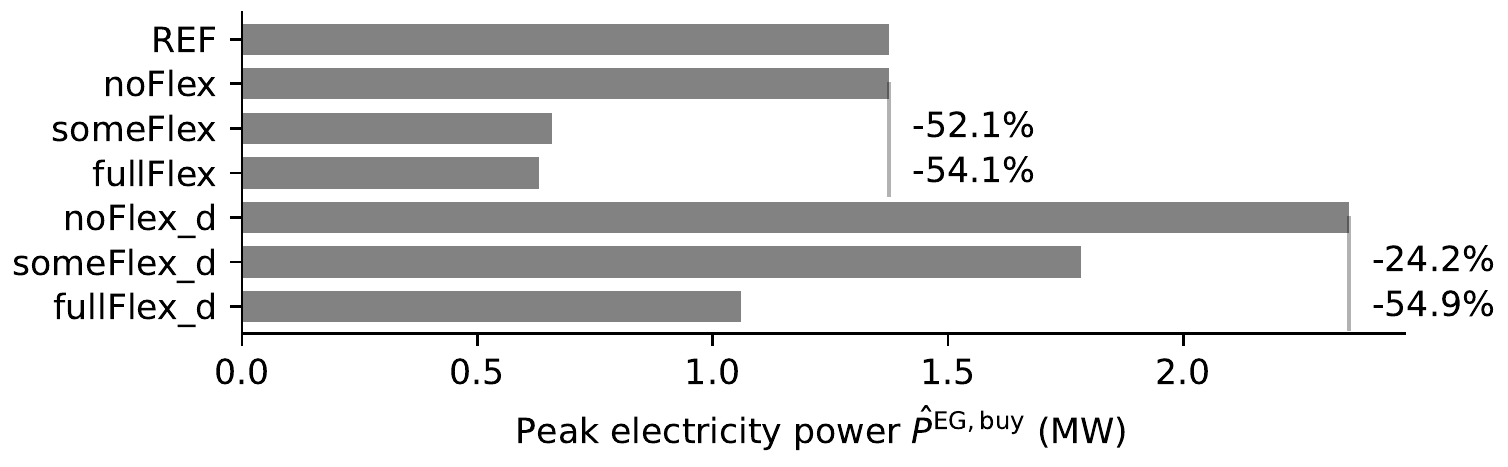} \put(1,27){\textbf{f)}} \end{overpic}
	\vskip 1mm {\color{lightgray}\hrule} \vskip 1mm
	\begin{overpic}[width=0.99\linewidth]{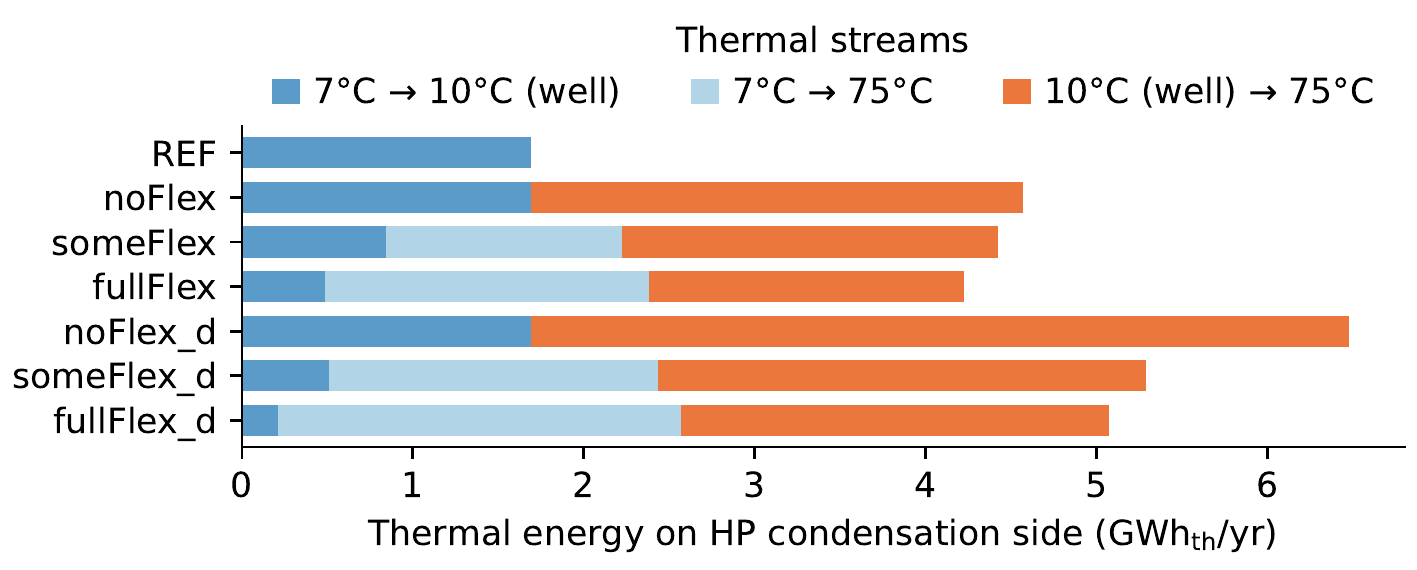} \put(1,36){\textbf{g)}} \end{overpic}
	\caption{Annual results of the \scen{c\_base} context.
		\textbf{a)}~Total annualized costs (TACs),
		\textbf{b)}~operating carbon emissions (CEs),
		\textbf{c)}~operating expenses (OpEx),
		\textbf{d)}~capital expenditures (CapEx),
		\textbf{e)}~electricity consumption (+) and supply ($-$),
		\textbf{f)}~electricity purchase peaks,
		\textbf{g)}~heat pump (HP) operation.
		\textbf{Note}:~Percentages refer to \scen{noFlex} or \scen{noFlex\_d}.
		\label{fig:balance_c_base}}
\end{figure}

\Cref{fig:sankey_c_base,fig:balance_c_base,fig:pareto,fig:WAP,fig:wap_slope,tab:capas_c_base,fig:v2x_cycles,fig:ts_balance,fig:load_violin,fig:heatmap_EG,fig:metric_bars} show the results of the base context \scen{c\_base}.

The Sankey plots in \Cref{fig:sankey_c_base} provide an overview of the resulting annual energy sums of the MES for each scenario.
It can be seen that the heat demands were partly electrified in the non-decarbonization scenarios and fully electrified in the decarbonization scenarios.

\begin{table}[!htb]
\centering
\caption{New capacities for the \scen{c\_base} context \label{tab:capas_c_base}}
\footnotesize
\begin{tabular}{r|ccccc}
\toprule
{} &    PV &    WT &    HP &   P2H &    TES \\
{} &(kW\textsubscript{p})&(kW\textsubscript{P})&(kW\textsubscript{th,cond})&(kW\textsubscript{th}) &(kWh\textsubscript{th}) \\
\midrule
\scen{noFlex} & 1,356 &     0 &   918 &   105 &      0 \\
\scen{someFlex} & 2,853 &     0 &   907 &   197 &      0 \\
\scen{fullFlex} & 3,028 &     0 &   760 &   222 &  3,216 \\
\scen{noFlex\_d} & 3,077 & 2,183 & 1,723 &   485 &      0 \\
\scen{someFlex\_d} & 3,077 & 2,140 & 1,049 &   863 &      0 \\
\scen{fullFlex\_d} & 3,077 & 2,344 &   899 & 1,082 & 14,675 \\
\bottomrule
\end{tabular}
\end{table}

\Cref{tab:capas_c_base} shows the resulting new capacities.
Additional PV capacity was installed in all non-REF scenarios.
In contrast to the decarbonization scenarios, in the non-decarbonization scenarios the upper limit of 3,077\,kW$_\mathrm{p}$ was not chosen.
WTs were only selected in the decarbonization scenarios ranging from 2,140\,kW$_\mathrm{p}$ to 2,344\,kW$_\mathrm{p}$.
HPs were built in all non-REF scenarios ranging from 760\,kW\textsubscript{th} in \scen{fullFlex} to 1,723\,kW\textsubscript{th} in \scen{noFlex\_d}.
P2Hs were built in all non-REF scenarios increasing from 105\,kW\textsubscript{th} in \scen{noFlex} to 1,082\,kW\textsubscript{th} in \scen{fullFlex\_d}.
TESs were optimal in \scen{fullFlex} (3,216\,kWh\textsubscript{th}) and \scen{fullFlex\_d} (14,675\,kWh$_\mathrm{th}$).
Whereas in \scen{full\allowbreak Flex}, the largest TES was built on the cooling temperature level, in \scen{fullFlex\_d} the largest was built on the process heat temperature level.
No BES or H\textsubscript{2}S was selected.

\Cref{fig:balance_c_base} displays the annual balances for costs, CEs, peak power, and energy streams and of all scenarios and its deviations compared to \scen{noFlex} and \scen{noFlex\_d}, respectively.
It can be seen in a) and b) that adding flexibility reduces TACs and CEs.
\scen{someFlex} has 10.4\% less TACs and 17.1\% less CEs than \scen{noFlex}.
\scen{fullFlex} has 11.7\% less TACs and 20.1\% less CEs than \scen{noFlex}.
In the decarbonization case (\scen{\dots\_d}), the TAC reductions were even more pronounced.
\scen{someFlex\_d} has 6.5\% and \scen{fullFlex\_d} 15.0\% less TACs than \scen{noFlex\_d}.
c) and d) show that \scen{someFlex} has 67,4\% higher CapEx but 19\% lower OpEx than \scen{noFlex}.
Similarly, \scen{someFlex\_d} has 78\% higher CapEx but 21.8\% lower OpEx than \scen{noFlex}.
Interestingly, \scen{someFlex\_d} has both lower CapEx and OpEx while \scen{fullFlex\_d} has 6.5\% higher CapEx and 26.7\% lower OpEx than \scen{noFlex\_d}.
From e) and f), we can see that while the purchased electricity $W^\mathrm{EG,buy}$ were comparable for all scenarios, the electricity purchase peak $\hat{P}^\mathrm{EG,buy}$ in \scen{someFlex} and \scen{fullFlex} were reduced by over 50\% compared to \scen{noFlex} and in \scen{some\allowbreak Flex\_d} and \scen{fullFlex\_d} by 24.2\% and 54.9\% reduced compared to \scen{noFlex\_d}, respectively.
From g) we see that while in \scen{REF}, as expected, the HP was only used in cooling mode, in the \scen{noFlex} scenarios, the HP was also used for heating.
In \scen{someFlex} and \scen{someFlex\_d} already most of the cooling energy being used for heating and going to the \scen{fullFlex} scenarios, this share was increased, so only a small fraction of the cooling energy is lost to well water.

\Cref{fig:ts_balance} indicates the electricity balance of one week in May for all scenarios of \scen{c\_base} and the day-ahead electricity price $c^\mathrm{elec}_t$.
In \scen{someFlex} and \scen{fullFlex}, fuel switches from natural gas (CHP) to electricity (P2H) are observable during times of high PV supply.
In all scenarios with flexibility, V2X activity can be seen in times of low PV and WT energy supply and/or high electricity prices.
High activities of flexible loads (HP, P2H, BEV) can be seen in times of high PV and WT energy supply and/or low electricity prices.
In \scen{fullFlex\_d}, own renewable energy is almost entirely used and not fed-in.
By comparing \scen{fullFlex} with \scen{noFlex}, it can be seen how the EG purchase profile is shaved.

\Cref{fig:v2x_cycles} presents the number of full V2X discharges per year.
Interestingly, the \scen{fullFlex} scenarios feature less discharges than the \scen{someFlex} scenarios.

\Cref{fig:pareto} a) depicts the pareto plot and the decarbonization costs $C^\mathrm{decarb}_s$ for the \scen{c\_base} context.
It can be seen that $C^\mathrm{decarb}_\scen{someFlex}$ were 2\% and $C^\mathrm{decarb}_\scen{fullFlex}$ 19\% lower than $C^\mathrm{decarb}_\scen{noFlex}$.

The added flexibility is also used for price-based and CEF-based DR, i.e., to optimize the electricity purchase against electricity prices and CEFs, respectively.
The EWAP and EWACEF are good indicators for this.
They are shown in \Cref{fig:WAP} a) for all scenarios together with the annual sum and the peak of the electricity purchase.
It can be seen that with increasing flexibility, electricity purchase decreases in volume ($W^\mathrm{EG,buy}$), peak ($\hat{P}^\mathrm{EG,buy}$), carbon intensity (EWACEF), and price (EWAP).
In other words, the flexibility allows for greener and cheaper purchases of less electricity.
It can also be seen that the dark blue arrow (indicating non-decarbonization scenarios) crosses the TWAP line since in \scen{someFlex} and \scen{fullFlex}, the EWAP is lower than the TWAP while in \scen{noFlex}, the EWAP is higher than the TWAP.
This shows that flexibility reduced the average paid price (EWAP) to below the average price (TWAP).

\Cref{fig:wap_slope} shows the gradients of line\textsubscript{TCER} and line\textsubscript{ECER} in \Cref{fig:WAP}.
As expected, the gradient in the decarbonization case is lower than in the non-decarbonization case.

\begin{figure}[h]
	\includegraphics[trim=0 1cm 0 0.2cm, clip, width=1.00\linewidth]{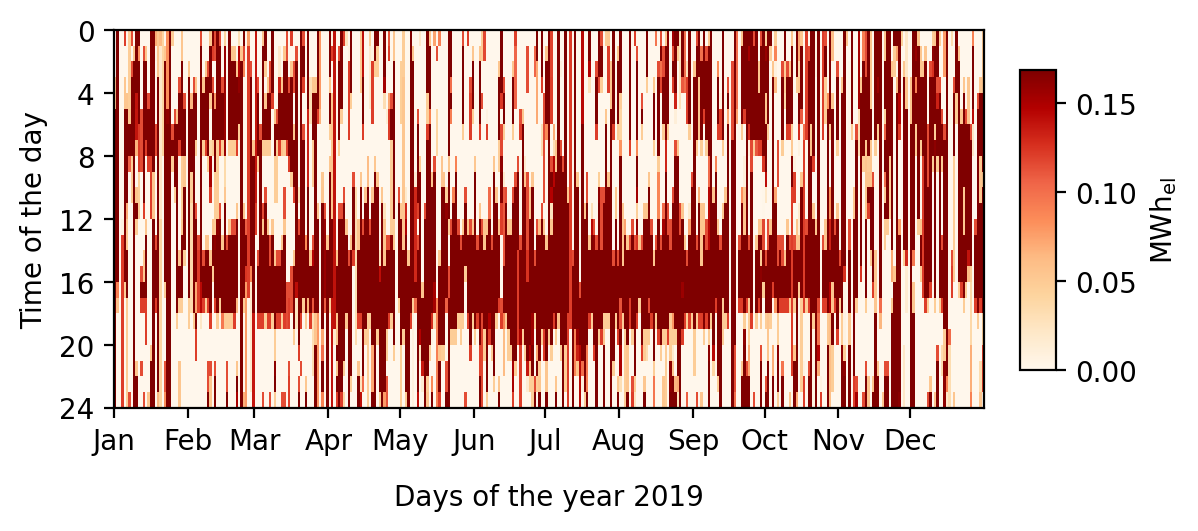}
	\includegraphics[trim=0 1cm 0 0.2cm, clip, width=0.979\linewidth]{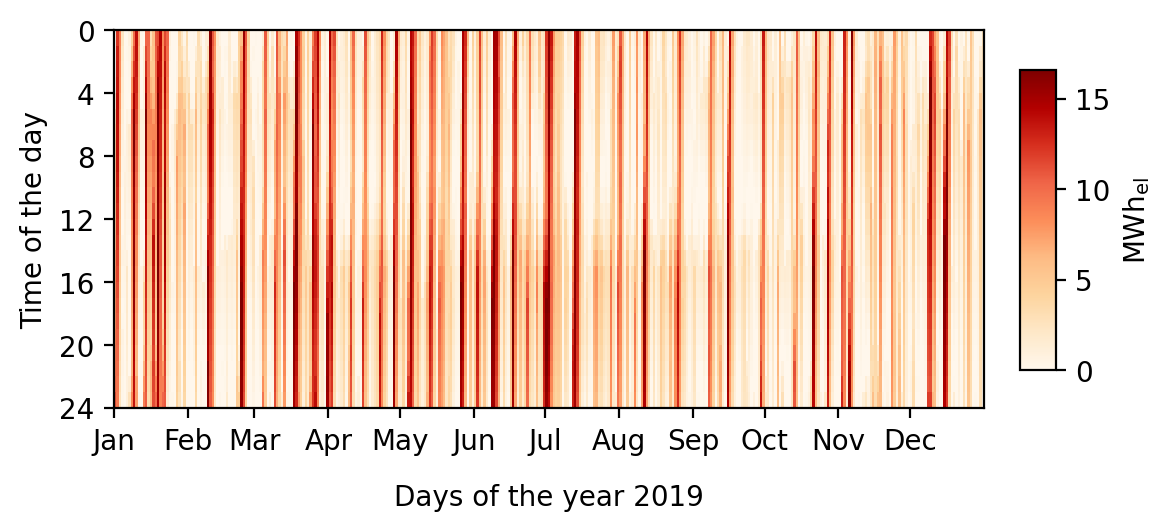}
	\includegraphics[trim=0 0cm 0 0.2cm, clip, width=0.992\linewidth]{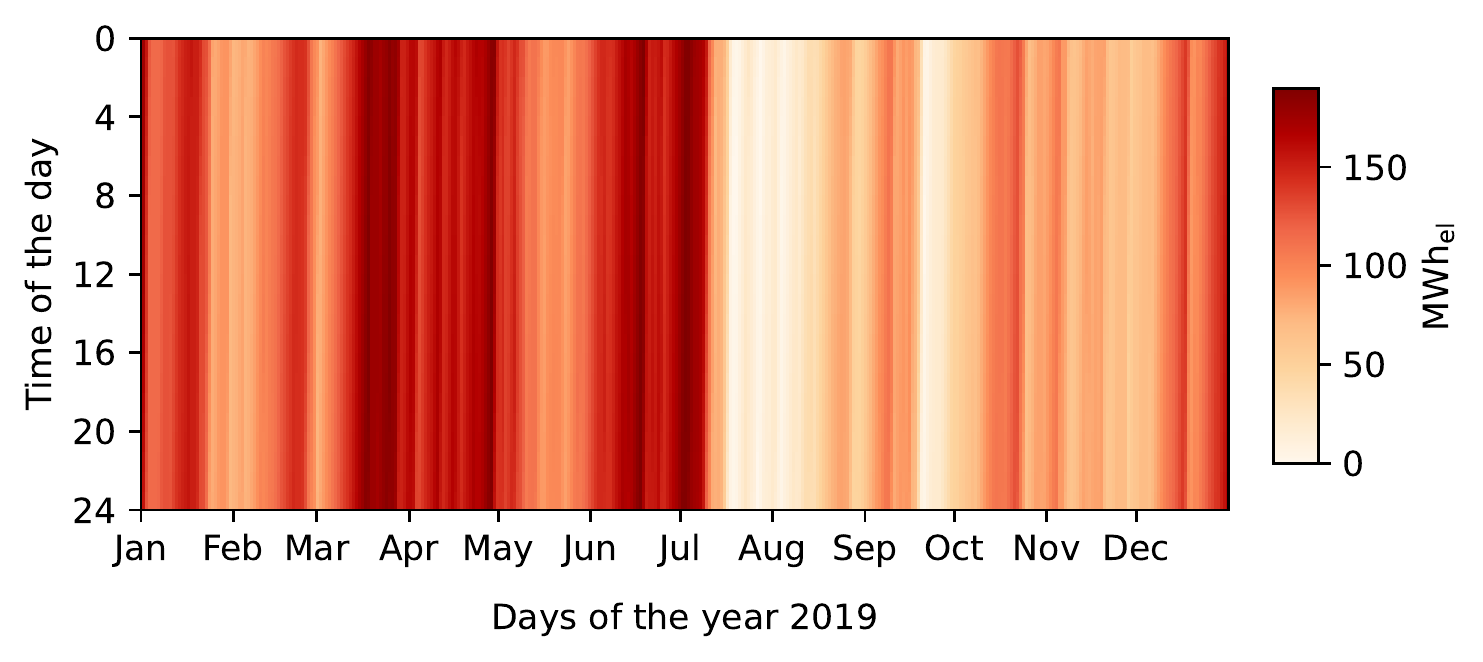}
	\caption{Energy filling level of BES (top), 97/75-TES (middle), and H\textsubscript{2}S (bottom) in scenario \scen{fullFlex\_d} of context \scen{c\_strict}.
	\label{fig:heatmap_H2}}
\end{figure}

\begin{figure}[!htb]
	\centering
	\scenbf{c\_strict}\par\vskip 2mm
	\begin{overpic}[width=0.99\linewidth]{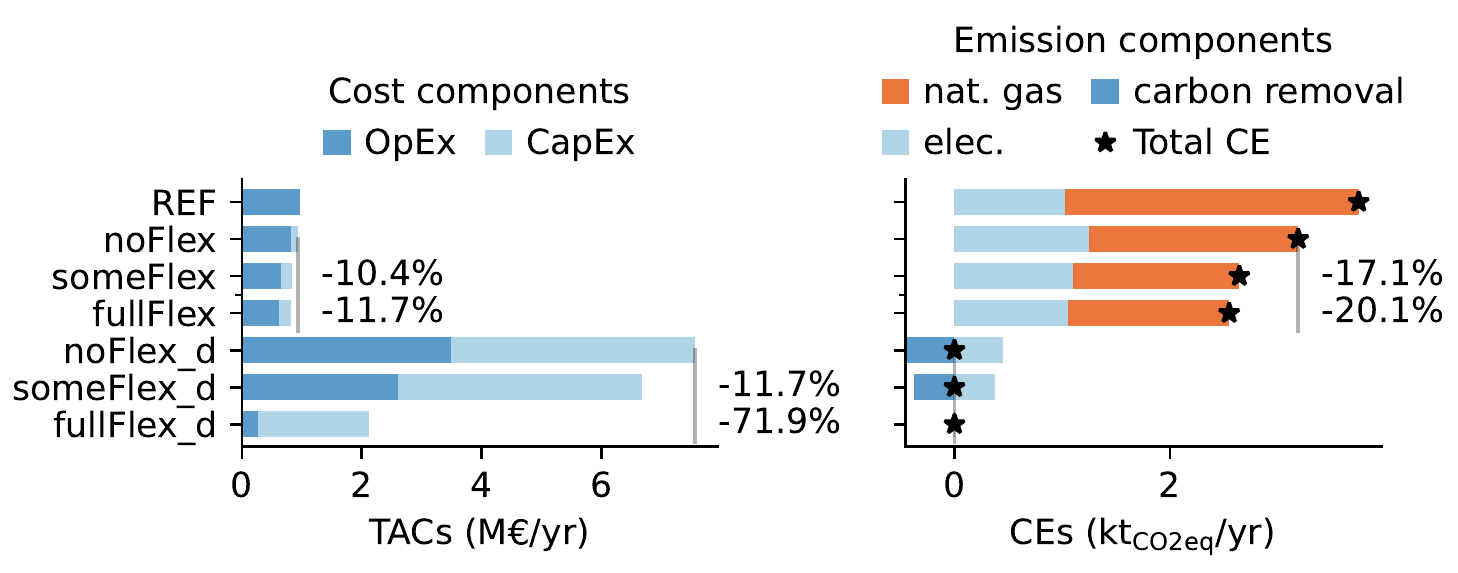} \put(1,35){\textbf{a)}} \put(55,35){\textbf{b)}} \end{overpic}
	\vskip 1mm {\color{lightgray}\hrule} \vskip 1mm
	\begin{overpic}[width=0.99\linewidth]{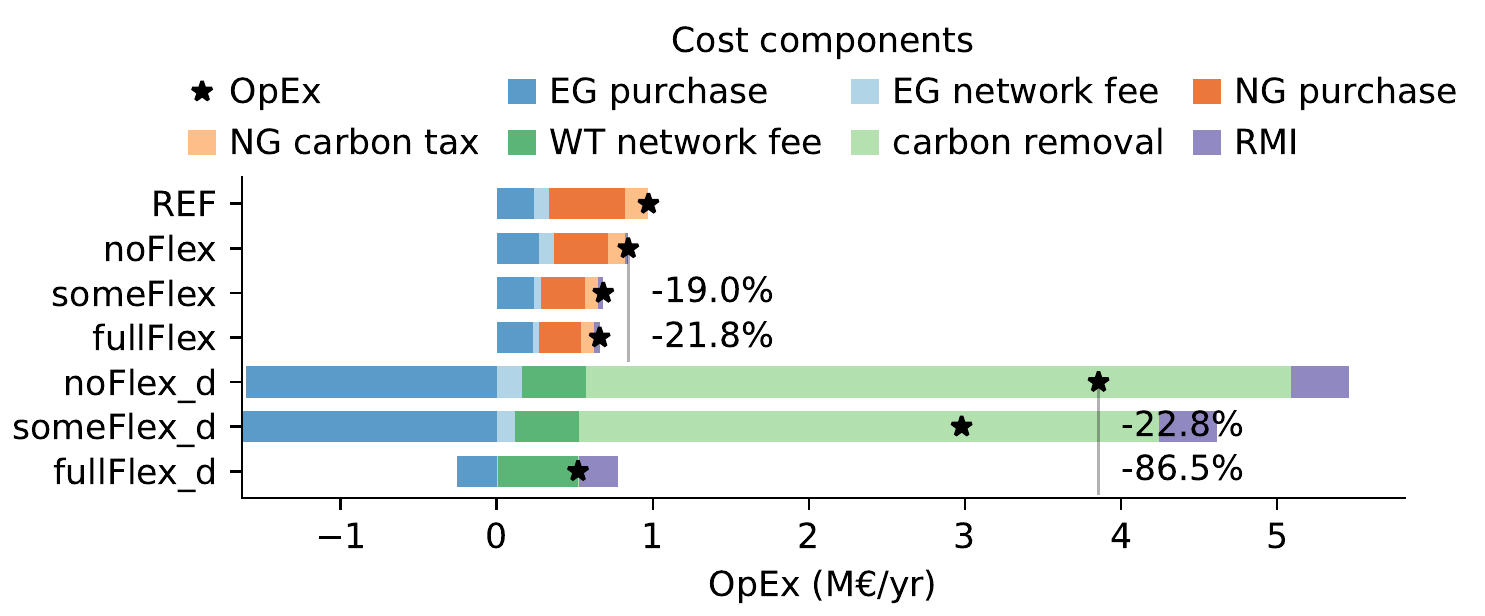} \put(1,38){\textbf{c)}} \end{overpic}
	\vskip 1mm {\color{lightgray}\hrule} \vskip 1mm
	\begin{overpic}[width=0.99\linewidth]{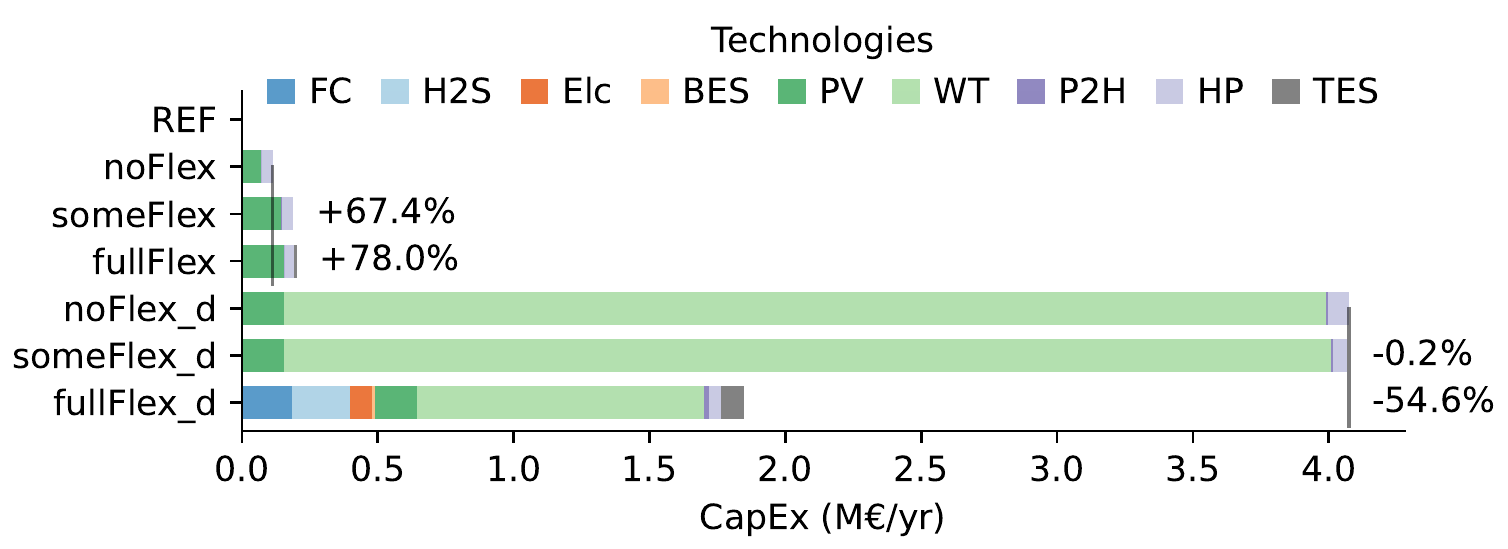} \put(1,32){\textbf{d)}} \end{overpic}
	\vskip 1mm {\color{lightgray}\hrule} \vskip 1mm
	\begin{overpic}[width=0.99\linewidth]{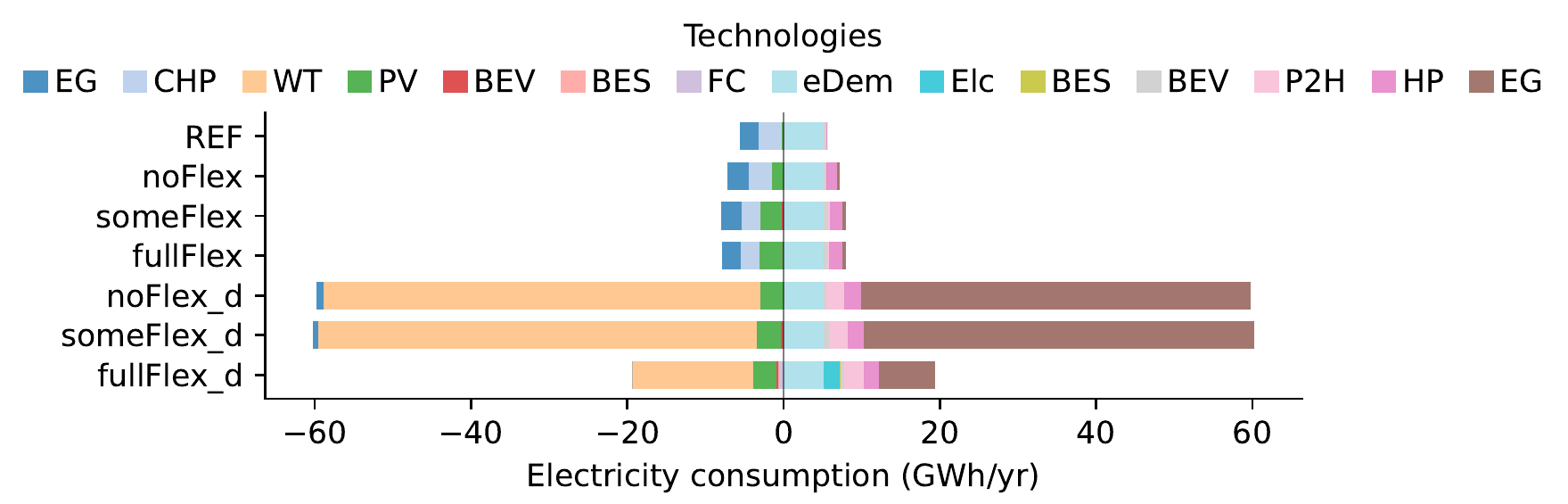} \put(1,30){\textbf{e)}} \end{overpic}
	\vskip 1mm {\color{lightgray}\hrule} \vskip 1mm
	\begin{overpic}[width=0.99\linewidth]{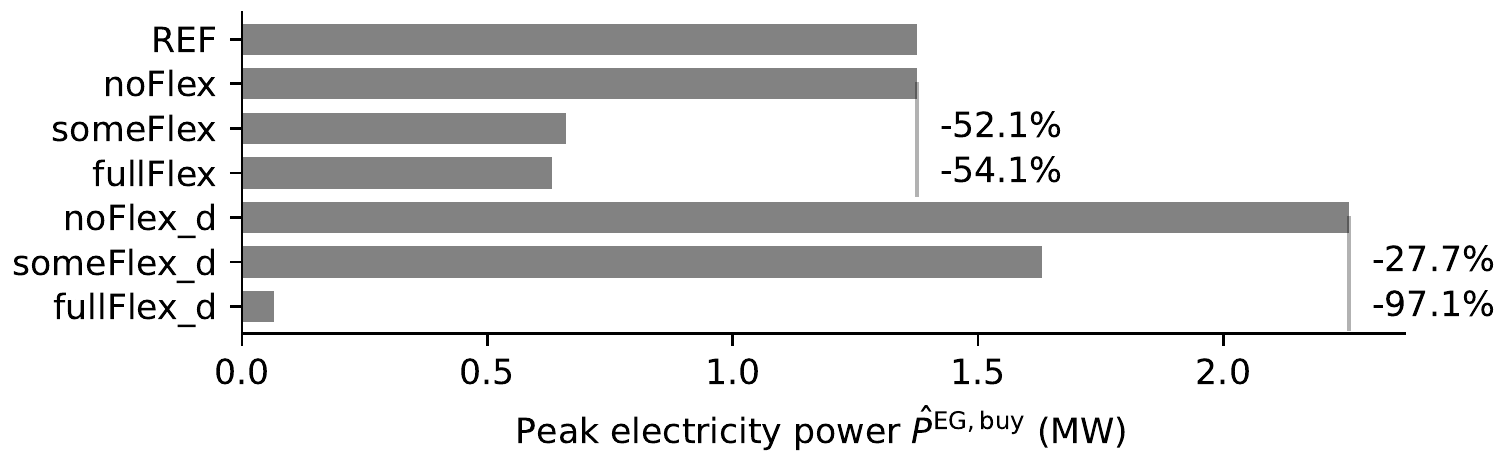} \put(1,26){\textbf{f)}} \end{overpic}
	\vskip 1mm {\color{lightgray}\hrule} \vskip 1mm
	\begin{overpic}[width=0.99\linewidth]{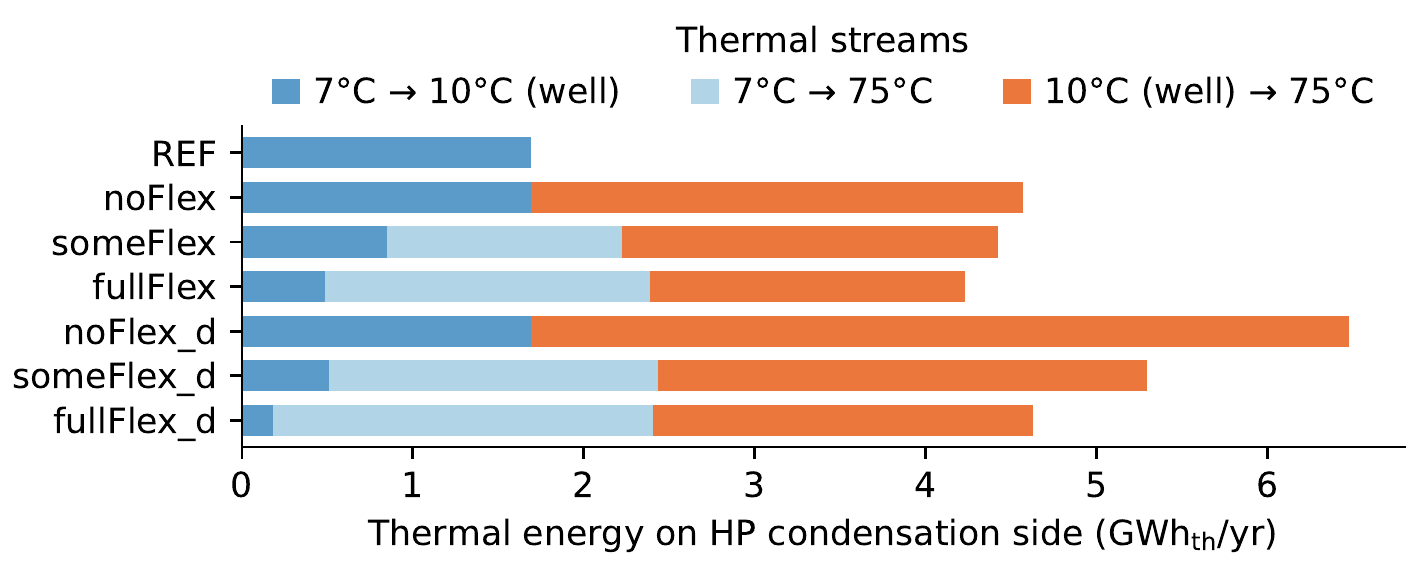} \put(1,36){\textbf{g)}} \end{overpic}
	\caption{Annual results of the \scen{c\_strict} context.
		\textbf{a)}~Total annualized costs (TAC),
		\textbf{b)}~operating carbon emissions (CEs),
		\textbf{c)}~operating expenses (OpEx),
		\textbf{d)}~capital expenditures (CapEx), 
		\textbf{e)}~electricity consumption (+) and supply ($-$), 
		\textbf{f)}~electricity purchase peaks, 
		\textbf{g)}~heat pump (HP) operation.
		\textbf{Note}:~Percentages refer to \scen{noFlex} or \scen{noFlex\_d}.
		\label{fig:balance_c_strict}}
\end{figure}

\subsection{Results of the strict decarbonization context \scen{c\_strict}}
The results of the \scen{c\_strict} context are shown in \Cref{fig:sankey_c_strict,fig:balance_c_strict,fig:pareto,tab:capn,fig:WAP,fig:wap_slope,fig:heatmap_H2,fig:load_violin,fig:heatmap_EG,fig:metric_bars}.
In \scen{c\_strict}, $C^\mathrm{decarb}_\scen{someFlex}$ were 12\% and $C^\mathrm{decarb}_\scen{fullFlex}$ 80\% lower than $C^\mathrm{decarb}_\scen{noFlex}$, see \Cref{fig:pareto} b).
Even with a carbon removal price of \euro10k\,t\,$_\mathrm{CO2eq}^{-1}$ in \scen{noFlex\_d} and \scen{someFlex\_d} significant amounts of electricity-based CEs were removed to reach carbon neutrality, while in \scen{full\allowbreak Flex\_d} carbon removal was not necessary, see \Cref{fig:balance_c_strict} b).
This demonstrates that strict decarbonization without energy flexibility is infeasible, at least when the 2019 power plant mix is assumed.
In \scen{fullFlex\_d}, strict decarbonization was possible without carbon removal by investing in BES, TES, and H\textsubscript{2}S technologies, see \Cref{fig:sankey_c_strict}, \Cref{fig:balance_c_strict} b), and \Cref{tab:capn}.
While the BES stored (mainly solar) energy for a few hours, the TES stored energy for a few days, and the H\textsubscript{2}S stored energy for multiple days or weeks, see \Cref{fig:heatmap_H2}.
The reason for this is that H\textsubscript{2} storage chains (Elc, H\textsubscript{2}S, FC) feature higher kW-costs and lower kWh-costs than BESs.
By comparing \Cref{fig:heatmap_H2} bottom and \Cref{fig:pv_wind_heatmap}, it can be seen that the H\textsubscript{2}S is completely discharged to cover several days of low solar and wind generation in the mid of July.
Through storage technologies (BES, TES, H\textsubscript{2}S) the WT capacity of \scen{noFlex\_d} and \scen{someFlex\_d} can be reduced from over 19\,MW$_\mathrm{p}$ to 5.3\,MW$_\mathrm{p}$, see \Cref{tab:capn}.

The long blue arrow in \Cref{fig:WAP} b) indicates a wide range of EWAP and EWACEF for the decarbonization scenarios.
Whereas in \scen{noFlex\_d} and \scen{someFlex\_d}, the metrics (EWAP, EWACEF, $\hat{P}^\mathrm{EG,buy}$, and W$^\mathrm{EG,buy}$) differed only slightly compared to \scen{c\_base}, in \scen{fullFlex\_d}, they were very low compared to the other scenarios.
This reveals that in \scen{fullFlex\_d}, the flexibility was not only used to optimize self-consumption and buy less electricity but also to buy in hours of lower prices and CEFs.

The \scen{fullFlex\_d} scenario features 71.9\% less TACs than \scen{noFlex\_d}.
What is striking is that through the flexibility, the CapEx could be lowered by 54.6\% mainly due to less WT capacity and the OpEx could be lowered by 86.5\% mainly due to avoiding carbon removal costs.

\subsection{Results of the scaled electricity price context \scen{c\_scaled}}
\begin{figure}[!htb]
	\centering
	\scenbf{c\_scaled}\par\vskip 2mm
	\begin{overpic}[width=0.99\linewidth]{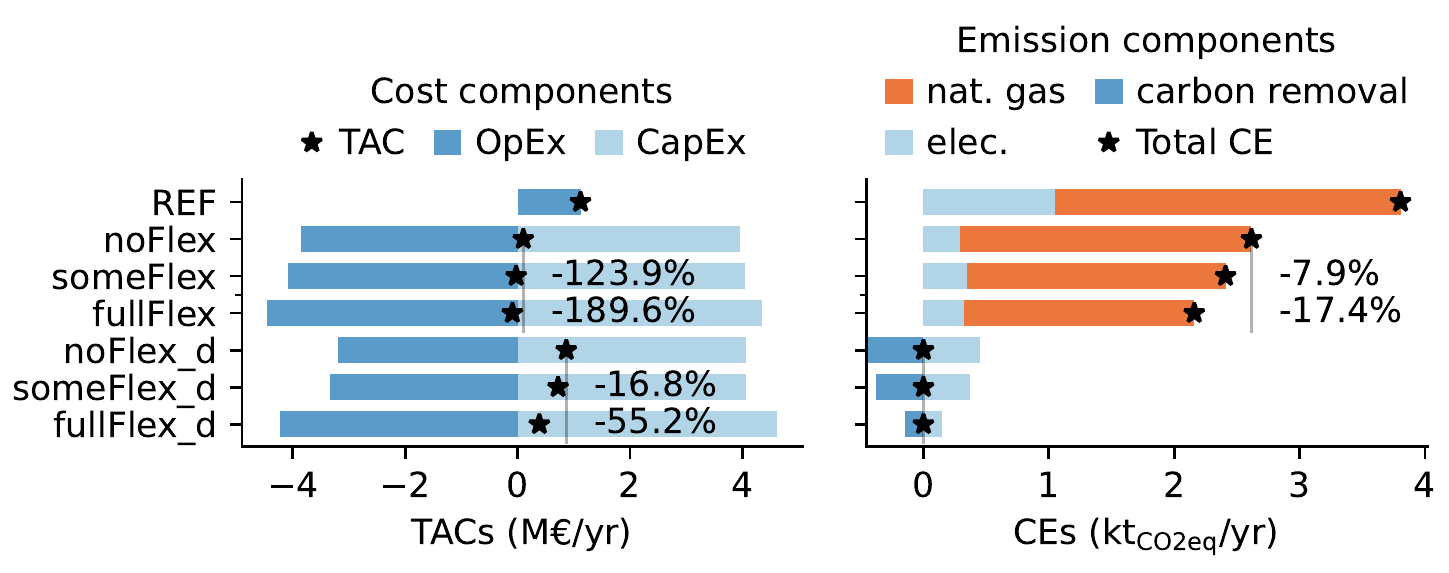} \put(1,35){\textbf{a)}} \put(55,35){\textbf{b)}} \end{overpic}
	\vskip 1mm {\color{lightgray}\hrule} \vskip 1mm
	\begin{overpic}[width=0.99\linewidth]{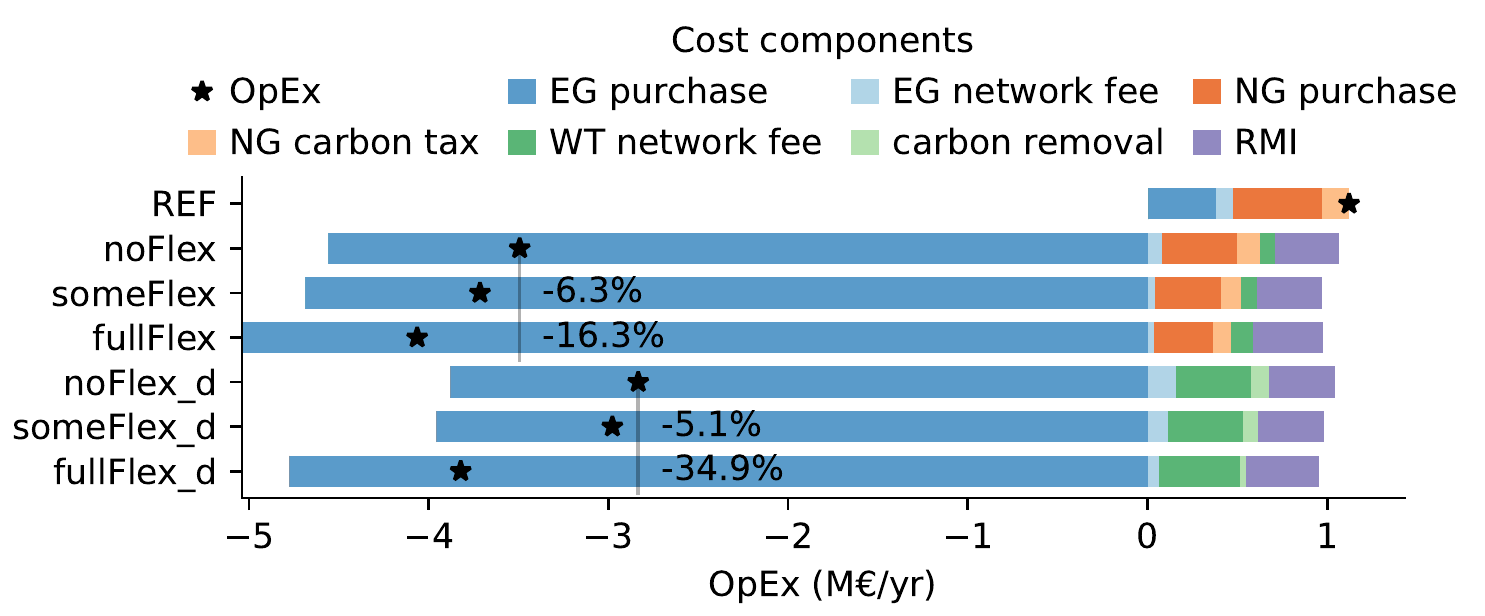} \put(1,38){\textbf{c)}} \end{overpic}
	\vskip 1mm {\color{lightgray}\hrule} \vskip 1mm
	\begin{overpic}[width=0.99\linewidth]{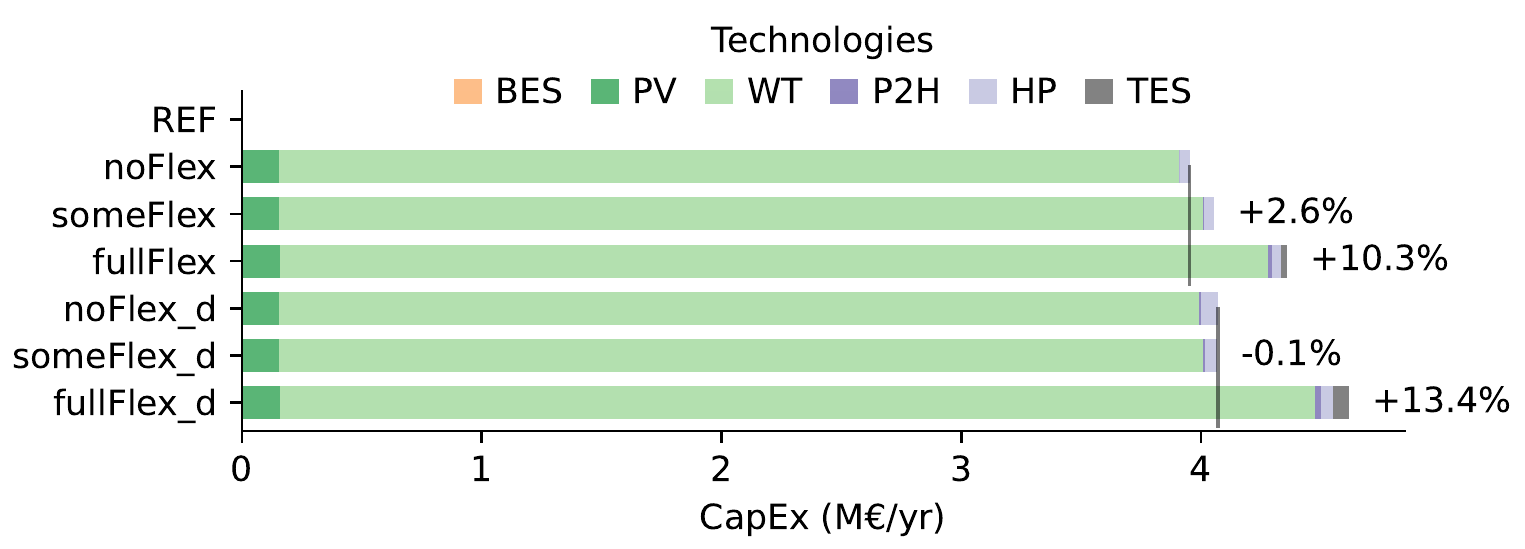} \put(1,32){\textbf{d)}} \end{overpic}
	\vskip 1mm {\color{lightgray}\hrule} \vskip 1mm
	\begin{overpic}[width=0.99\linewidth]{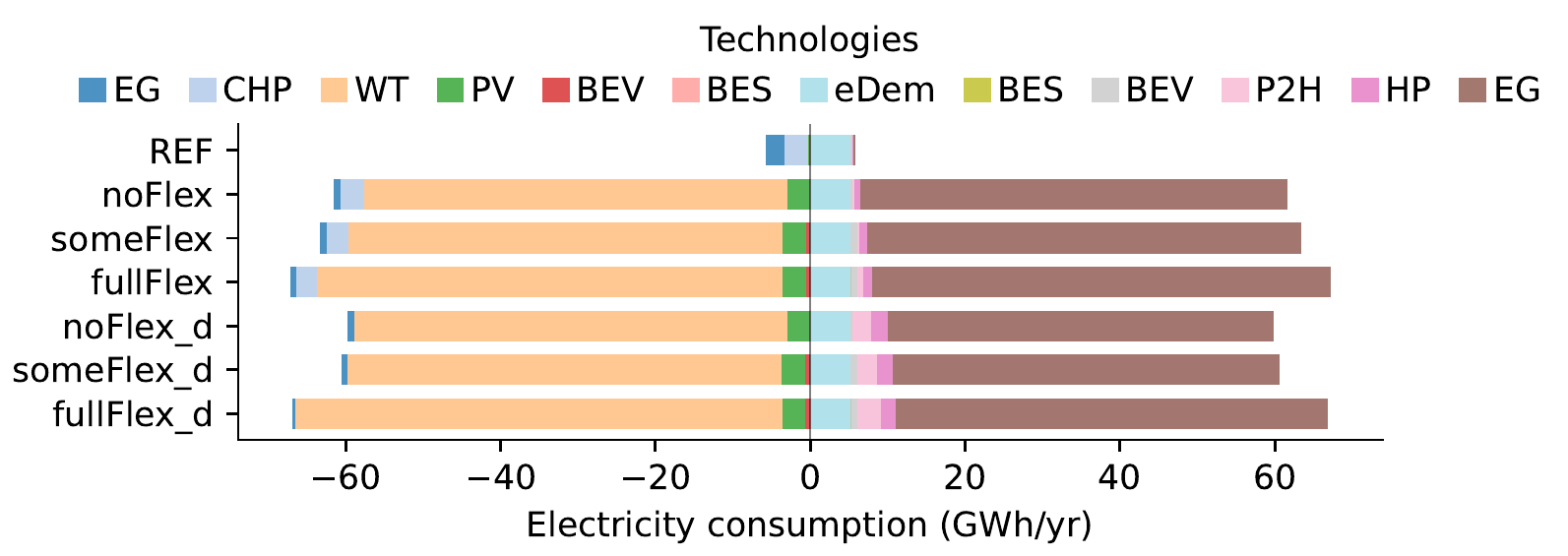} \put(1,32){\textbf{e)}} \end{overpic}
	\vskip 1mm {\color{lightgray}\hrule} \vskip 1mm
	\begin{overpic}[width=0.99\linewidth]{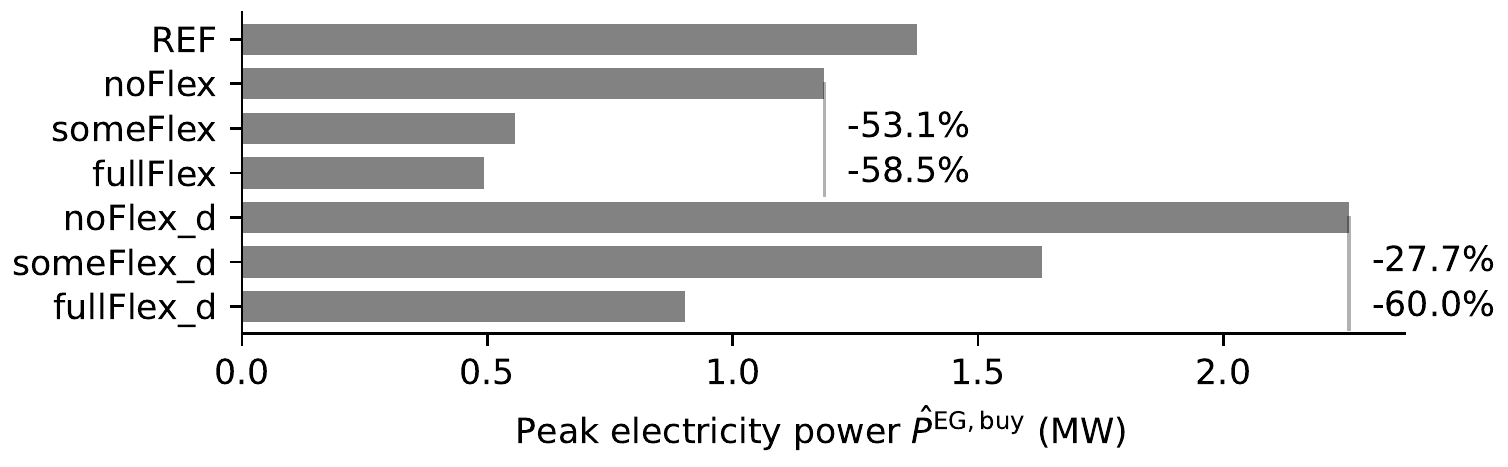} \put(1,26){\textbf{f)}} \end{overpic}
	\vskip 1mm {\color{lightgray}\hrule} \vskip 1mm
	\begin{overpic}[width=0.99\linewidth]{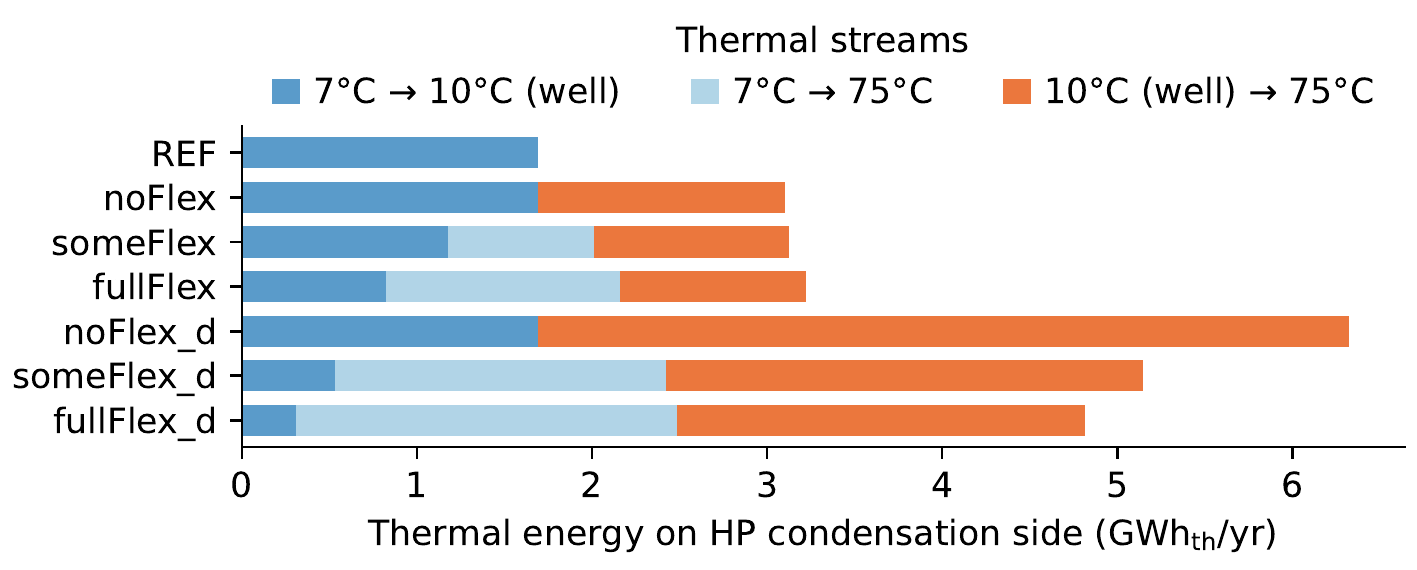} \put(1,36){\textbf{g)}} \end{overpic}
	\caption{Annual results of the \scen{c\_scaled} context.
		\textbf{a)}~Total annualized costs (TAC),
		\textbf{b)}~operating carbon emissions (CEs),
		\textbf{c)}~operating expenses (OpEx),
		\textbf{d)}~capital expenditures (CapEx), 
		\textbf{e)}~electricity consumption (+) and supply ($-$), 
		\textbf{f)}~electricity purchase peaks, 
		\textbf{g)}~heat pump (HP) operation.
		\textbf{Note}: Percentages refer to \scen{noFlex} or \scen{noFlex\_d}.
		\label{fig:balance_c_scaled}}
\end{figure}

The results of the \scen{c\_scaled} context are shown in \Cref{fig:sankey_c_scaled,fig:balance_c_scaled,fig:pareto,tab:capn,fig:WAP,fig:wap_slope,fig:load_violin,fig:heatmap_EG,fig:metric_bars}.
In \scen{c\_scaled}, $C^\mathrm{decarb}_\scen{someFlex}$ were 2\% and $C^\mathrm{decarb}_\scen{fullFlex}$ 37\% lower than $C^\mathrm{decarb}_\scen{noFlex}$, see \Cref{fig:pareto} c).
In contrast to \scen{c\_base}, in \scen{c\_scaled} a (small) BES was selected in \scen{fullFlex} and the maximum possible PV capacity of 3.077\,MW$_\mathrm{p}$ was chosen in the non-decarbonization scenarios (\scen{noFlex}, \scen{someFlex}, and \scen{fullFlex}), see \Cref{tab:capn}.
TACs of the non-decarbonization scenarios were significantly reduced compared to the other contexts, showing even negative TAC values for \scen{someFlex} and \scen{fullFlex}, see \Cref{fig:pareto} c).
I.e., assuming the level and fluctuation of the electricity prices of the year 2021, the company's MES (together with offshore WTs) can be seen as a business model with an above 10\% internal return rate, provided that the WT installation is feasible.
However, WT capacity is limited by the maximum feed-in of 20\,MWh/h, which can be seen in \Cref{fig:load_violin}.
As a result, WT capacities of 19--22\,MW$_\mathrm{p}$ were selected (see \Cref{tab:capn})
even though most of it was sold on the day-ahead market.
\Cref{fig:WAP} c) shows that adding flexibility reduced EWAP, EWACEF, $\hat{P}^\mathrm{EG,buy}$, and W$^\mathrm{EG,buy}$.
Due to the high WT capacities, the level of W$^\mathrm{EG,buy}$ for all optimized scenarios was low  compared to \scen{REF}.
From \Cref{fig:wap_slope} we can see that, for \scen{c\_scaled}, both $\omega$ values are over 300\%, i.e, in the decarbonization and non-decarbonization case, ECER were more than three times the TCER, or in other words, the modification of $\bm{P}^\mathrm{EG,buy}_t$ in \scen{c\_scaled} proportionately reduced significantly more costs than CEs.

\subsection{Discussion of the results} \label{sec:discussion}
When comparing all scenarios and contexts, the following results are striking:
\begin{enumerate}

	\item \label{itm:reduced_TACs_and_CEs}
	In all analyzed contexts, adding flexibility significantly reduced TACs and CEs.
	TAC reductions were between 10.4 and 189.6\%.
	However, the high reduction of 189.6\% was based on the already very low TAC of \euro0.10M/yr in the \scen{c\_scaled} \scen{noFlex} scenario.
	The second-highest relative TAC decrease appeared in the decarbonization case of \scen{c\_strict} with 71.9\%.
	This was because \scen{c\_strict} \scen{fullFlex\_d} was the only scenario reaching carbon neutrality without carbon removal since flexibility increased the self-consumption which eliminated the need for expensive carbon removal.
	While the \scen{someFlex} scenarios showed 7.9--17.1\% CE reductions compared to \scen{noFlex}, the \scen{fullFlex} scenarios showed 17.1--20.1\%.

	\item \label{itm:decarb_costs}
	Through storage investments and the smart/bidirectional charging of existing powered industrial trucks (\scen{fullFlex}), decarbonization costs decreased by 19\% in \scen{c\_base} (assuming $c^\mathrm{DAC}$=\allowbreak\euro222\,t\,$_\mathrm{CO2eq}^{-1}$ and 2019 electricity prices).
	This number increased to 37\% in \scen{c\_scaled} (assuming $c^\mathrm{DAC}$=\allowbreak\euro222\,t\,$_\mathrm{CO2eq}^{-1}$ and 2019 electricity prices scaled to 2021 levels), and further increased to 80\% in \scen{c\_\allowbreak strict} (assuming $c^\mathrm{DAC}$=\allowbreak\euro10k\,t\,$_\mathrm{CO2eq}^{-1}$ and 2019 electricity prices).

	\item \label{itm:realistic_decarb}
	With current CEFs, the MES could technically only achieve strict Scope \RomanNumeralCaps{1} and \RomanNumeralCaps{2} net-zero CEs (\scen{c\_strict}) when using the full flexibility potential (\scen{full\allowbreak Flex\_d}) of different mutually complementary energy storage technologies, one of them being H$_2$S which aligns with the results of Petkov and Gabrielli~\cite{Petkov2020}.
	However, due to the high CapEx, it seems rather unrealistic, even if the full flexibility potential is considered.
	In contrast, if assuming electricity grid decarbonization, which we indirectly did in \scen{c\_base} and \scen{c\_scaled} with a considerable price of \euro222\,t$^{-1}_\mathrm{CO2eq}$ on electricity-based CEs, full decarbonization is realistic, especially when considering the full flexibility potential.
	When comparing \scen{fullFlex\_d} to \scen{REF}, TACs increased by 55\% in \scen{c\_base} and decreased by 65\% in \scen{c\_scaled}.
	However, the economic benefits of \scen{c\_scaled} mainly stem from wind power feed-in, see \Cref{fig:load_violin}.
	In any circumstance, the results showed that both the existing and the added flexibility through storage and sector coupling is highly valuable especially for the decarbonization and can be used for multiple purposes including price and CEF-based DR, peak shaving, and self-consumption optimization.

	\item \label{itm:all_metrics_reduced}
	In all cases, $\hat{P}^\mathrm{EG,buy}$, W$^\mathrm{EG,buy}$, $\pi$-rate, and $\varepsilon$-rate reduced with increasing flexibility (see \Cref{fig:metric_bars}), i.e, price and CEF-based DR is used together with peak shaving and self-consumption optimization.
	All scenarios except the \scen{c\_strict} \scen{fullFlex\_d} feature an $\varepsilon$-rate above 100\%, i.e., the average accounted CEF (i.e., EWACEF) was above the average CEF (i.e., TWACEF).
	In the case of \scen{REF}, the reason for this was the high electrical and thermal energy demand at business hours, where CEFs tend to be above average.
	\scen{REF} and \scen{noFlex(\_d)} of all contexts featured $\pi$ and $\varepsilon$-rates of above 100\%, i.e., without flexibility use, their electricity purchase profiles are less market-serving than a flat load profile.

	All dark blue arrows in \Cref{fig:WAP} crossed the TWAP line from above, i.e., by using flexibility in the non-decarbonization case, the $\pi$-rate declined from above 100\% to below 100\%, or, put differently, the average paid price (i.e., EWAP) fell below the average price (i.e., TWAP).

	\item \label{itm:valuable_metrics}
	The results in \Cref{fig:WAP} show that the DR evaluation metrics proposed in \Cref{sec:eval_DR}
	are simple, tangible and useful to compare and visualize the intensity of price and CEF-based DR in a meaningful way.
	The metrics EWAP and EWACEF are especially tangible since they have the same unit as the prices and CEFs, and can be compared to the TWAP and TWACEF, respectively.
	The metrics $\pi$-rate and $\varepsilon$-rate stand out when comparing scenarios under different contexts.
	Due to the normalization with TWAP and TWACEF, they are independent from the level of prices and CEFs, respectively.
	However, when evaluating price-based and CEF-based DR, $\hat{P}^\mathrm{EG,buy}$ and W$^\mathrm{EG,buy}$ should also be considered to account for interactions with peak shaving and self-consumption optimization.
	The metrics TCER and ECER measure the dominance of prices versus CEFs as signal for DR and can be interpreted as gradients in a price-CEF diagram.
	
	\item \label{itm:2022}
	As expected, the results of \scen{c\_scaled} demonstrate that high and highly fluctuating electricity prices incentivize the activation of existing flexibility as well as investments in vRES and energy storage systems.
	E.g., while the V2X activity was similar in \scen{c\_base} and \scen{c\_\allowbreak strict}, it was up to 2.7x higher in \scen{c\_scaled} with 498 full V2X discharges per year (=1.36 per day) in \scen{some\allowbreak Flex\_d}, see \Cref{fig:v2x_cycles}.
	Since energy shock scenarios, like in the year 2022, might not be a once-in-a-century event in a time of political instability, it might be interesting to quantify the value of being a flexumer in such a price shock scenario.

\end{enumerate}

%% file: chapter/7_concl.tex
\section{Conclusions} \label{sec:concl}
This study quantified the value of flexibility for decarbonizing a production company's MES within a detailed optimal design and operation case study considering and evaluating price-based and CEF-based DR.
Based on the case study results discussed in \Cref{sec:discussion}, the following conclusions can be drawn:
\begin{itemize}
	\item Production companies can significantly reduce TACs, CEs, and decarbonization costs by using the energy flexibility of their MES (\Cref{itm:reduced_TACs_and_CEs,itm:decarb_costs}).
	
	\item A strict net-zero CE MES is only achieved using energy flexibility.
	To achieve net-zero CEs in a cost-effective manner, comparative advantages of energy storage systems are used (\Cref{itm:realistic_decarb}).
	
	\item Independently from electricity prices and decarbonization ambitions, less, greener, and cheaper electricity is purchased when using energy flexibility (\Cref{itm:all_metrics_reduced}).

	\item The flexibility metrics defined in \Cref{sec:eval_DR} are simple, tangible, and effective to evaluate the intensity of DR based on prices and CEFs (\Cref{itm:valuable_metrics}).
	
	\item High electricity prices encourage vRES investments and are, due to market mechanisms, typically accompanied by increased price volatility which incentivizes operational flexibility through energy storage investments and price-based DR (\Cref{itm:2022}).
	
\end{itemize}
The findings of this study underline the significant potential of energy flexibility to reduce costs, CEs, and fossil fuel dependency for manufacturing companies.
They illustrate the importance of considering future electricity price fluctuations, decarbonization goals, and the company's assessment of future carbon removal within the planning of sustainable MESs.
Also, they demonstrate the multifaceted techno-economic interactions of modern sustainable MESs and, therefore, highlight the importance of integrated models for planning and flexibility potential evaluation.

%% file: chapter/credits.tex
\section*{Acknowledgments}
This research was performed as part of the MeSSO Research Group at the Munster Technological University (MTU) and in relation to the project WIN4climate as part of the National Climate Initiative financed by the Federal Ministry for Economic Affairs and Climate Action (BMWK) on the basis of a decision by the German Bundestag (No. 03KF0094A).
It was additionally funded by the MTU Risam scholarship scheme.

\section*{Declaration of competing interest}
The authors declare that they have no known competing financial interests or personal relationships that could have appeared to influence the work reported in this paper.

\section*{CRediT authorship contribution statement}
\textbf{\textbf{Markus Fleschutz:}} Conceptualization, Methodology, Software, Validation, Formal analysis, Investigation, Data Curation, Writing - original draft, Writing - review \& editing, Visualization.
\textbf{\textbf{Markus Bohlayer:}} Methodology, Writing - review \& editing.
\textbf{\textbf{Marco Braun:}} Supervision, Project administration, Funding acquisition.
\textbf{Michael D. Murphy:} Supervision, Conceptualization, Writing - review \& editing, Funding acquisition.

%% file: chapter/appendix.tex
\section{Detailed results}

\begin{figure}[!htb]
	\centering
	\includegraphics[width=0.9\linewidth]{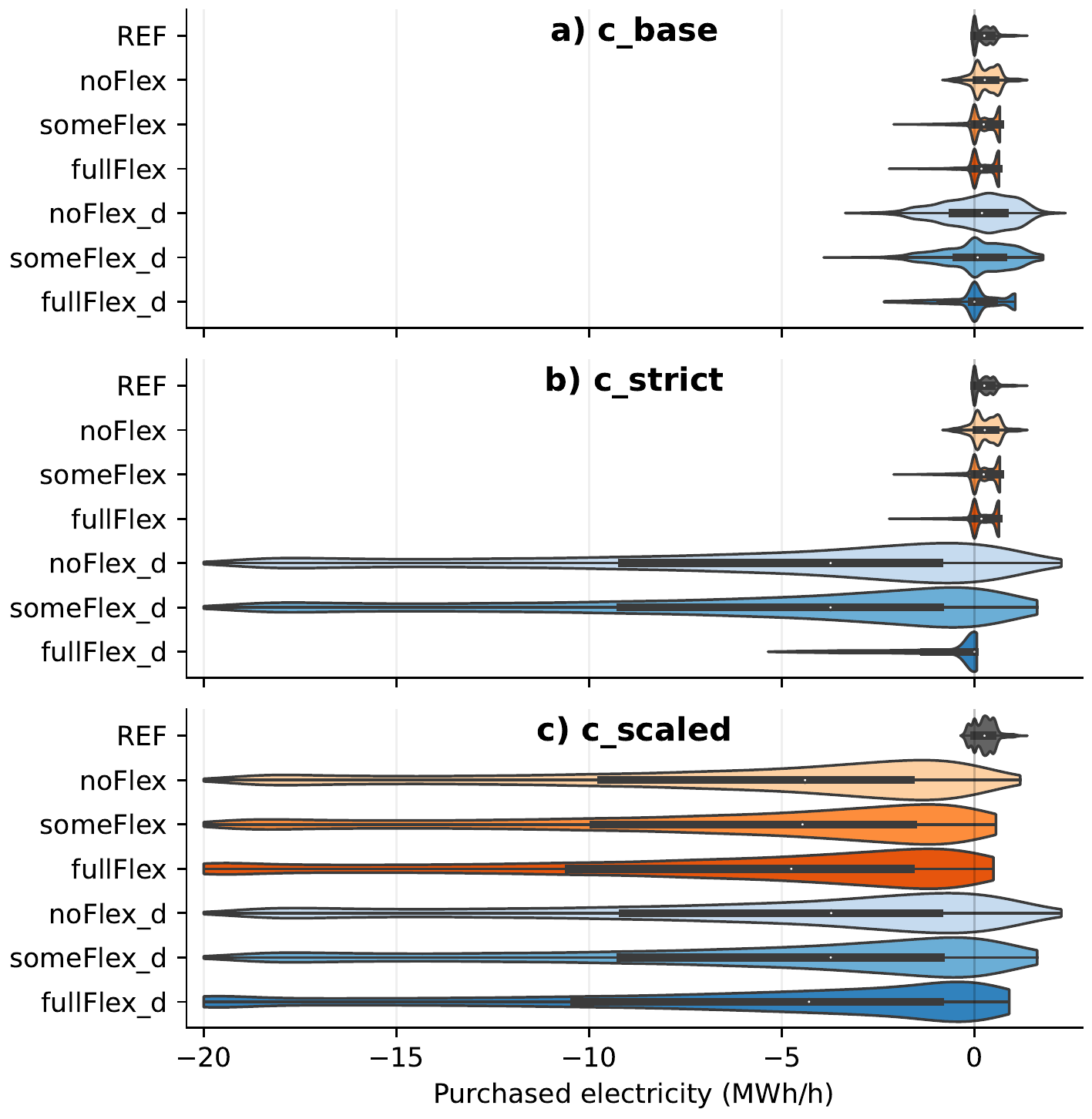}
	\caption{Violin plot of purchased (+) and sold ($-$) electricity.
		\label{fig:load_violin}}
\end{figure}

\begin{figure}[!htb]
	\centering
	\includegraphics[width=0.9\linewidth]{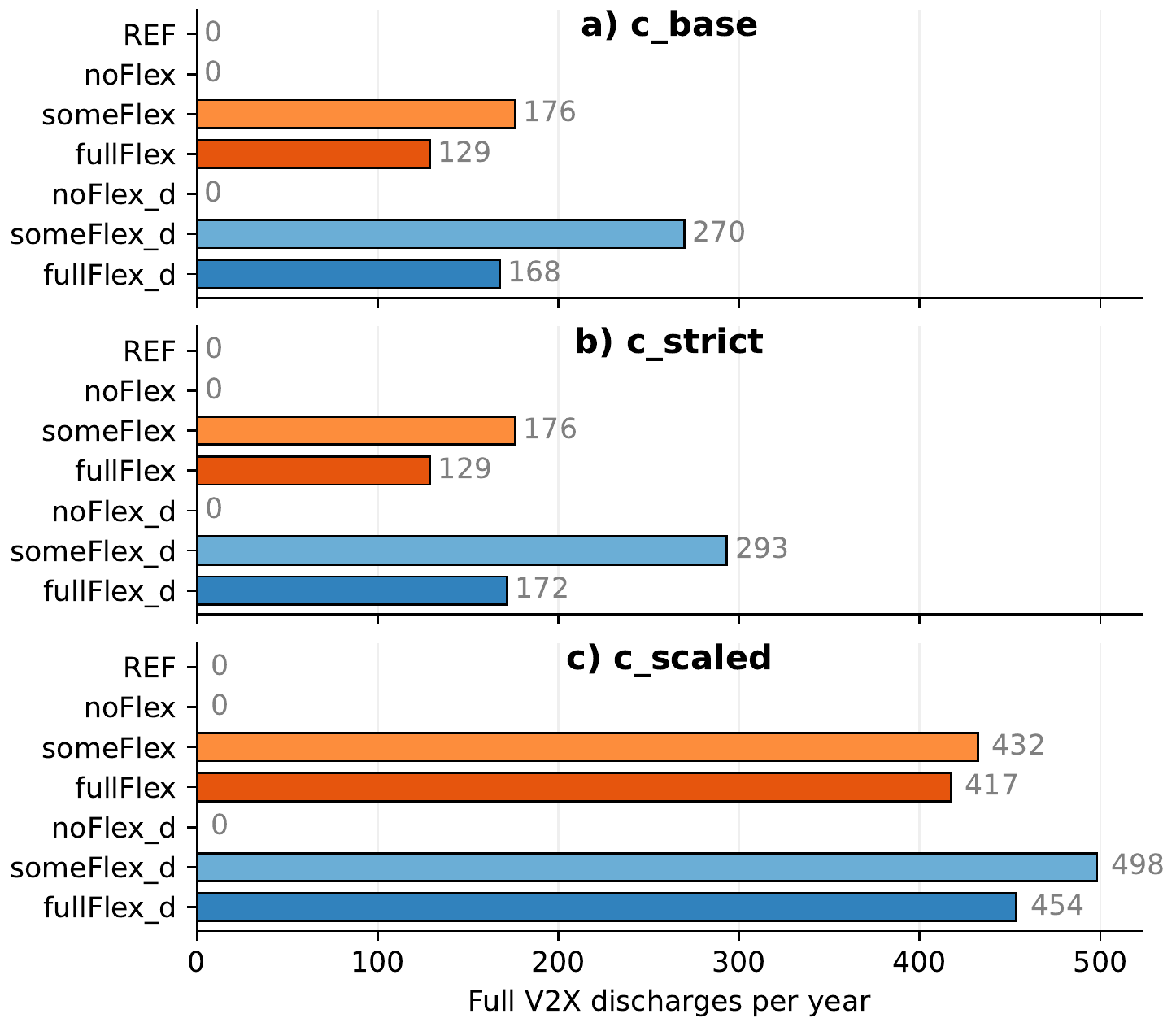}
	\caption{Number of full V2X discharges per year of average BEV battery (i.e. total annual energy V2X discharge divided by the usable battery capacity).
		\label{fig:v2x_cycles}}
\end{figure}

\begin{table*}[!htb]
	\centering
	\caption{
		Nominal capacities of new technologies in kW or kWh (see base unit of \Cref{tab:tech_params}).
		Colors highlight differences between scenarios.
		All-zero columns were removed.
		\label{tab:capn}}
	\includegraphics[width=1\linewidth]{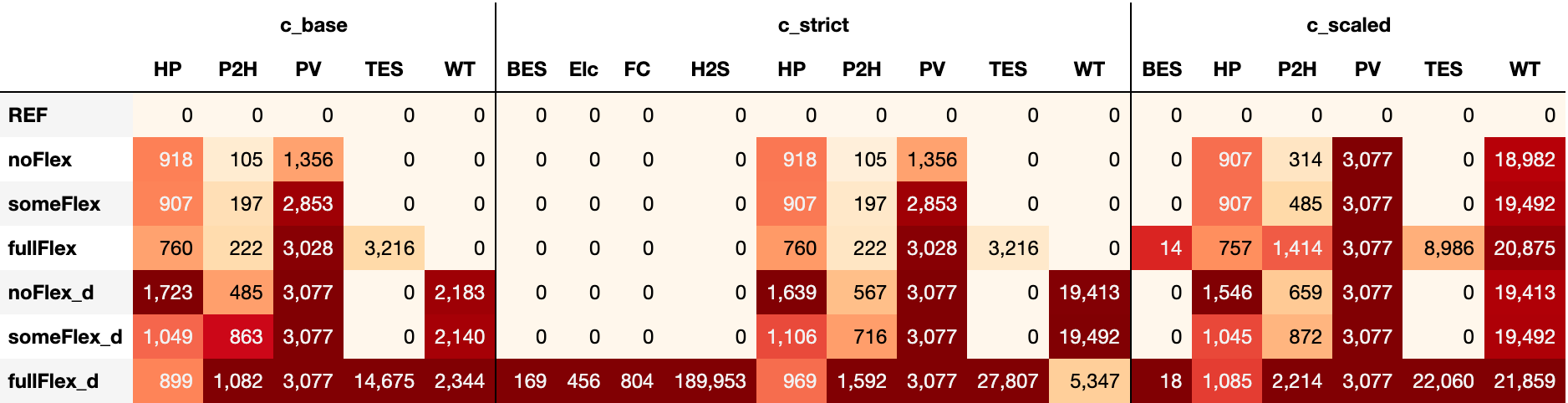}
\end{table*}

\begin{figure*}[!htb]
	\centering
	\includegraphics[width=1\linewidth]{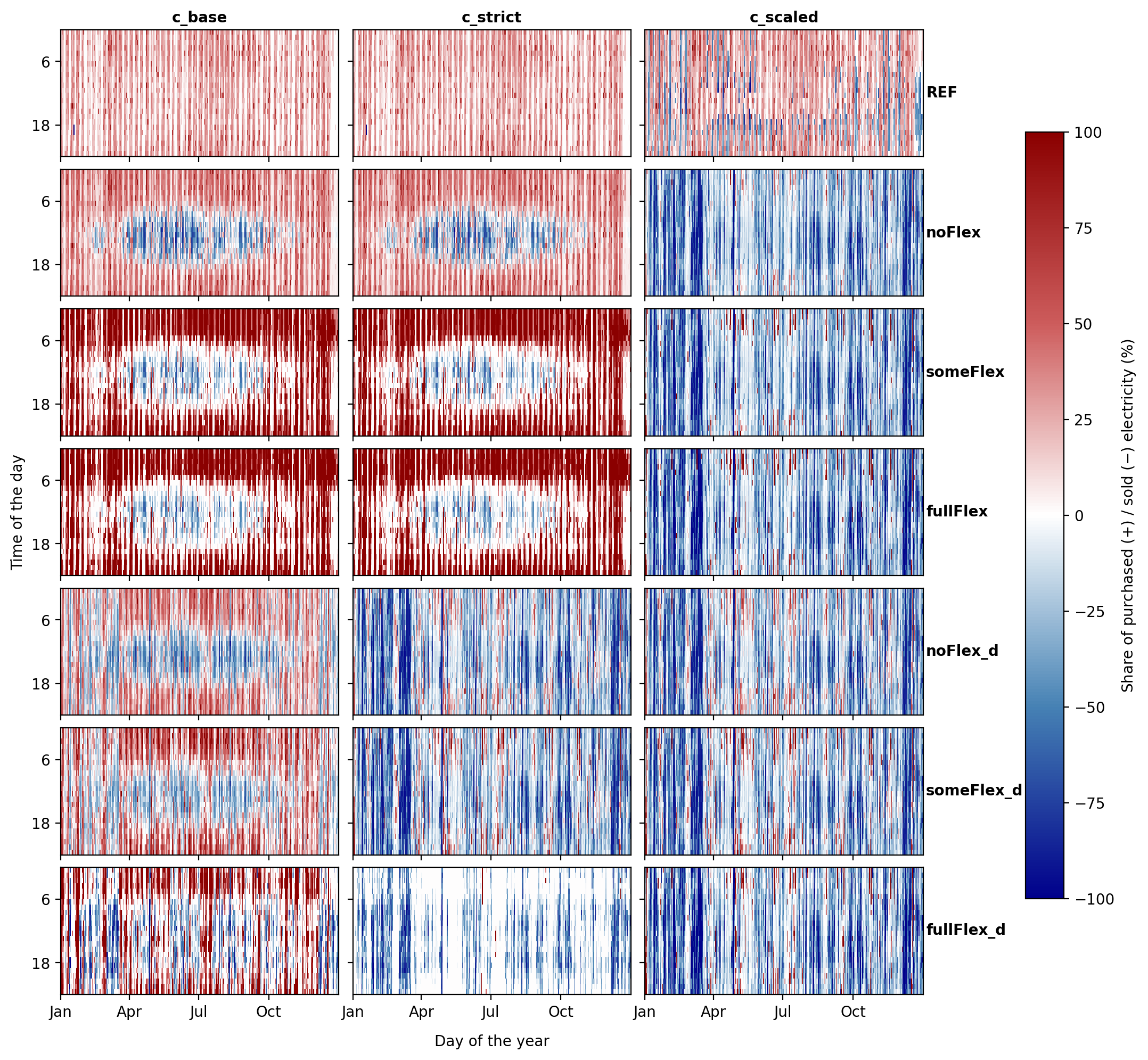}
	\caption{Heatmaps of purchased (+) and sold ($-$) electricity in \% based on maximum purchased and sold electricity, respectively.
		\label{fig:heatmap_EG}}
\end{figure*}

\begin{figure*}[!h]
	\centering
	\includegraphics[trim=0.1cm 0.1cm 0.1cm 0.1cm, clip, width=0.85\linewidth]{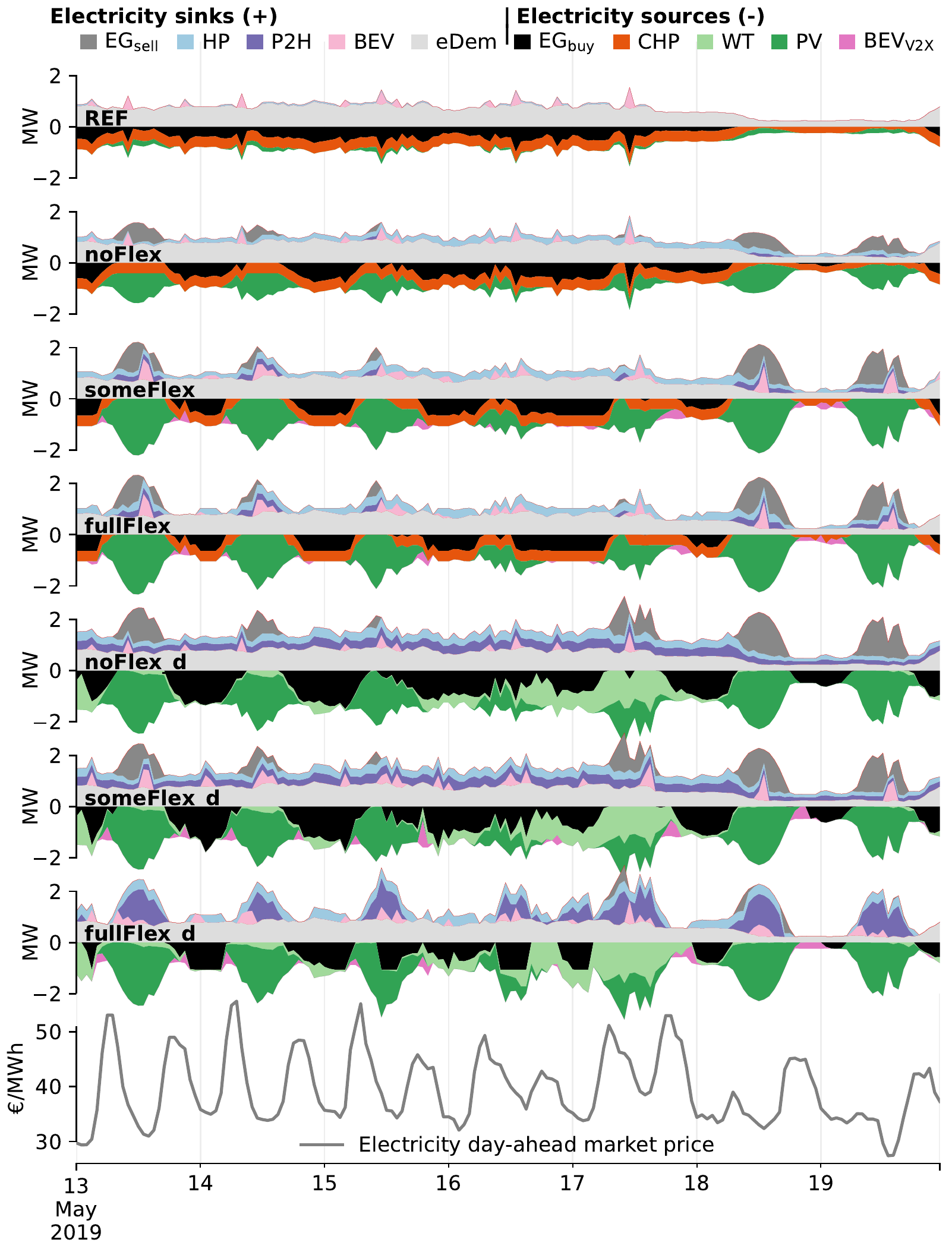}
	\caption{Electricity balance of the \scen{c\_base} context for one week (Monday--Sunday) in May 2019.
		\label{fig:ts_balance}}
\end{figure*}

\begin{figure*}[!htb]
	\centering
	\scenbf{c\_base}\par\vskip 4mm
	\begin{overpic}[width=0.46\linewidth]{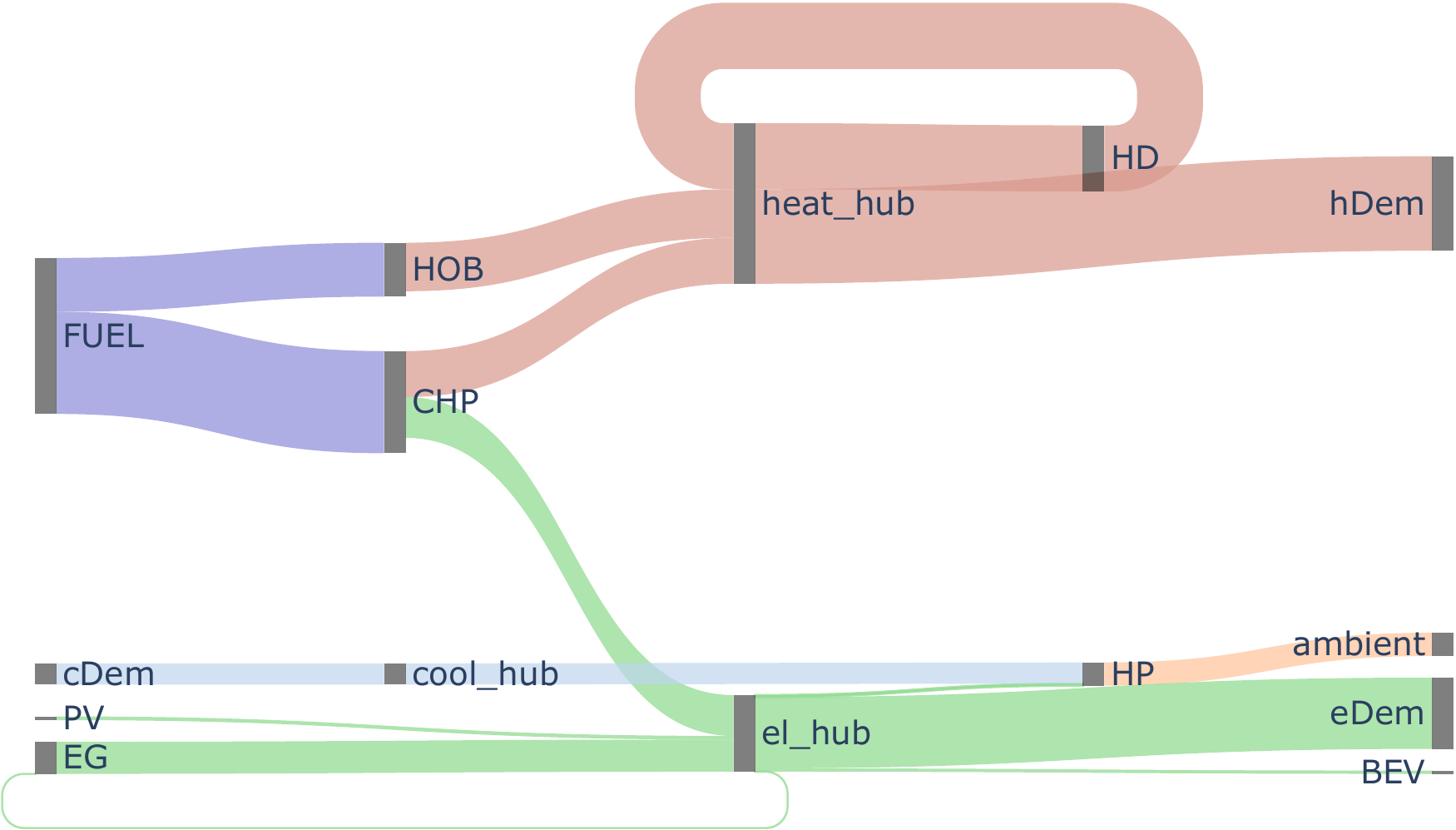} \put(75,25){\scenbf{REF}} \end{overpic}
	\begin{overpic}[width=0.46\linewidth]{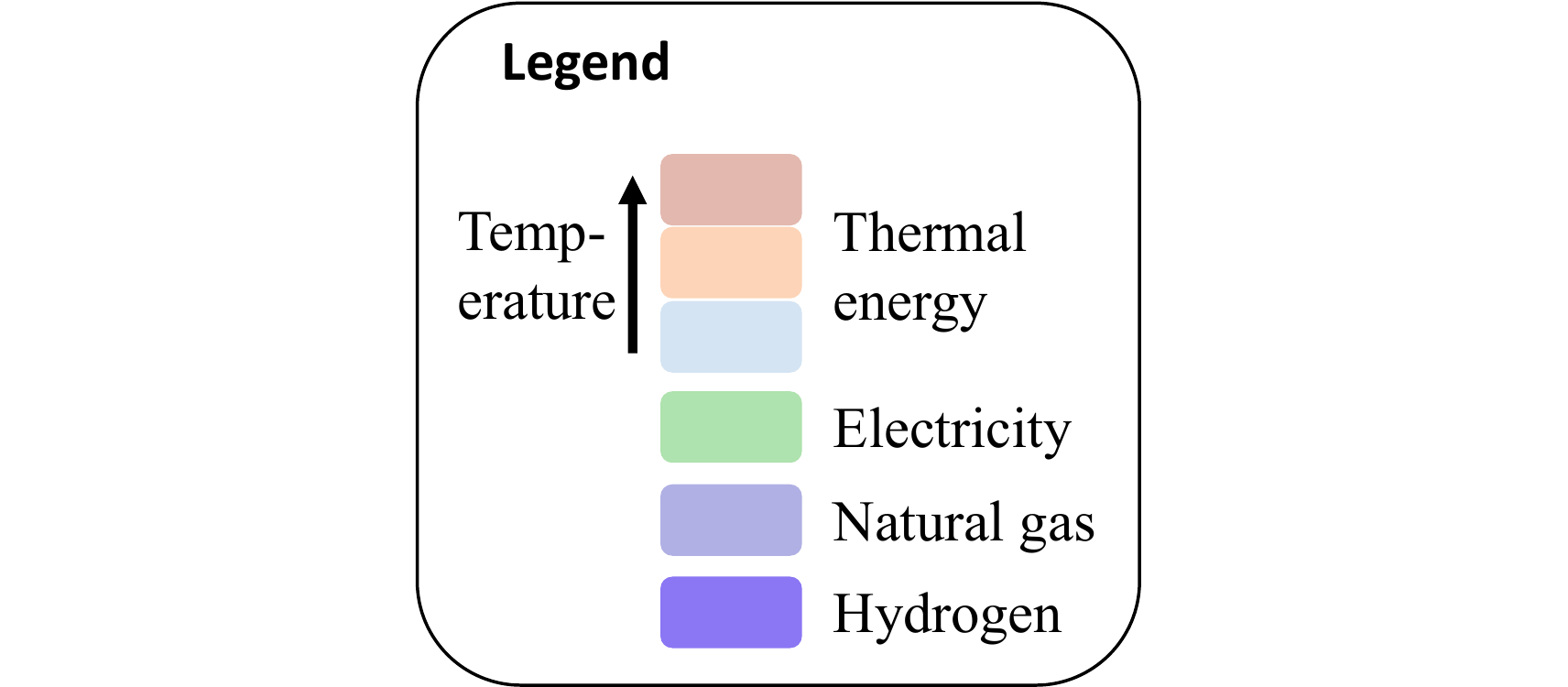} \end{overpic}
	\vskip 3mm
	\begin{overpic}[width=0.46\linewidth]{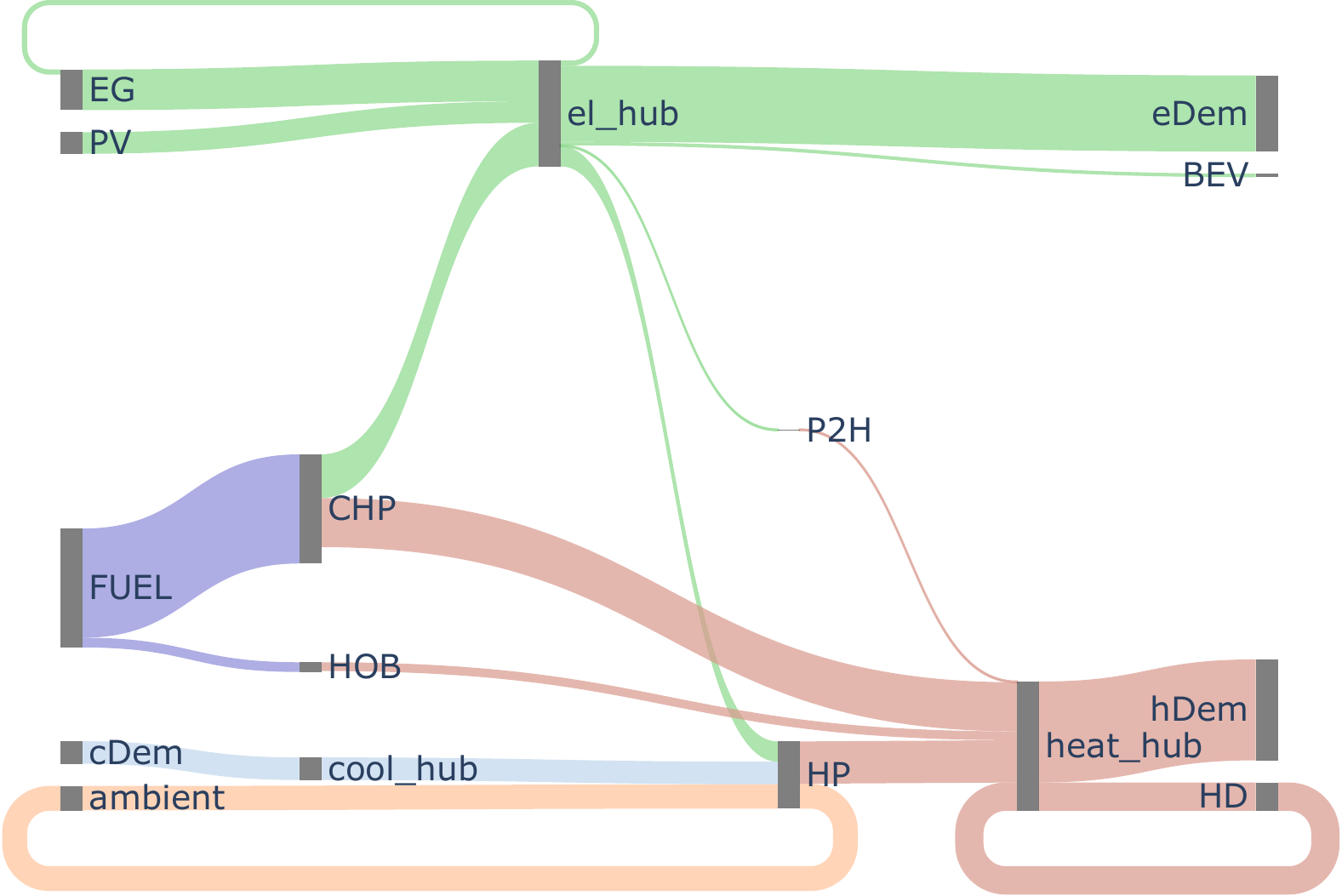} \put(75,25){\scenbf{noFlex}} \end{overpic}
	\begin{overpic}[width=0.46\linewidth]{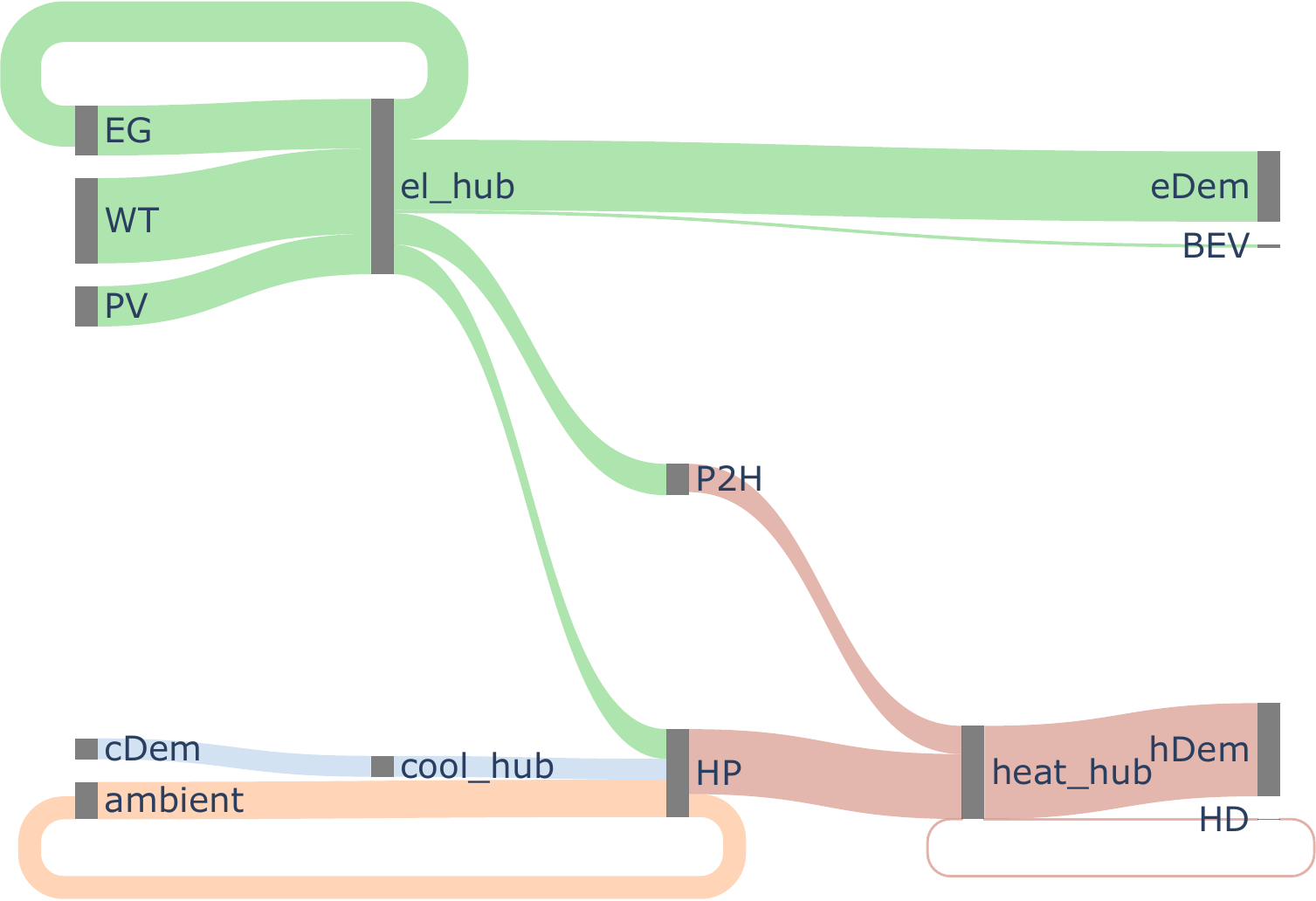} \put(75,25){\scenbf{no\_Flex\_d}} \end{overpic}
	\vskip 3mm
	\begin{overpic}[width=0.46\linewidth]{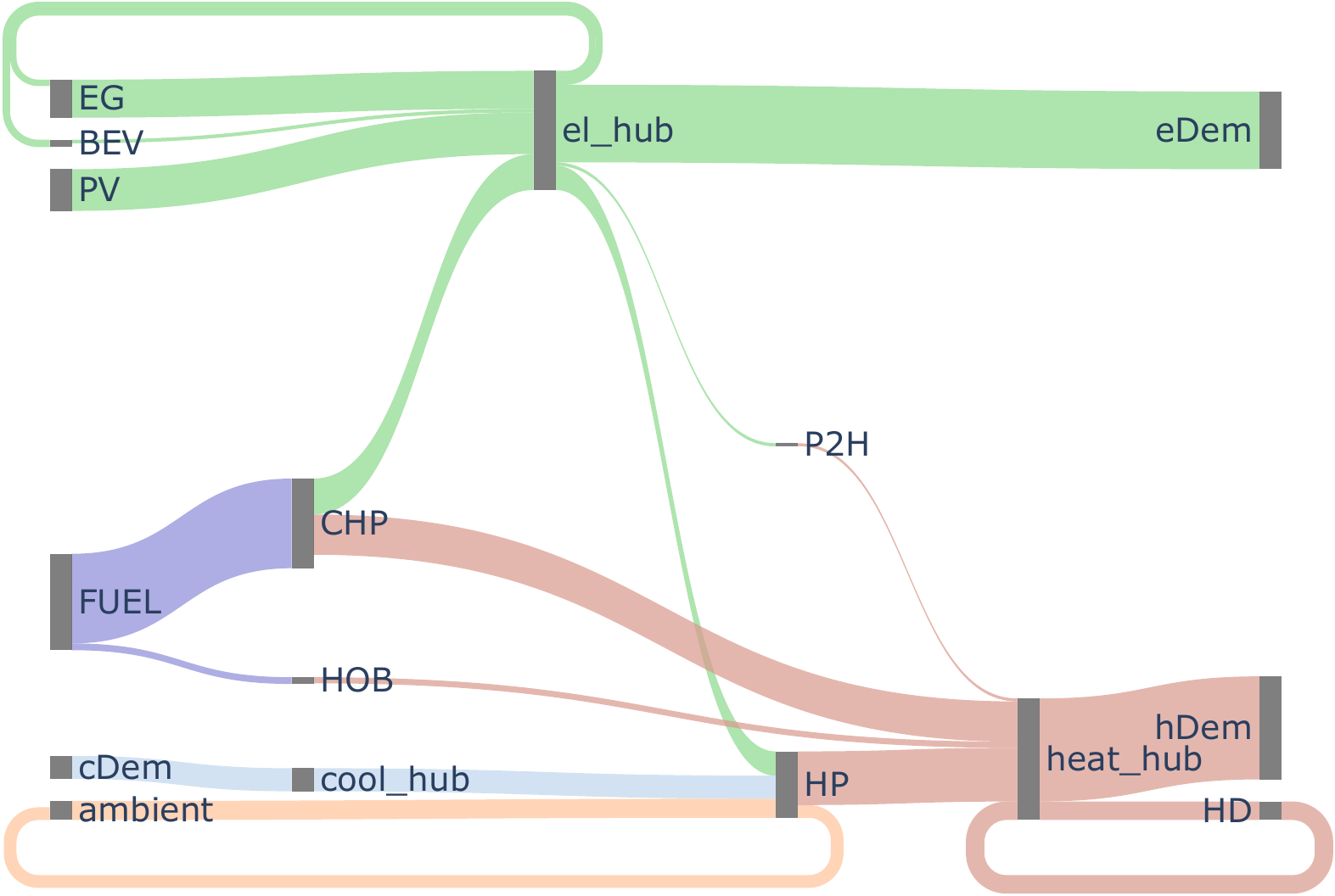} \put(75,25){\scenbf{someFlex}} \end{overpic}
	\begin{overpic}[width=0.46\linewidth]{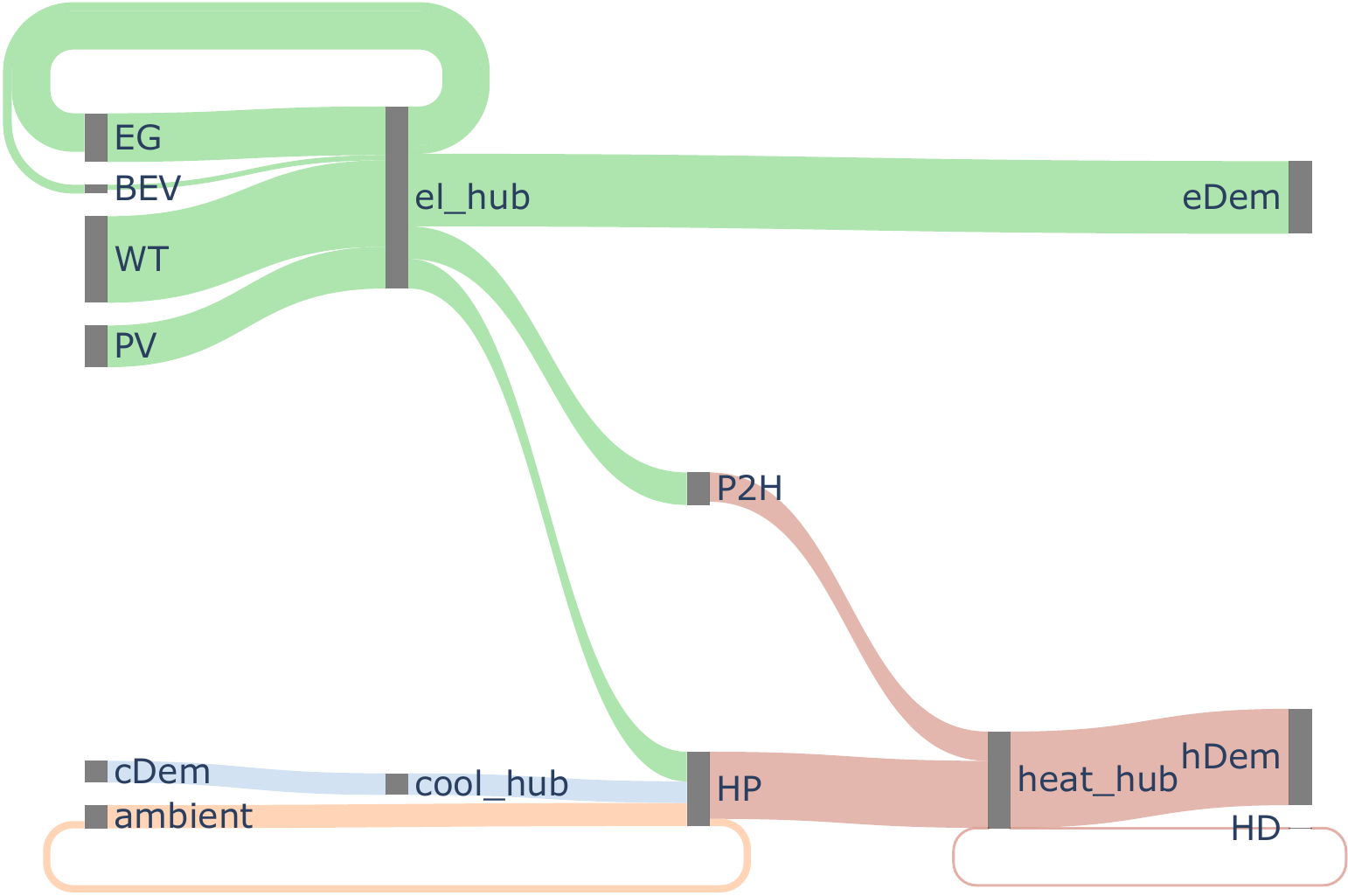} \put(75,25){\scenbf{someFlex\_d}} \end{overpic}
	\vskip 3mm
	\begin{overpic}[width=0.46\linewidth]{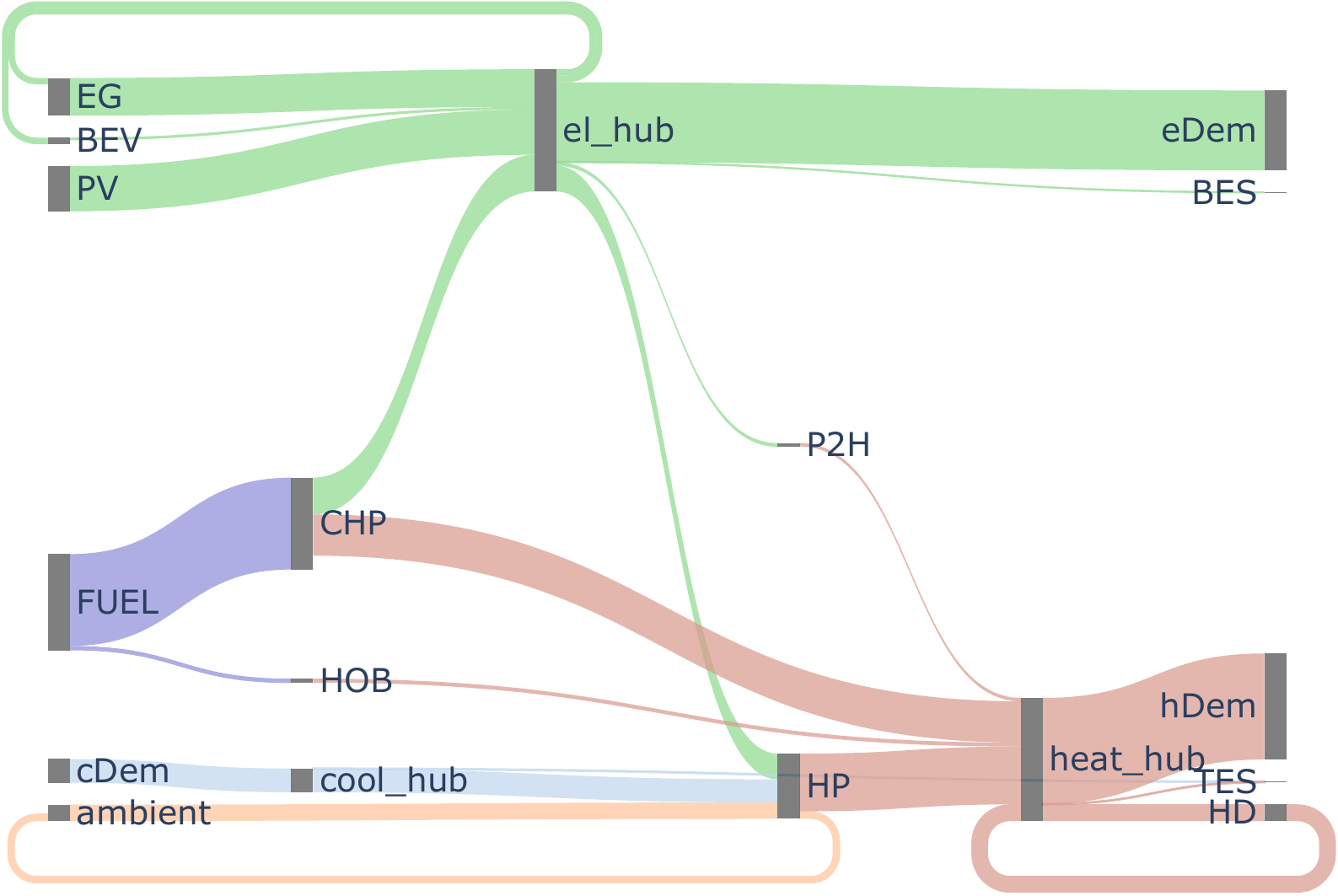} \put(75,25){\scenbf{fullFlex}} \end{overpic}
	\begin{overpic}[width=0.46\linewidth]{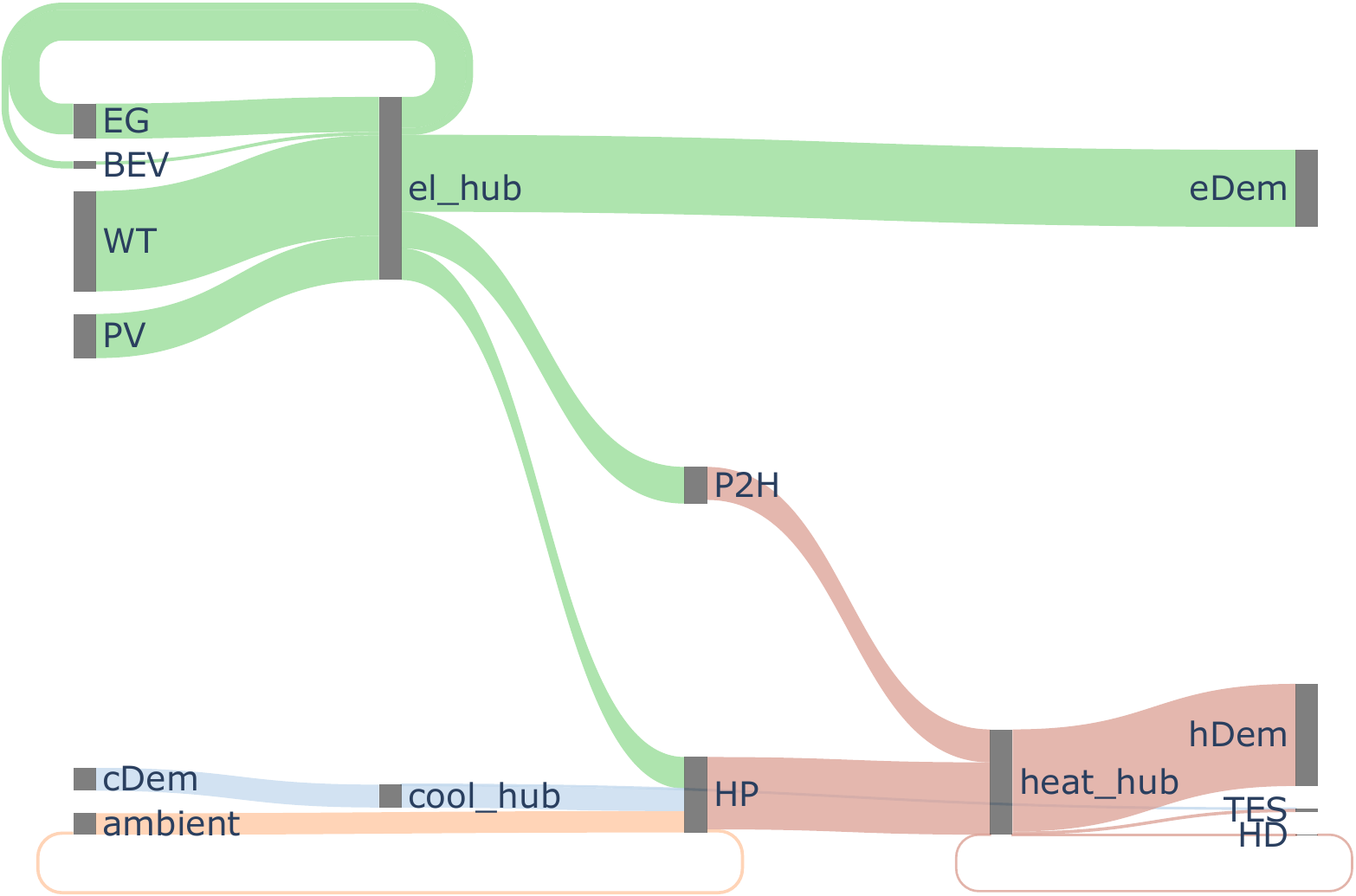} \put(75,25){\scenbf{fullFlex\_d}} \end{overpic}
	\caption{Sankey diagrams of annual energy sums for the \scen{c\_base} context.
		\label{fig:sankey_c_base}}
\end{figure*}

\begin{figure*}[!htb]
	\centering
	\scenbf{c\_strict}\par\vskip 4mm
	\begin{overpic}[width=0.46\linewidth]{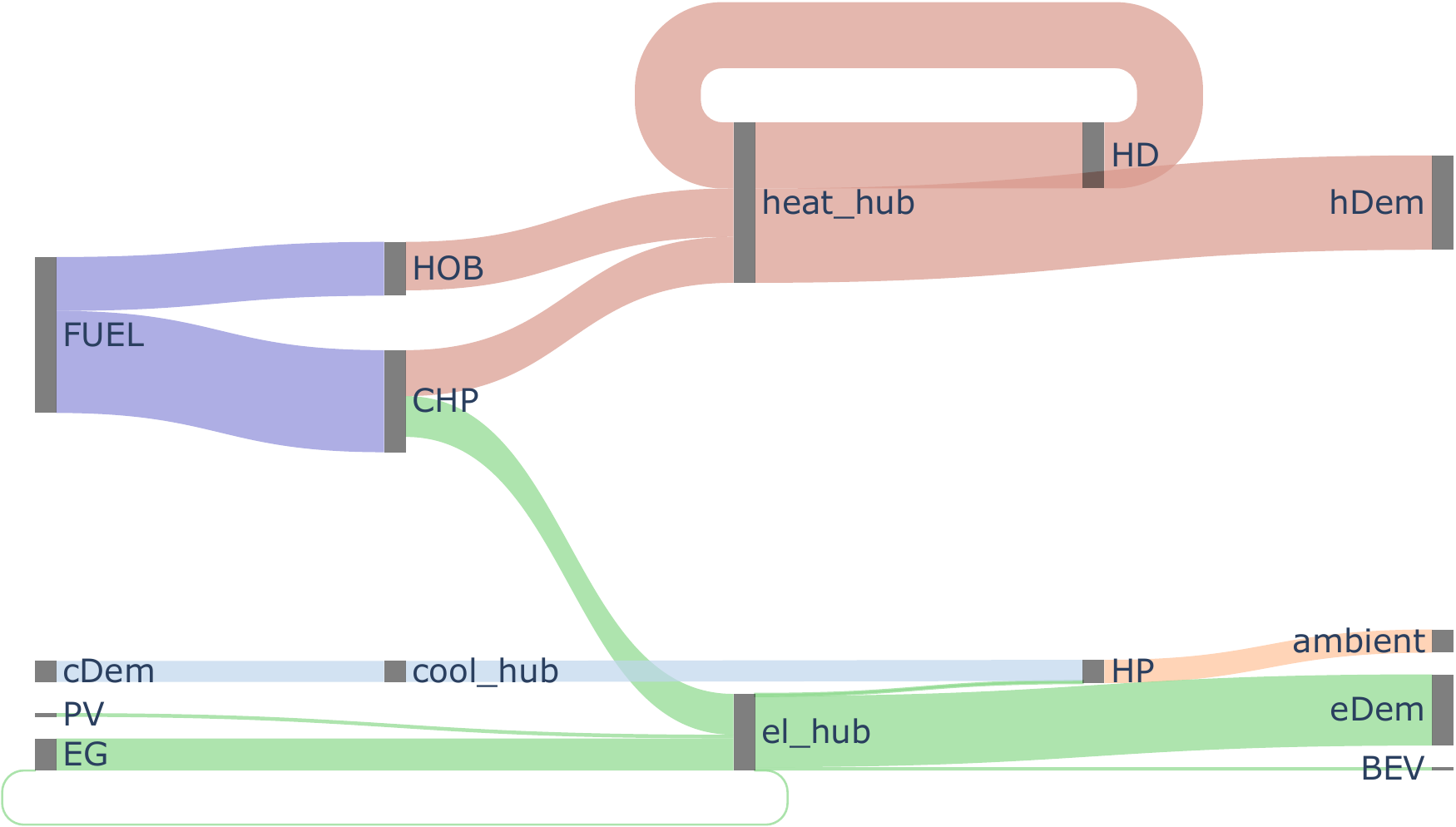} \put(75,25){\scenbf{REF}} \end{overpic}
	\begin{overpic}[width=0.46\linewidth]{pics/sankey_legend} \end{overpic}
	\vskip 3mm
	\begin{overpic}[width=0.46\linewidth]{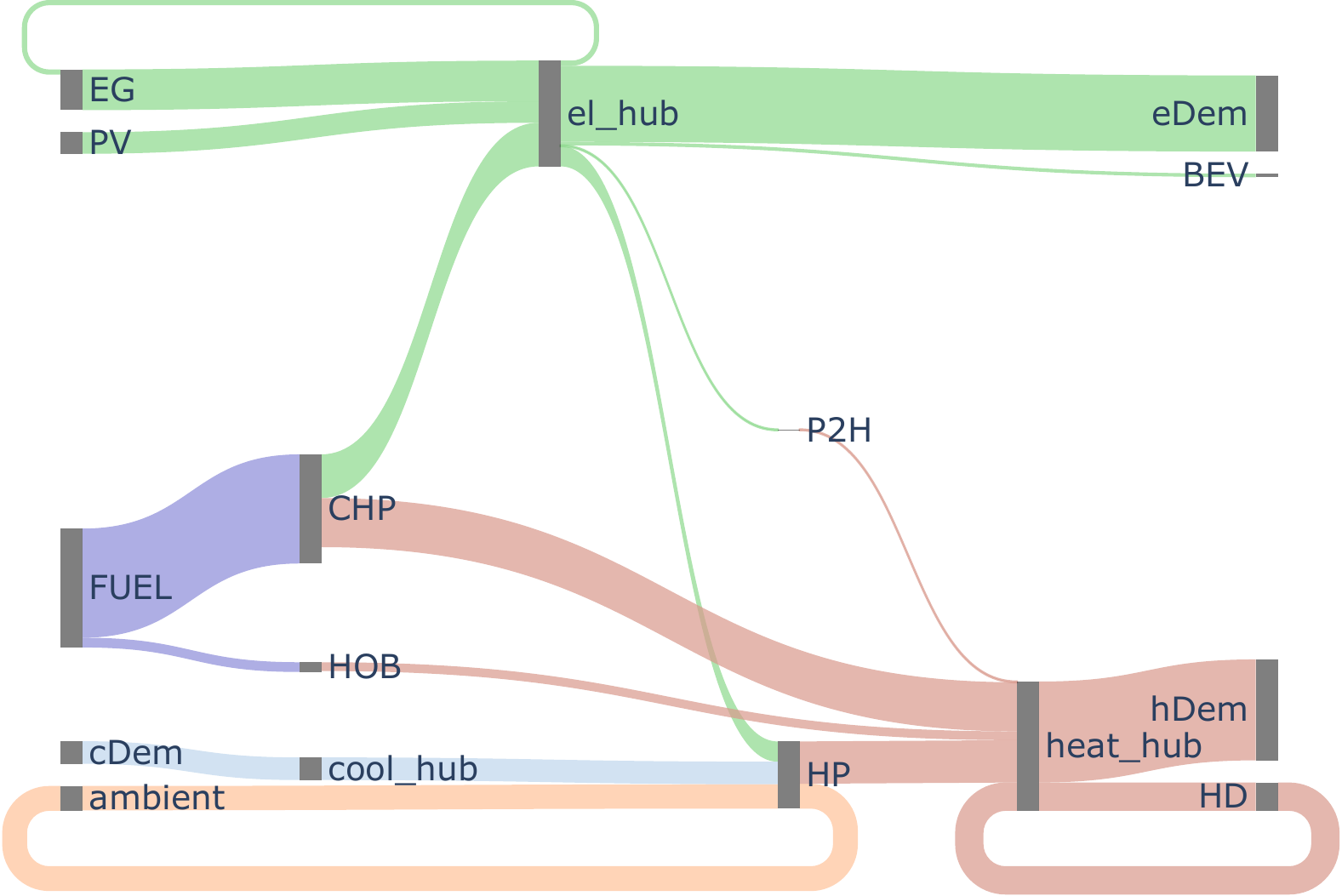} \put(75,25){\scenbf{noFlex}} \end{overpic}
	\begin{overpic}[width=0.46\linewidth]{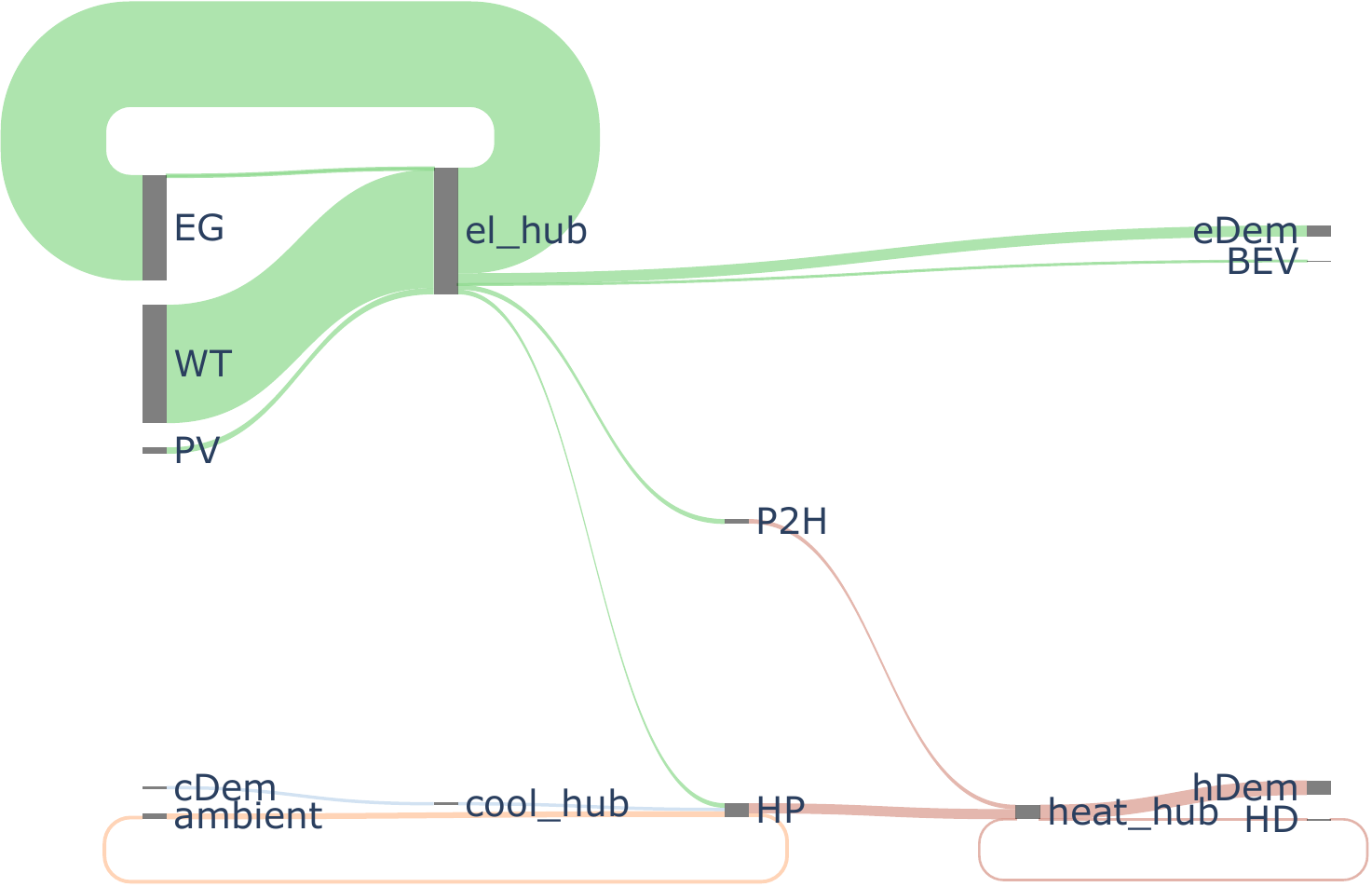} \put(75,25){\scenbf{no\_Flex\_d}} \end{overpic}
	\vskip 3mm
	\begin{overpic}[width=0.46\linewidth]{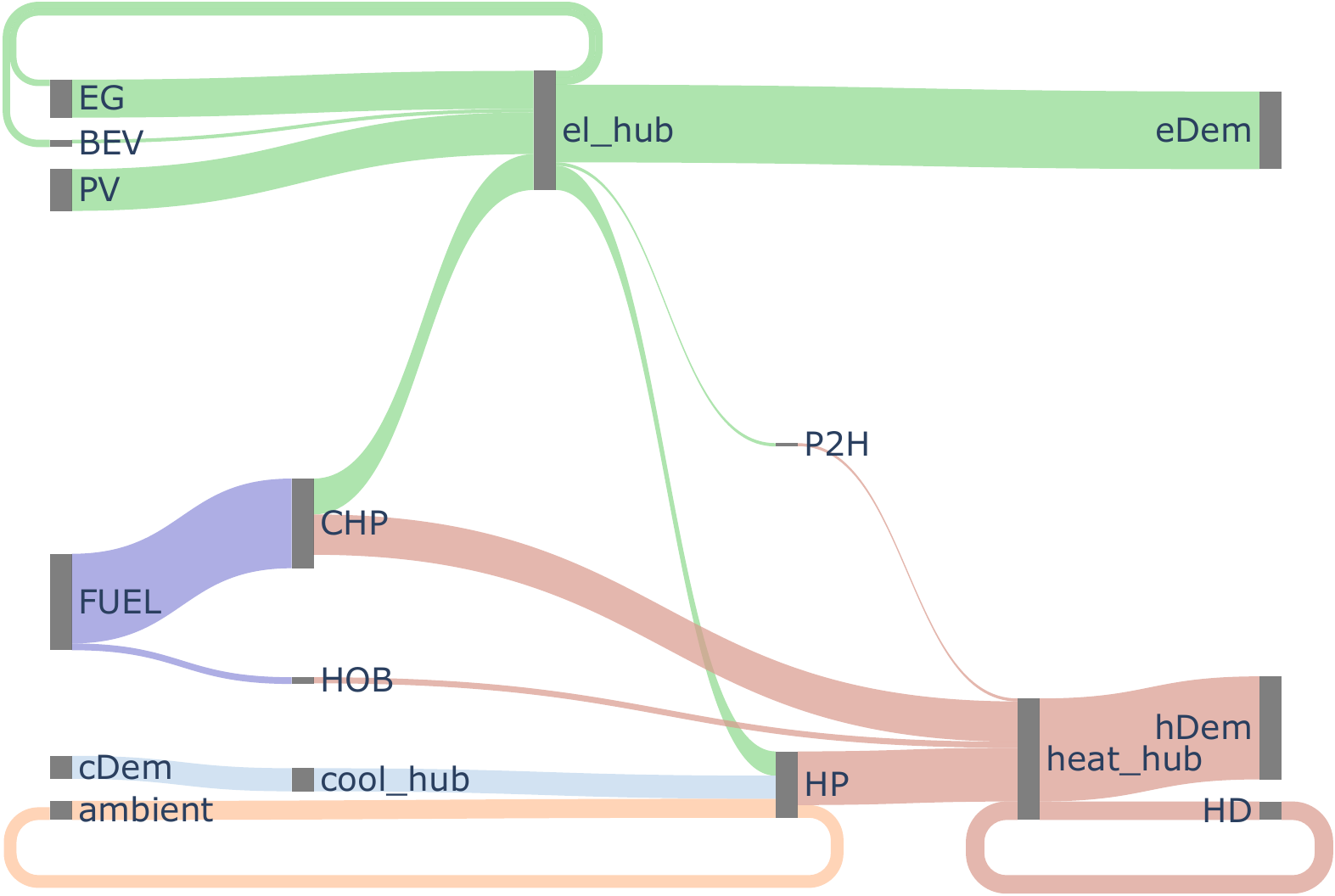} \put(75,25){\scenbf{someFlex}} \end{overpic}
	\begin{overpic}[width=0.46\linewidth]{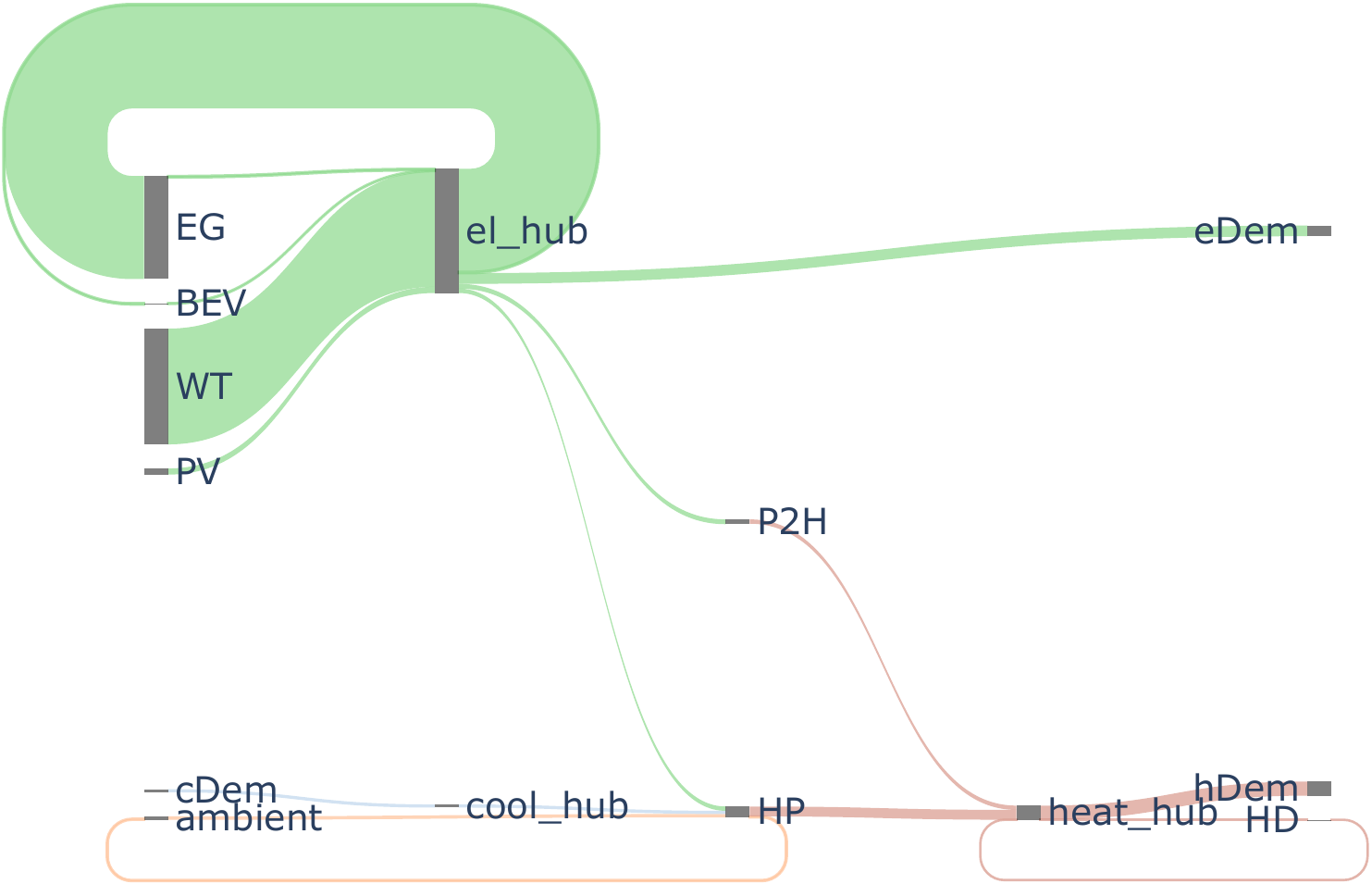} \put(75,25){\scenbf{someFlex\_d}} \end{overpic}
	\vskip 3mm
	\begin{overpic}[width=0.46\linewidth]{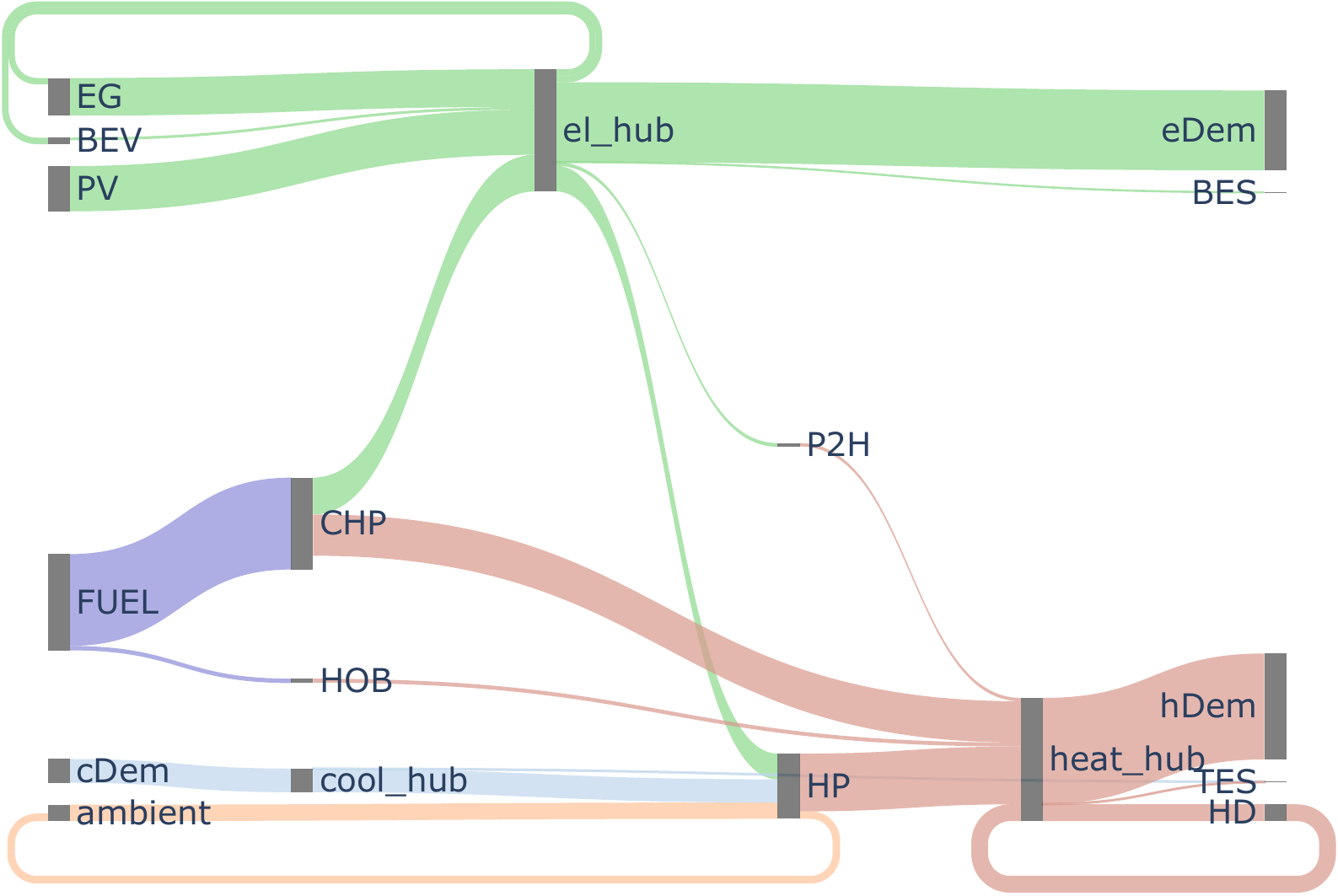} \put(75,25){\scenbf{fullFlex}} \end{overpic}
	\begin{overpic}[width=0.46\linewidth]{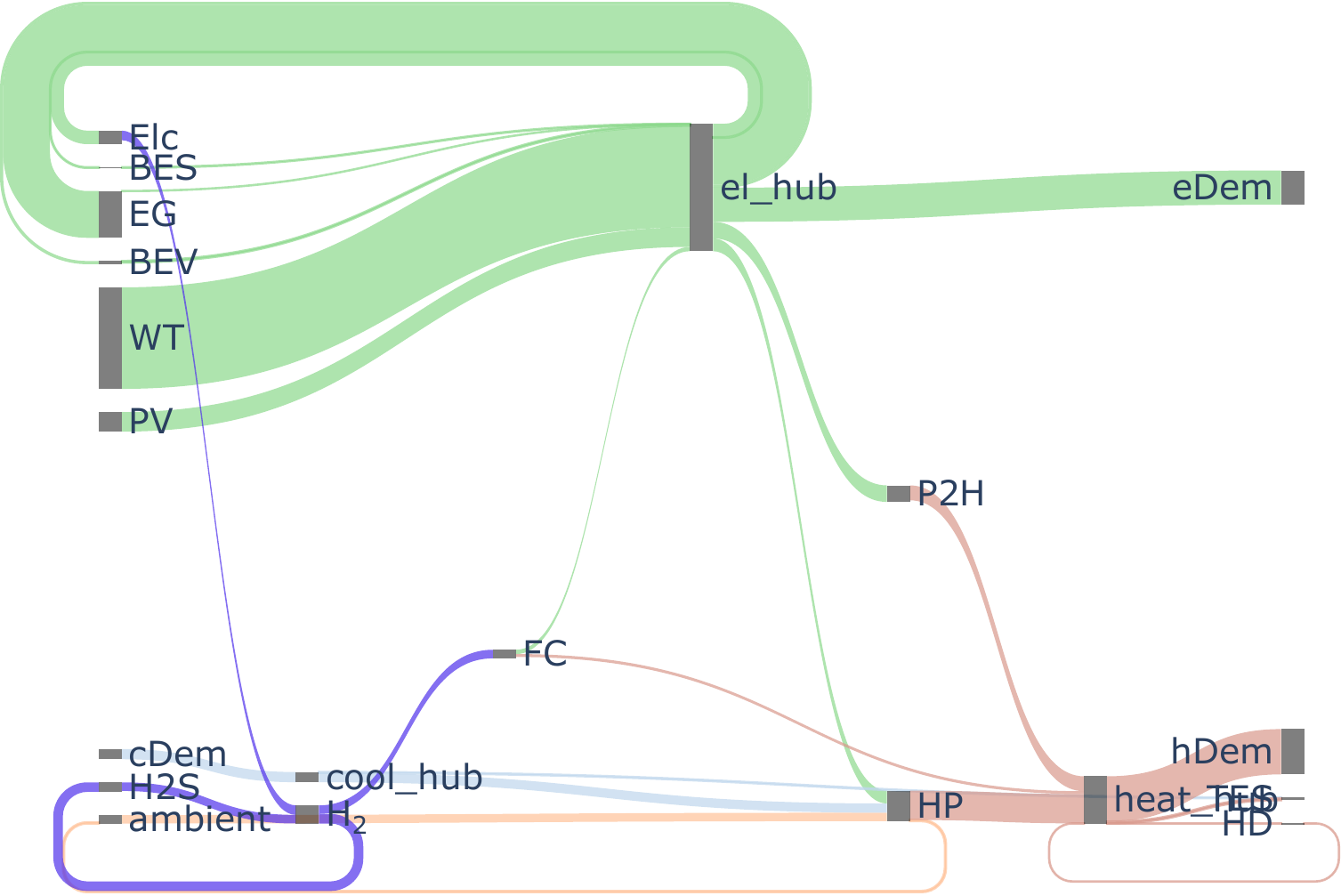} \put(75,25){\scenbf{fullFlex\_d}} \end{overpic}
	\caption{Sankey diagrams of annual energy sums for the \scen{c\_strict} context.
		\label{fig:sankey_c_strict}}
\end{figure*}

\begin{figure*}[!htb]
	\centering
	\scenbf{c\_scaled}\par\vskip 4mm
	\begin{overpic}[width=0.46\linewidth]{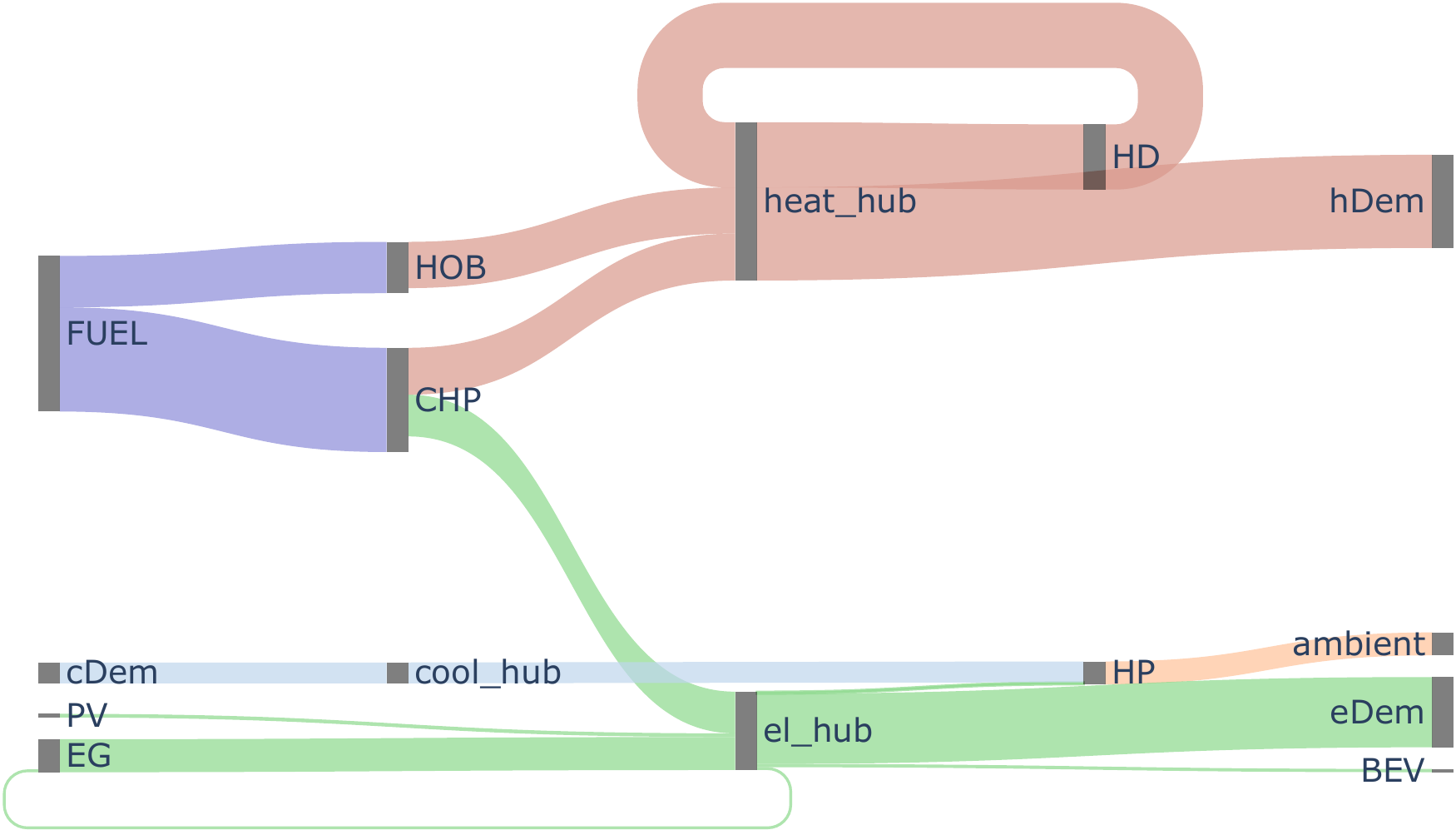} \put(75,25){\scenbf{REF}} \end{overpic}
	\begin{overpic}[width=0.46\linewidth]{pics/sankey_legend} \end{overpic}
	\vskip 3mm
	\begin{overpic}[width=0.46\linewidth]{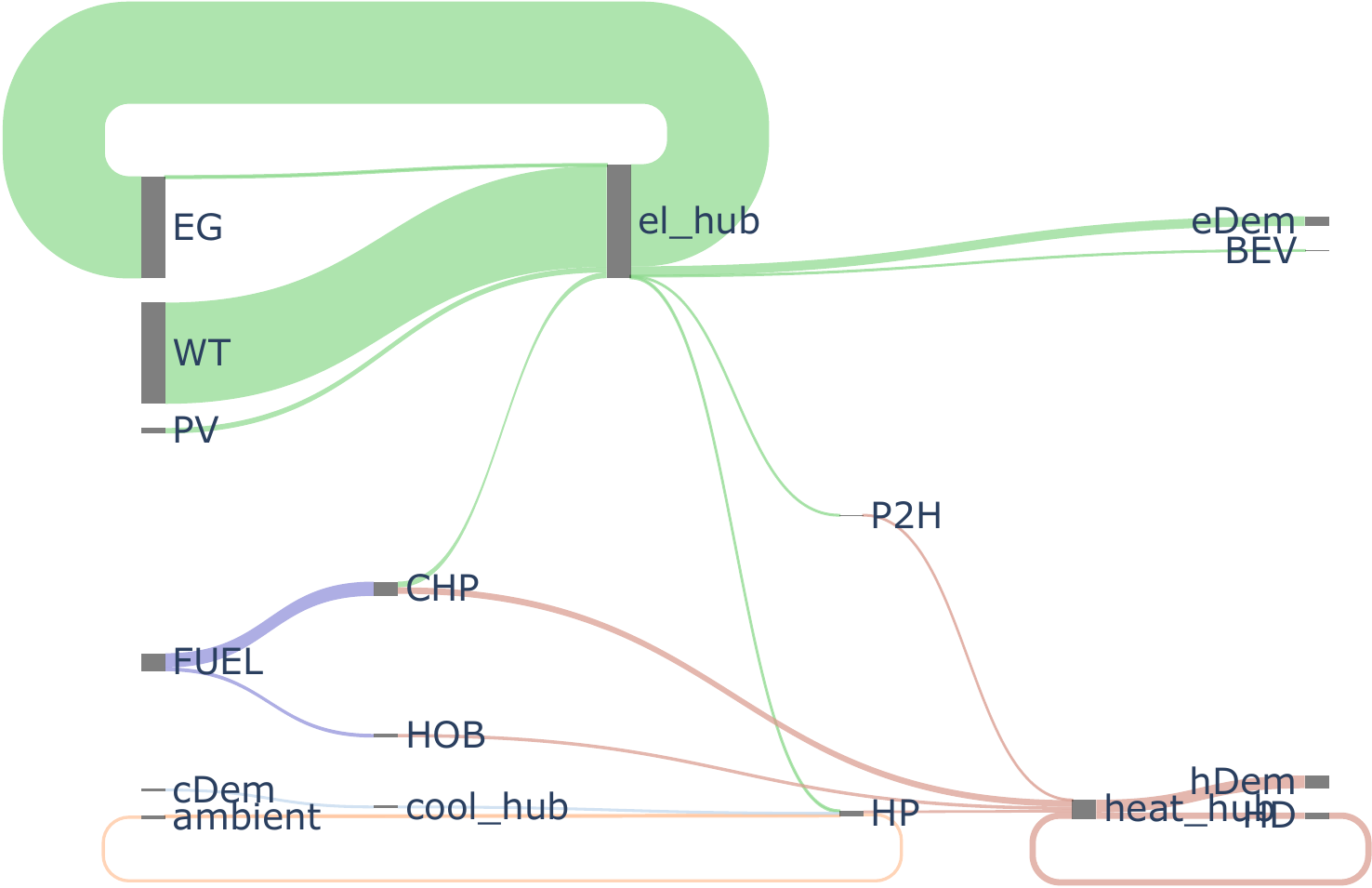} \put(75,25){\scenbf{noFlex}} \end{overpic}
	\begin{overpic}[width=0.46\linewidth]{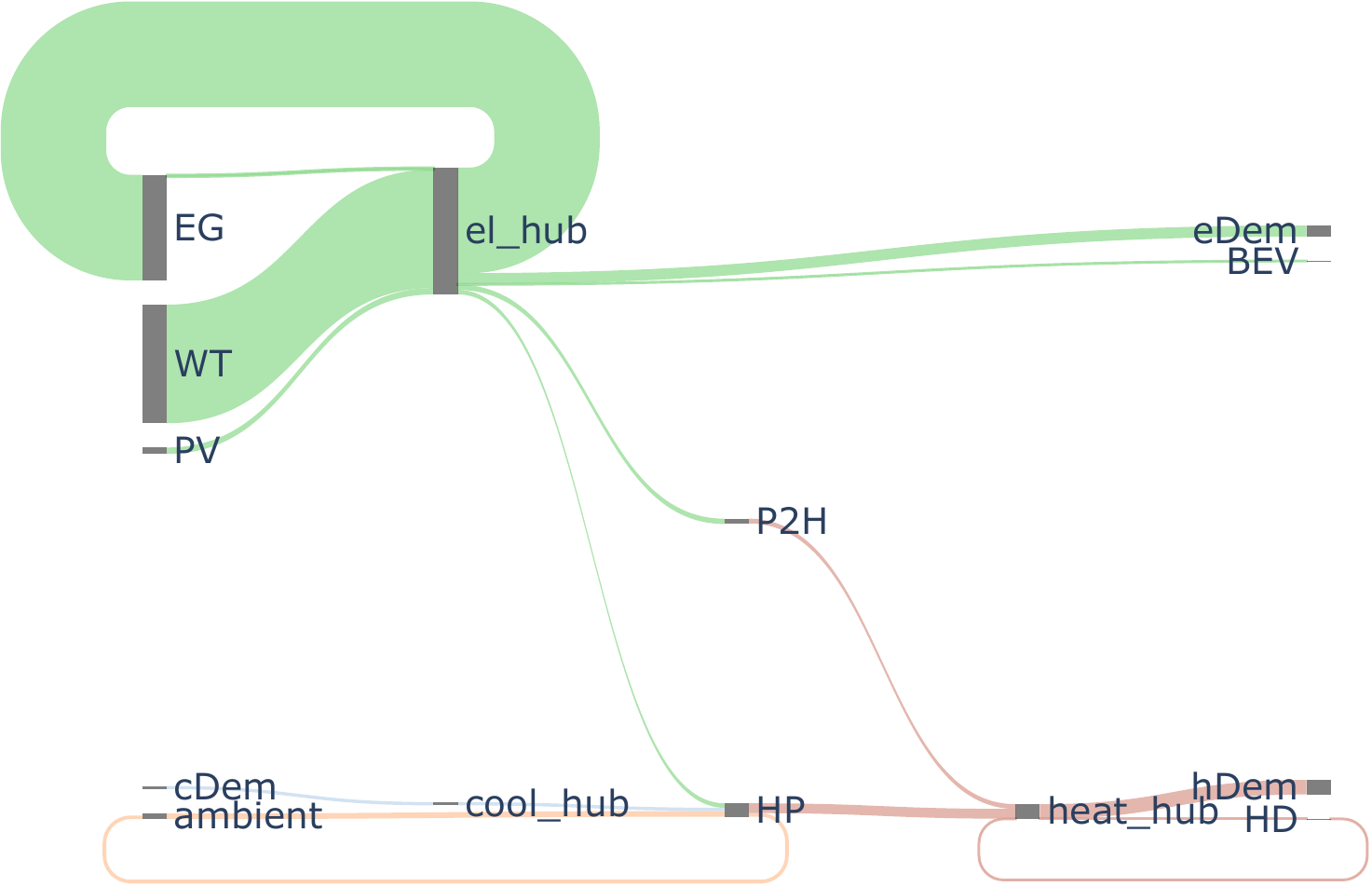} \put(75,25){\scenbf{no\_Flex\_d}} \end{overpic}
	\vskip 3mm
	\begin{overpic}[width=0.46\linewidth]{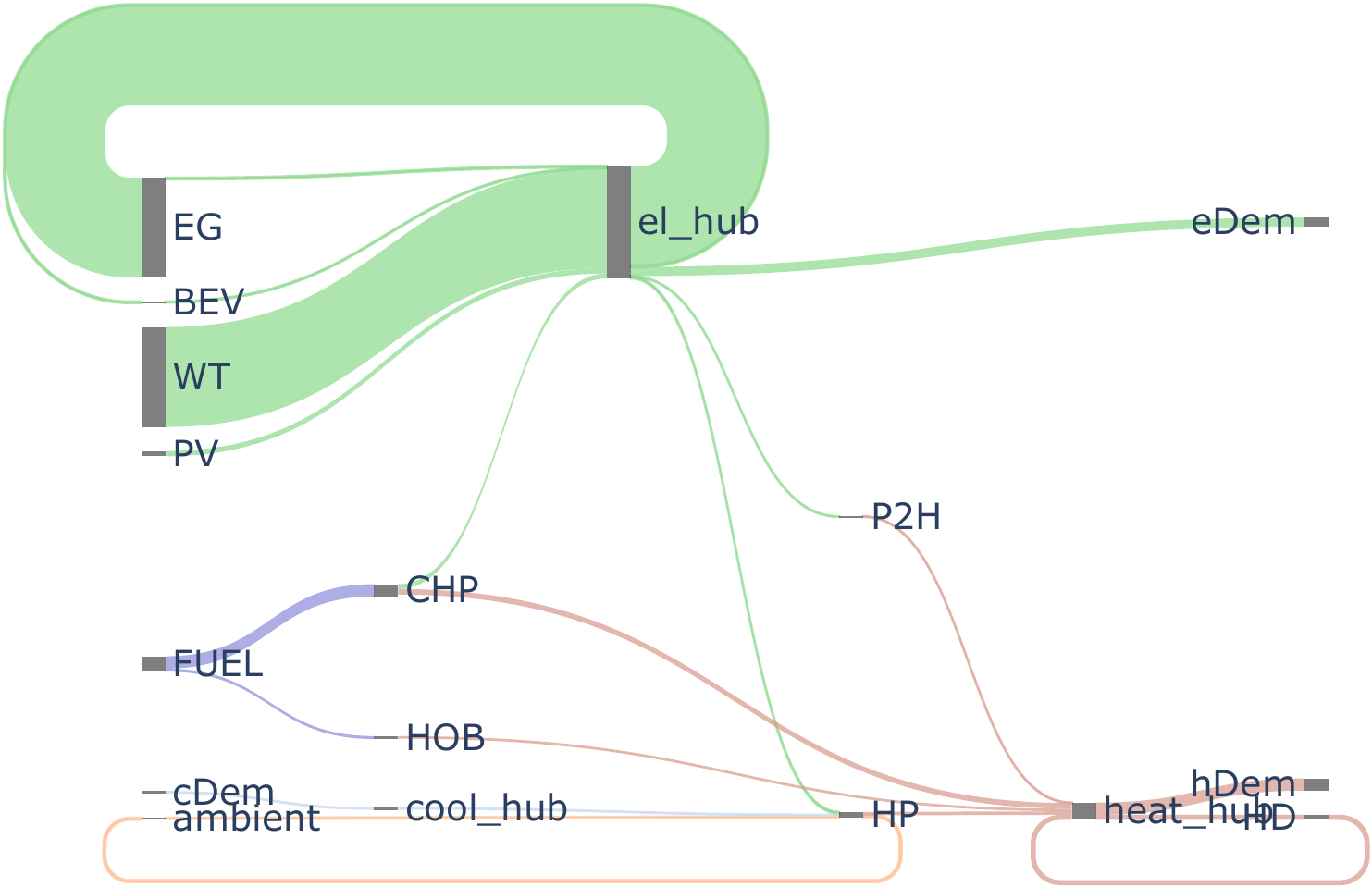} \put(75,25){\scenbf{someFlex}} \end{overpic}
	\begin{overpic}[width=0.46\linewidth]{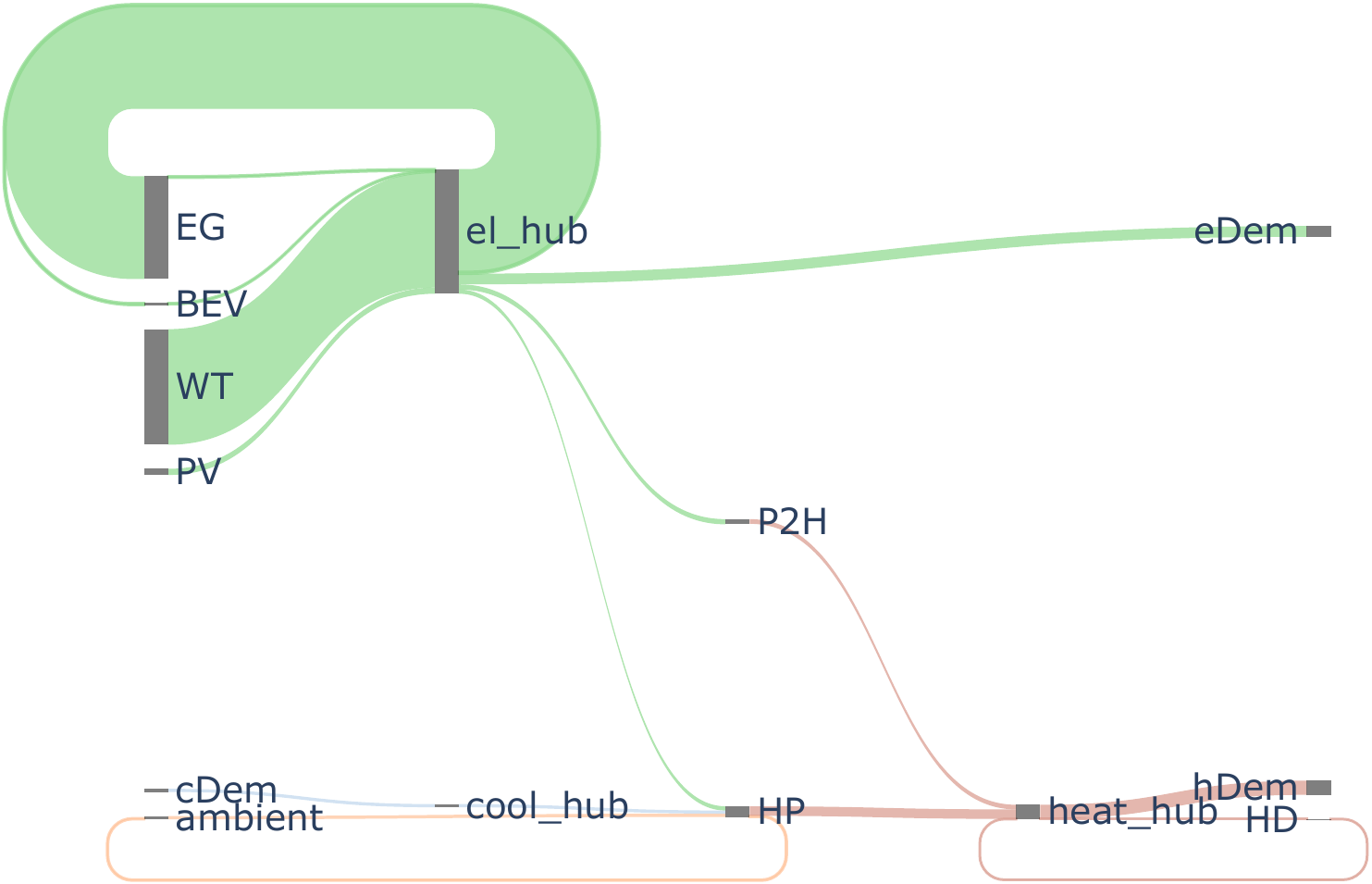} \put(75,25){\scenbf{someFlex\_d}} \end{overpic}
	\vskip 3mm
	\begin{overpic}[width=0.46\linewidth]{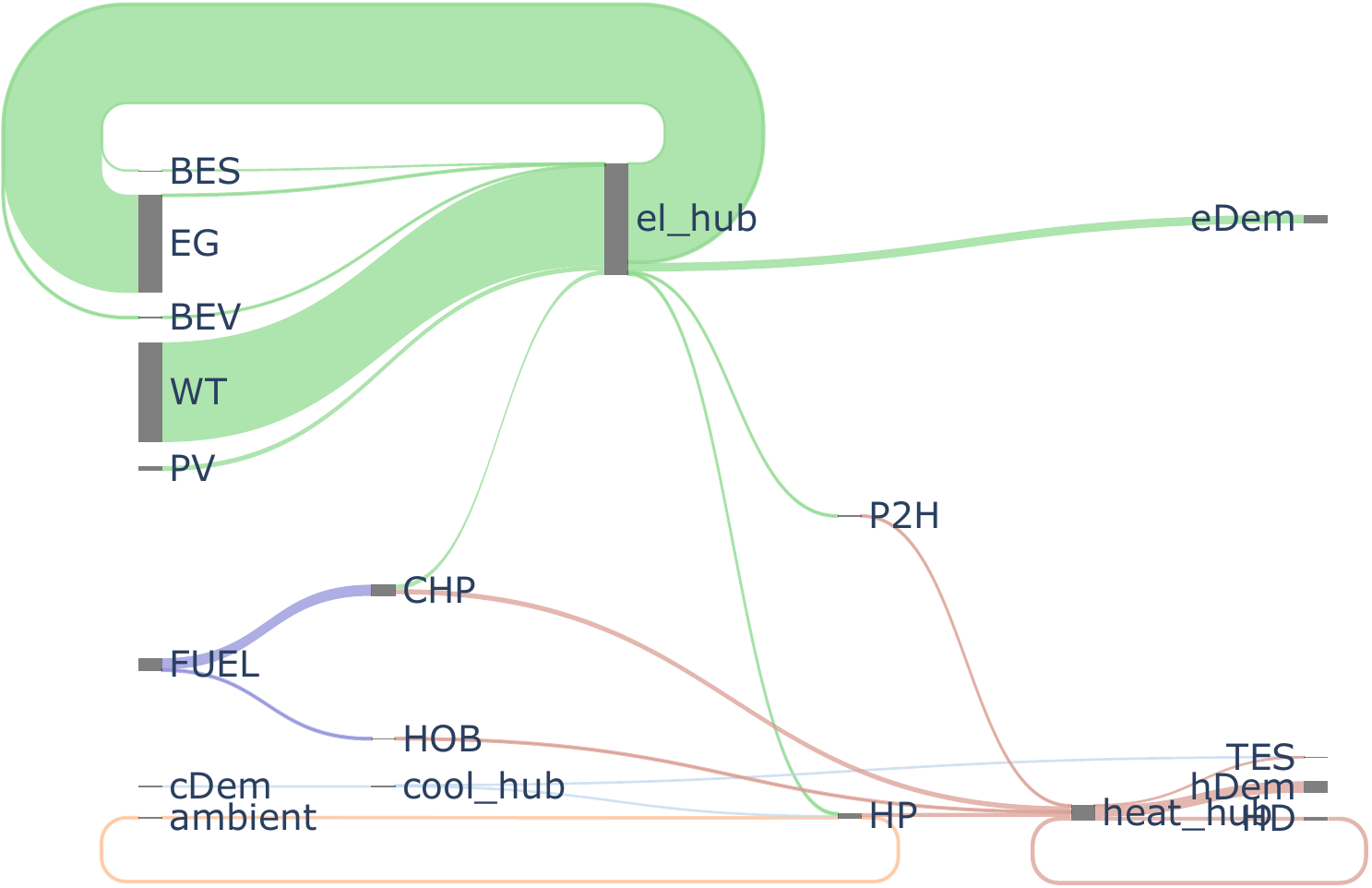} \put(75,25){\scenbf{fullFlex}} \end{overpic}
	\begin{overpic}[width=0.46\linewidth]{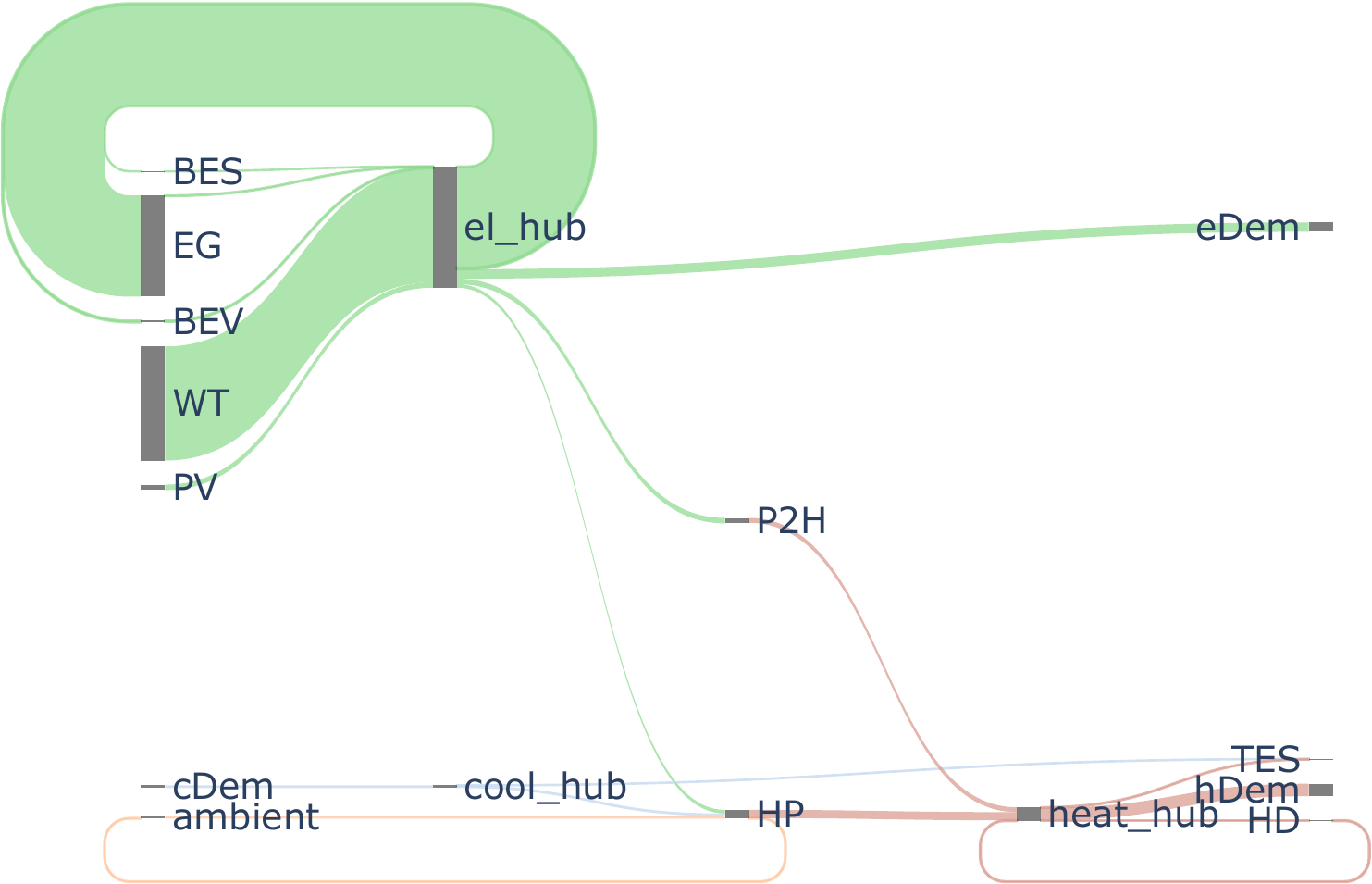} \put(75,25){\scenbf{fullFlex\_d}} \end{overpic}
	\caption{Sankey diagrams of annual energy sums for the \scen{c\_scaled} context.
		\label{fig:sankey_c_scaled}}
\end{figure*}